%% file: Supergauge18.tex
\documentclass[aps,12pt,tightenlinesletterpaper,superscriptaddress,secnumarabic,nofootinbib]{revtex4}
\usepackage[dvips]{graphicx}

\DeclareRobustCommand{\baselinestretch{1.0}}

\usepackage{amsmath}
\usepackage{bm}
\usepackage{epsfig}
\usepackage{verbatim}
\usepackage{psfrag}
\usepackage{bm}
\usepackage{cancel}

\allowdisplaybreaks[1]

\newcommand{\be}{\begin{equation}}
\newcommand{\ee}{\end{equation}}
\newcommand{\bea}{\begin{eqnarray}}
\newcommand{\eea}{\end{eqnarray}}
\newcommand{\ba}{\begin{array}}
\newcommand{\ea}{\end{array}}

\newcommand{\vect}[1]{\mathbf{#1}}

\newcommand{\abs}[1]{\left\lvert #1\right\rvert}
\newcommand{\boldgamma}{\mbox{\boldmath$\gamma$}}
\newcommand{\diracslash}[1]{#1\!\!\!/}
\newcommand{\pd}[2]{\frac{\partial #1}{\partial #2}}
\newcommand{\mcdot}{\!\cdot\!}

\begin{document}
\preprint{}
\vfill
\preprint{}
\begin{flushright}
{\tt NPAC-09-11}
\end{flushright}
\vspace{-3.0cm}
\title{\Large Supergauge interactions and electroweak baryogenesis}
\vspace{4.0cm}
\author{Daniel J. H. Chung}
\email{danielchung@wisc.edu}
\author{Bj\"orn Garbrecht}
\email{bjorn@physics.wisc.edu}
\affiliation{
{\small University of Wisconsin-Madison, Department of Physics} \\
{\small 1150 University Avenue, Madison, WI 53706, USA}}
\author{Michael. J. Ramsey-Musolf}
\email{mjrm@physics.wisc.edu}
\affiliation{
{\small University of Wisconsin-Madison, Department of Physics} \\
{\small 1150 University Avenue, Madison, WI 53706, USA}}
\affiliation{
{\small Kellogg Radiation Laboratory, California Institute of Technology} \\
{\small 1200 E. California Blvd., Pasadena, CA, 91125, USA}}
\author{Sean Tulin}
\email{tulin@caltech.edu}
\affiliation{
{\small California Institute of Technology} \\
{\small 1200 E. California Blvd., Pasadena, CA, 91125, USA}}

%\date{\today}
%%%%%%%%%%%%%%%%%%%%%%%%%%%%%%%%%%%%%%%%%%%%%%%%%%%%%%%%%%%%%%%%%%%%
\begin{abstract}
We present a complete treatment of the diffusion processes for supersymmetric
electroweak baryogenesis that characterizes transport dynamics ahead of the phase
transition bubble wall within the symmetric phase. In particular,
we generalize existing approaches to distinguish between chemical
potentials of particles and their superpartners. This allows us
to test the assumption of superequilibrium (equal chemical potentials
for particles and sparticles) that has usually been made in
earlier studies. We show that in the Minimal Supersymmetric Standard Model, superequilibrium is generically
maintained -- even in the absence of fast supergauge interactions -- due to the
presence of Yukawa interactions. We
provide both analytic arguments as well as illustrative numerical examples.
We also extend the latter to regions where analytical
approximations are not available since down-type Yukawa couplings
or supergauge interactions only incompletely equilibrate. We further comment on
cases of broken superequilibrium wherein a heavy superpartner decouples from the electroweak 
plasma, causing a kinematic bottleneck in the chain of equilibrating reactions. Such situations may be relevant for baryogenesis within extensions of the MSSM.
We also provide a compendium of inputs required to characterize the symmetric phase transport dynamics. 

\end{abstract}
%%%%%%%%%%%%%%%%%%%%%%%%%%%%%%%%%%%%%%%%%%%%%%%%%%%%%%%%%%%%%%%%%%%%
\maketitle

\newpage
\tableofcontents
%%%%%%%%%%%%%%%%%%%%%%%%%%%%%%%%%%%%%%%%%%%%%%%%%%%%%%%%%%%%%%%%%%%%%
\newpage
\section{Introduction}
\label{sec:intro}
%%%%%%%%%%%%%%%%%%%%%%%%%%%%%%%%%%%%%%%%%%%%%%%%%%%%%%%%%%%%%%%%%%%%%
Electroweak baryogenesis (EWB) remains one of the most attractive and
testable scenarios for explaining the baryon asymmetry of the universe
(BAU). The BAU is characterized by the baryon number-to-entropy
density ratio $Y_B \equiv n_B/s$, where $n_B$ is the baryon number density and $s$ is the entropy density.
The value for $Y_B$ obtained from
analysis of light element abundances in the context of Big Bang
Nucleosynthesis (BBN) is consistent with the value extracted by the
WMAP collaboration from acoustic peaks in the cosmic microwave
background anisotropy:
\be
\begin{array}{ccccccc}
6.7 \times 10^{-11} &< & Y_B &< &9.2 \times 10^{-11} & (95\% \; \textrm{C.L.}) & \textrm{BBN} \; \textnormal{\cite{Yao:2006px}} \\
8.36 \times 10^{-11} &< &Y_B &< &9.32 \times 10^{-11} & (95\% \; \textrm{C.L.}) & \textrm{CMB} \; \textnormal{\cite{Yao:2006px,Dunkley:2008ie}} . \end{array}
\ee
As observed by
Sakharov~\cite{Sakharov:1967dj}, a particle physics explanation for
this tiny number, corresponding to roughly five percent of the cosmic
energy density, requires three ingredients (assuming a
matter-antimatter symmetric universe at the end of inflation): (1)
violation of baryon number, $B$; (2) violation of both $C$ and $CP$
symmetry; and (3) a departure from thermal equilibrium.\footnote{This
criterion can be evaded if $CPT$ invariance is broken.}

In EWB, these ingredients come into play if the scalar (Higgs) sector
of the theory gives rise to a strongly first order electroweak phase
transition (EWPT) at temperatures $T\sim 100$ GeV. Such a phase
transition proceeds via bubble nucleation, wherein regions of broken
electroweak symmetry emerge in a background of unbroken electroweak
symmetry. $C$ and $CP$-violating interactions of fields near the
bubble wall lead to the creation of left-handed charge that is injected
into the unbroken phase, where electroweak sphalerons convert it into
baryon number. The bubbles expand into regions of $n_B\not=0$,
freezing it in because sphaleron transitions are quenched in the
bubble interiors. A strong first order EWPT is required in order to
sufficiently quench the sphalerons inside the bubble, thereby
preventing wash out of the captured baryon number density.

Although the Standard Model (SM) in principle contains all the necessary
ingredients for EWB, the effects of SM $CP$-violation are too suppressed
to generate sufficient left-handed charge during the process of electroweak
symmetry-breaking. Moreover, the LEP II lower bound on the mass of the
SM Higgs boson, $m_h\geq 114.4$ GeV is too high to allow for a strong first
order EWPT~\footnote{Numerical studies indicate that in a SM universe,
electroweak symmetry breaking occurs through a smooth cross over rather
than through a phase transition~\cite{Laine:1998qk}.}. Consequently, EWB can be viable only
in the presence of new physics at the electroweak scale. In particular,
augmenting the scalar sector of the SM can lead to a strong first order
EWPT consistent with a SM-like Higgs scalar that is heavier than the direct
search lower bound. The possibilities for doing so encompass both
supersymmetric and non-supersymmetric scenarios, and in either case, searches
for new scalars at the Large Hadron Collider could provide important tests (for recent work, see Refs.~\cite{Menon:2009mz,Barger:2008jx,Noble:2007kk,Carena:2008mj,Barger:2007im,Balazs:2007pf,Profumo:2007wc} and references therein). Similarly, the presence of new $CP$-violating interactions,
the effects of which are not suppressed by light quark Yukawa couplings and small mixing
angles as it is the case for the CKM mechanism~\cite{Jarlskog:1985ht}, could lead to sufficient left-handed
charge generation during a first order EWPT. Experimental searches for the
permanent electric dipole moments (EDMs) of the electron, neutron and
neutral atoms with enhanced sensitivity could uncover the existence of such
interactions~\cite{Pospelov:2005pr,Erler:2004cx,Ramsey-Musolf:2006vr}.
In addition, even though other evidence for $CP$-violation may be
provided by B-physics \cite{Baek:1999qy,Hewett:2004tv} (although not
for the minimal field content of MSSM \cite{Murayama:2002xk}), a
direct measurement of the parameters fixing the relevant $CP$-violating
physics will most likely require a collider beyond LHC such as the ILC
\cite{Murayama:2002xk,Barger:2001nu}.
In light of these prospective experimental searches for the
ingredients needed for EWB, it is important to refine the theoretical
apparatus for relating their results to the baryon asymmetry.

The left-handed charge density, that biases weak sphaleron transitions
and that is therefore of central relevance for the computation
of the baryon asymmetry, is given by the sum
of all charge densities of left-handed quarks and leptons of
all generations,
\begin{equation}
n_{\rm left}=\sum\limits_{i=1}^3(q_i+\ell_i)\,.
\end{equation}
The number densities are understood to be the sum of both isospin components and
of the three colors for the quarks. Moreover, we use the word
{\it density} as a short hand
expression for charge number density (the zero component of the vector current, which is
the difference of particle and antiparticle
number densities), and denote the densities by the symbols that also 
represent the particular particles.
As outlined above, the left handed density gets converted into a baryon density
$n_B$ through weak sphaleron transitions. The following formula describes
baryon generation and washout ahead of the bubble wall~\cite{Huet:1995sh}:
\begin{equation}
\label{nB:sphal}
n_{B}=-3\frac{\Gamma_{\rm ws}}{v_{\rm w}}
\int\limits_{-\infty}^0 dz\; n_{\rm left}(z)
{\rm e}^{\frac{15}{4} \frac{\Gamma_{\rm ws}}{v_{\rm w}}z}\,,
\end{equation}
where $v_w$ is the bubble wall velocity and $z$ is the spatial coordinate perpendicular to the wall in the
frame where the wall is at rest. Negative values of $z$ correspond
to the symmetric electroweak phase (bubble exterior), positive values to the broken phase (bubble interior).
Because the weak sphaleron rate, $\Gamma_{\rm ws}$, is much slower than the rates for both the creation of $n_\mathrm{left}$ and its diffusion ahead of the bubble wall, 
application of Eq.~(\ref{nB:sphal})  is usually decoupled
from the network of diffusion equations, a simplification that we also
adapt here.
Eq.~(\ref{nB:sphal}) underlines the essential need of accurate theoretical methods
of determining $n_{\rm left}$ in order to make quantitative predictions for
$Y_B$.

The importance of diffusion for EWB has been
emphasized in Refs.~\cite{Cohen:1994ss,Huet:1995sh,Joyce:1994zn}.
Due to scatterings with the thermal bath, the $CP$-violating density
$n_{\rm left}$ is not only localized at the bubble wall, but it is also transported to the region
ahead of the bubble wall. Therefore, there remains a larger amount
of time for weak sphaleron processes to turn $n_{\rm left}$ into
the baryon asymmetry, before the it is captured by the bubble inside of which
sphaleron transitions are quenched.

The purpose of the present paper is twofold: First, we present a
more detailed discussion on the derivation of the diffusion equations
and the computation of the interaction rates that enter these.
This supplements our recent
publications~\cite{Chung:2008aya,Chung:2009cb}.
Second, we extend the network of diffusion equations to distinguish
between particle and sparticle chemical potentials.
In earlier publications (see Refs.~\cite{Huet:1995sh,Chung:2008aya,Chung:2009cb} and references therein), it has been assumed that the chemical
potentials for particles and their superpartners are identical,
a situation that we refer to as superequilibrium.
Here, we provide numerical
evidence that superequilibrium holds in most regions of parameter
space. On the other hand, the generalization presented here also allows 
for a computation of the baryon asymmetry in 
parametric regions where superequilibrium does not hold.
As for the source of $CP$-violation,
although we focus in this paper on the MSSM as an illustrative
case, our methods can be applied to any supersymmetric scenario ({\em
e.g.}, the \lq\lq next-to-minimal" Supersymmetric Standard Model).

\subsection{Existing approaches to diffusion}

In order to compute the left-handed charge,
we need to derive and solve a coupled
set of transport equations for the densities of
particles that couple directly or indirectly to the $CP$-violating
sources~\cite{Huet:1995sh,Joyce:1994zn}:
\be
\label{eq:trans1}
\partial_\mu\ j_r^\mu = -\sum_s \Gamma_{rs}\: n_s + S_r^{\;
\cancel{CP}}\ \ \ ,
\ee
where $j_r^\mu$ is the current density for
particle species $r$, $\Gamma_{rs}$ are transport coefficients that
couple the evolution of species $r$ to the number densities $n_s$
of other species $s$, and $S_r^{\; \cancel{CP}}$ is a $CP$-violating
source term for the species $r$.

In earlier treatments on diffusion for EWB, it is usually assumed that
the only relevant Yukawa coupling is the one of the top quark, as
it is much larger than Yukawa couplings of the first generations but also
much larger than third-generation couplings of down-type quarks and
leptons~\cite{Huet:1995sh}. While we agree with the general framework
for the diffusion equations and the strategies for analytical solutions
that is described Ref.~\cite{Huet:1995sh} and followed in most subsequent work,
in two recent publications~\cite{Chung:2008aya,Chung:2009cb},
we have shown that the ratio between the Yukawa
couplings does not directly answer the question of their relevance.
Rather, the  timescale associated with the interactions induced by these Yukawa couplings
has to be compared to the inverse diffusion length
$\Gamma_{\rm diff}^{-1}$ that is characteristic
for the EWPT. In supersymmetric scenarios, the Yukawa couplings of down-type fermions and their superpartners grows with $\tan\beta$, thereby enhancing the equilibration rate for these reactions relative to diffusion.
We find that for $\tan\beta\stackrel{>}{{}_\sim}5$
($\tan\beta\stackrel{>}{{}_\sim}15$) interactions between bottom- \mbox{($\tau$-)}~particles and the Higgs sector can in general equilibrate on diffusion
time-scales.
This observation induces important qualitative and quantitative changes
to the description of the diffusion process in EWBG:
\begin{itemize}
\item
When bottom quark Yukawa couplings are in equilibrium, no net
chemical potential associated with the axial charge of left handed third generation fermions arises. As a consequence, the production of
densities of first generation quark densities through strong sphaleron
processes (thermal ${\rm SU}(3)$ instantons) is suppressed.
\item
The sign of the baryon asymmetry depends on the sparticle mass spectrum
and on $\tan\beta$, and it is therefore not uniquely given in terms
of the $CP$-violating phase. In particular, in parametric regions where
$\tau$-Yukawa couplings are negligible but bottom-Yukawa interactions
equilibrate, the sign changes according to whether the right-handed sbottom-particle is heavier than the right handed stop or not.
\item
When also $\tau$-Yukawa interactions are in equilibrium, there are
substantial contributions from third generation leptons to
$n_{\rm left}$.
\end{itemize}

\subsection{Supergauge interactions and diffusion}

The particular step towards a complete treatment of the diffusion
dynamics in the symmetric phase that we take in this paper is to
generalize the diffusion equations to allow for different chemical
potentials for particles and their superpartners, thereby accounting for
possible deviations from superequilibrium.
This is of importance because
in supersymmetric scenarios the
$CP$-violating sources generally involve interactions between
supersymmetric particles and the space-time varying Higgs vacuum
expectation values, leading to non-vanishing densities for the
superpartners. Since the electroweak sphalerons feed on a net left-handed
charge for quarks and leptons, supersymmetric interactions must
efficiently transfer the non-vanishing superpartner densities into an
asymmetry involving left-handed SM fermions.

In the MSSM, the most important source of left-handed charge is CP-violation in the Higgsino-gaugino sector that gives rise to a non-vanishing Higgsino density (see, {\em e.g.}, Refs.~\cite{Riotto:1998zb,ewbstop,Lee:2004we} and references therein). The source $S_{\tilde H}^{\; \cancel{CP}}$ requires a non-vanishing phase between the supersymmetric $\mu$ parameter and the SUSY-breaking gaugino mass parameters, $M_{1,2}$. Its magnitude is largest when the difference between $\mu$ and either $M_1$ or $M_2$ is small compared to their magnitudes, leading to so-called resonant electroweak baryogenesis. In previous work, it is usually assumed that supergauge interactions , such as ${\widetilde H} \, {\widetilde W}\leftrightarrow H_{u,d}$, are sufficiently fast, that once $S_{\tilde H}^{\; \cancel{CP}}$ generates a non-vanishing Higgsino density, the latter immediately converts into a non-vanishing Higgs boson density. Under this assumption of gaugino-mediated superequilibrium, one may work with a total density $H$ for the Higgs bosons and their superpartners. Yukawa interactions convert the $H$ density into that for SM quarks and leptons, which are again assumed to be in equilibrium with their superpartners.

These assumptions are indeed well justified when the supergauge interactions are in equilibrium. By not distinguishing
between chemical potentials of particles and their
superpartners, superequilibrium has been implicitly
imposed in producing the numerical results presented in
Refs.~\cite{Chung:2008aya,Chung:2009cb}. In this paper, we
do distinguish between particle and sparticle chemical potentials
and take accurate account of the finite interaction rates that
tend to establish superequilibrium. In doing so, we show that:
\begin{itemize}
\item  Superequilibrium holds in most of the relevant MSSM parameter space. We produce both analytic arguments to 
illustrate the reasons why and numerical studies for parameter space regions where the analytic arguments break down.
\item Superequilibrium yet may be maintained
when supergauge interactions are slow, either because the gauginos
are heavy and decouple from the plasma or because the corresponding three-body interactions
are kinematically forbidden.  This preservation of superequilibrium arises through a chain of reactions involving Yukawa interactions.
\item In the gaugino decoupling regime, the chain of Yukawa reactions may be broken due to the presence of addtional heavy (s)particles, leading to a departure from superequilibrium. The assumption of superequilibrium in this case may lead to an unrealistic prediction for $Y_B$. Such scenarios may be relevant in extensions of the MSSM that do not require light gauginos for the existence of significant CP-violating sources in the transport equations (\ref{eq:trans1}).
\end{itemize}

%As a result, we produce the numerical
%evidence that superequilibrium holds in most of the relevant MSSM parameter space.
%In addition, we point out that 
%By relaxing we present here the numerical
%evidence that this procedure is justified.
%
%
%, leading to the requisite left-handed charge asymmetry. The %corresponding transport equation for $H$ thus has the simple form
%\be
%\label{eq:higgstrans1}
%\partial_\mu j^\mu_H = -\Gamma_H \frac{H}{k_H} -\Gamma_Y\left( \frac{H}{k_H}-\frac{T}{k_T} +\frac{Q}{k_Q}\right) + S_{\tilde H}^{\; \cancel{CP}} +\cdots
%\ee
%where $Q$ and $T$ and the number densities for the third generation left- and right-handed quark supermultiplets, respectively, $k_{H,Q,T}$ are statistical weights, and the $+\cdots$ denote terms involving other quark and lepton supermultiplets that are suppressed by their Yukawa couplings. The term proportional to $\Gamma_H$ favors a relaxation of any non-vanishing Higgs supermultiplet density, while the term proportional to $\Gamma_Y$ induces a transfer of that density into the corresponding quark supermultiplet densities. Analogous transport equations apply to the quark supermultiplet densities. 

\subsection{Outline of this paper}

The plan of this paper is as follows: Section~\ref{sec:DiffTran},
culminates in the full network of Boltzmann equations that
describe diffusion, which
we present in Section~\ref{sec:Boltzmann}. Leading to this,
we discuss how these equations may be derived within the
closed time path formalism (Section~\ref{Sec:ThreeBody}),
list the relevant interactions within the MSSM
(Section~\ref{section:InteractionLagrangian}), introduce
the fully thermally averaged three-body supergauge interactions
(Section~\ref{section:Supergauge}), discuss the thermally averaged
Yukawa and triscalar interactions (Section~\ref{section:YukTri})
and the particular source and relaxation rates for the
asymmetry (Section~\ref{sec:sourcerelax}).
Additional inputs needed are thermal masses
(Section~\ref{sec:ThMass}) and diffusion constants
(Section~\ref{sec:DiffCon}).
Some simplifications in
the approximation of thermal effects in the averaged
interaction rates are discussed and justified in
Section~\ref{sec:particlehole}.

In Section~\ref{sec:analytic}, we
present approximate analytical solutions to the Boltzmann equations.
While in Section~\ref{Sec:Analytical}, a brief review of the
discussion in Refs.~\cite{Chung:2008aya,Chung:2009cb} is provided,
in Section~\ref{section:superconditions} we go beyond that and
show how superequilibrium can be maintained even in case when supergauge
interactions are quenched (e.g. through large gaugino masses or for
kinematic reasons).

In Section~\ref{sec:numerical}, we provide the numerical evidence for
the preceding discussions. An illustrative point in parameter space
is presented, where the analytical approximation is justified and
yields reasonably accurate predictions for the densities ahead of the
wall (Section~\ref{sec:NumYukSGEq}). By variation of $\tan\beta$,
the effect of changing the strength of down-type Yukawa couplings is
investigated in Section~\ref{sec:NumTanBeta}. We refer the reader to Fig.~\ref{fig:supergauge}  which illustrate this $\tan\beta$-dependence on the relationships between chemical potentials for particles and their superpartners, and to Fig.~\ref{fig:yukawa}, which shows the corresponding impact on the relationships between left- and right-handed Standard Model fermion chemical potentials. These figures also illustrate the impact of Yukawa-induced superequilibrium that would persist if the supergauge interaction rates were set to zero. How superequilibrium
is broken and how it can be maintained in the absence of supergauge 
interactions is exemplified in Section~\ref{sec:NumNoSG}. Fig.~\ref{fig:BAU:TANBETA} summarizes the final impact on the baryon asymmetry of several of these features. There we give $Y_B$ as a function of $\tan\beta$ that results from the full computation with our benchmark input parameters and compare to the results that would have been obtained had we neglected the presence of supergauge interactions, the third generation lepton contributions, or departures from superequilibrium that arise at large $\tan\beta$.
Conclusions are drawn in Section~\ref{sec:conclude}.

\section{Diffusion transport equations}
\label{sec:DiffTran}

In this Section, we discuss in detail the derivation of diffusion
transport equations. In comparison to earlier treatments, we generalize
these equations to distinguish between particle densities and the
densities of their superpartners.

\subsection{Three-body rates: general formalism}
\label{Sec:ThreeBody}

We derive the diffusion transport equations for EWB using the closed time path (CTP) Schwinger-Dyson equations.
Although one may use conventional kinetic theory for this purpose,
we adopt the CTP framework as it allows one to
systematically include higher-order corrections and the effects associated with
departure from adiabatic quantum evolution (for a detailed review of the CTP
framework as applied to EWB, see our earlier work in
Refs.~\cite{Lee:2004we,Cirigliano:2006wh}). The CTP transport equations
for bosons and fermions have the form~\cite{kadanoff-baym,Riotto:1998zb}:
\begin{equation}
\begin{split}
\pd{n_B}{X_0} (X) +{\mbox{\boldmath$\nabla$}}\mcdot\vect{j}_B(X) 
= \int d^3 z\int_{-\infty}^{X_0} dz_0\
\Bigl[ \Sigma_B^>(X,z) G^<(z,X)&-G^>(X,z)\Sigma_B^<(z,X)\\
+G^<(X,z) \Sigma_B^>(z,X) &- \Sigma_B^<(X,z) G^>(z,X)\Bigr]\,,
\label{eq:scalar1}
\end{split}
\end{equation}
\begin{equation}
\begin{split}
\pd{n_F}{X_0} (X) + {\mbox{\boldmath$\nabla$}}\mcdot\vect{j}_F(X) =  
-\int d^3 z\int_{-\infty}^{X_0} dz_0\
{\rm Tr}\Bigl[ \Sigma_F^>(X,z) S^<(z,X)&-S^>(X,z)\Sigma_F^<(z,X)\\
+S^<(X,z) \Sigma_F^>(z,X) &- \Sigma_F^<(X,z) S^>(z,X)\Bigr]\,,
\label{eq:fermion1}
\end{split}
\end{equation}
where $j^\mu_B=(n_B, {\mbox{\boldmath$j_B$}})$ and $j^\mu_F=(n_F, {\mbox{\boldmath$j_F$}})$ are the boson and fermion current densities, respectively. The functions $G^{>,\ <}(x,y)$ and $S^{>,\ <}(x,y)$ are elements of the $2\times 2$ matrix of CTP boson and fermion propagators:
\begin{eqnarray}
G^{ab}(x,y) &=& \langle  T_{\cal P}\, \left[ \phi^a (x) \phi^{b\, \dagger} (y) 
\right] \rangle\,,
\label{eq:gfb}
\\
S^{ab}(x,y) &=& \langle  T_{\cal P}\, \left[ \psi (x)^a \overline{\psi}^b (y) 
\right] \rangle\,,
\label{eq:gff}
\end{eqnarray}
where $\langle ... \rangle $ denotes an average over the physical state
of the system, $T_{\cal P}$ is a path ordering operator, and the indices $a,b$ denote the branch of a closed time integration path running from $-\infty$ to $+\infty$ (the \lq\lq $+$" branch) and back to $-\infty$ (the \lq\lq $-$" branch) the fields inhabit. In the case of the bosonic Greens functions, one has
\begin{eqnarray}
G^{++}(x,y) &\equiv&  G^{t}(x,y) =
 \langle  T \, \left[ \phi (x) \phi^\dagger (y) \right] \rangle\,,
\\
G^{+-}(x,y) &\equiv&  G^<(x,y) =
 \langle \phi^\dagger (y) \phi(x) \rangle\,, 
\\
G^{-+}(x,y) &\equiv&  G^>(x,y) =
\langle \phi (x) \phi^\dagger (y) \rangle\,, 
\\
G^{--}(x,y) &\equiv&  G^{\bar{t}}(x,y) =
 \langle \bar{T} \, \left[ \phi (x) \phi^\dagger (y) \right] \rangle \,,
\end{eqnarray}
while the corresponding expressions for the elements of $S^{ab}(x,y)$ contain the appropriate factors of $-1$ to account for fermion anti-commutation relations. The self energy functions $\Sigma_B^{>,\ <}(x,y)$ and $\Sigma_F^{>,\ <}(x,y)$ give the corresponding one particle irreducible corrections to the free inverse CTP propagators. 

The source terms on the RHS of Eqns.~(\ref{eq:scalar1}) and (\ref{eq:fermion1}) can be obtained by computing $\Sigma_B^{>,\ <}(x,y)$ and $\Sigma_F^{>,\ <}(x,y)$ order-by-order in perturbation theory. In general, doing so requires knowledge of the non-equilibrium distribution functions. However, the presence of a hierarchy of scales allows one to expand these functions about their equilibrium values in powers of appropriate scale ratios. As discussed in Refs.~\cite{Lee:2004we,Cirigliano:2006wh}, these scales include a decoherence time, $\tau_d$, associated with the departure from adiabatic dynamics; a plasma time, $\tau_p$, associated with mixing between degenerate states in the plasma; and an intrinsic quasiparticle evolution time, $\tau_\mathrm{int}$, associated with the time evolution of a state of a given energy. For the dynamics of the electroweak plasma, one finds that $\tau_\mathrm{int} \ll \tau_p \ll \tau_d$, leading to a natural expansion in the scale ratios: $\varepsilon_d \equiv \tau_\mathrm{int}/\tau_d \sim v_w k_\mathrm{eff}/\omega$ and $\varepsilon _p \equiv \tau_\mathrm{int}/\tau_p\sim \Gamma_p/\omega$ with $v_w$ being the bubble wall expansion velocity, $k_\mathrm{eff}^{-1}$ being an effective length scale (such as the wall thickness), $\Gamma_p$ being a thermal quasiparticle \lq\lq damping rate" in the plasma, and $\omega$ being the quasiparticle frequency. Here, we include in the damping rate any an process involving emission and absorption from the thermal bath. In addition, the plasma is relatively dilute, so that the ratio of chemical potentials to temperature, $\varepsilon_\mu=\mu/T$, provides an additional expansion parameter. 

In terms of the $\varepsilon$ parameters, the leading contributions to the RHS of Eqns.~(\ref{eq:scalar1}) and (\ref{eq:fermion1}) occur at $\mathcal{O}(\varepsilon^2)$. Specifically, the $CP$-violating sources include effects of order $\varepsilon_d\times\varepsilon_p$ or $\varepsilon_d^2$, while the $CP$-conserving sources arise at order $\varepsilon_p\times\varepsilon_\mu$. In this context, the supergauge interactions generate terms of the latter type. As discussed in Refs.~\cite{Lee:2004we,Cirigliano:2006wh}, the presence of the scale hierarchies embodied in the $\varepsilon$ parameters allows us to adopt the quasiparticle ansatz for the CTP Green functions and to work near chemical and kinetic equilibrium when computing these terms.
To this end, we compute the the terms on the right hand side of
the transport equations~(\ref{eq:scalar1},\ref{eq:fermion1}) following the procedure used in Ref.~\cite{Cirigliano:2006wh} for the calculation of the $\Gamma_Y$-type terms. The bosonic CTP Green functions entering the computation are given by 
\begin{eqnarray}
G_i^> (x,y) &=& \int \frac{d^4 k}{(2 \pi)^4} \Big(1 + f_B(k_0,\mu_i) \Big)
\ \rho_i(k_0,{\bf k})\,,
\label{eq:gf>}
 \\
G_i^< (x,y) &=& \int \frac{d^4 k}{(2 \pi)^4} \  f_B(k_0,\mu_i) 
\ \rho_i(k_0,{\bf k})\,,
\label{eq:gf<}
\end{eqnarray}
with spectral functions 
\bea
\label{eq:spectral1}
 \rho_i({k}_0, \vect{k}) &= & \pi/\omega_{\bf k} \left[ \delta(k^0 - \omega_{\bf k})
- \delta(k^0 + \omega_{\bf k})\right]\\ 
\nonumber
 &=&  {i\over 2\omega_k}\biggl[
\left({1\over
 {k}_0-\omega_k+i\epsilon}-{1\over {k}_0+\omega_k+i\epsilon}\right)
 -\left({1\over {k}_0-\omega_k-i\epsilon}-
{1\over {k}_0+\omega_k-i\epsilon}\right)\biggr]\  \ \ , 
\eea
where $\omega_{\bf k} = 
\sqrt{{|\bf k|}^2 + m^2}$. The $ \rho_i({k}_0, \vect{k})$  can be   
appropriately modified to take into account collision-broadening and
thermal masses. Using the expansion in $\varepsilon$ parameters introduced above, it suffices to take
the distribution functions to be close to the equilibrium
form
\begin{equation}
f_B(k_0,\mu_i) = n_B(k_0,\mu_i) + {\cal O}(\varepsilon_{\rm d}/\varepsilon_p) \,,
\label{eq:distr}
\end{equation}
where $ n_B (k_0,\mu_i) = 1/[e^{(k_0 - \mu_i)/T} - 1] $ and $\mu_i$ is a  local chemical potential .  Here we neglect the terms of order $\varepsilon_{\rm d}/\varepsilon_p\sim v_w k_\mathrm{eff}/\Gamma_p \ll 1$. 

Similar expressions are obtained for the fermion Green functions:
\bea
S^>(x,y)& = & \langle \psi(x){\bar\psi}(y)\rangle\,,\\ 
S^<(x,y) & = & -\langle {\bar\psi}(y)\psi(x)\rangle\,,
\eea
which  can be expressed 
as 
\be
\label{eq:slambdafree}
S^\lambda(x,y)=\int {d^4k\over (2\pi)^4} e^{-i{k}\cdot(x-y)}
f_F^\lambda({k}_0,\mu)\rho(k_0,\vect{k})\left(\diracslash{k}+m\right)
\,,
\ee
where $\lambda$ denotes either ``$>$" or ``$<$" and 
and with the functions
\begin{subequations}
\begin{align}
f_F^>(k_0,\mu)&=1-n_F(k_0-\mu)\,, \\
f_F^<(k_0,\mu)&= -n_F(k_0-\mu)\,,
\end{align}
\end{subequations}
with $ n_F (k_0,\mu_i) = 1/[e^{(k_0 - \mu_i)/T} + 1] $.

\subsection{MSSM interaction Lagrangian}
\label{section:InteractionLagrangian}

In order to calculate the transport coefficients in the Boltzmann equations, we first identify the relevant interactions in the MSSM Lagrangian.  These interactions, denoted by $\mathcal{L}_{int}$, can be divided into three classes:
\be
\mathcal{L}_{int} = \mathcal{L}_{M} + \mathcal{L}_{Y} + \mathcal{L}_{\widetilde{V}} \;.
\ee
Bilinear interactions that arise when the neutral Higgs bosons acquire vacuum expectation values (vevs) are
\begin{align}
\mathcal{L}_M = & - y_t \: \widetilde{t}_R^* \:  \widetilde{t}_L \: \left(A_t \: v_u + \mu^* v_d \right) - y_t \: v_u \: \bar{t}_R \: P_L \: t_L  \label{eq:LintM} \\
&- y_b\: \widetilde{b}_R^*\: \widetilde{b}_L\:
\left(
A_b \: v_d + \mu^* v_u
\right)
-y_b \:v_d\: \bar{b}_R\: P_L\: b_L
\notag\\
&- y_\tau\: \widetilde{\tau}_R^*\: \widetilde{\tau}_L\:
\left(
A_\tau \: v_d + \mu^* v_u
\right)
-y_\tau \:v_d\: \bar{\tau}_R\: P_L\: \tau_L
\notag\\
& - \frac{g_1}{\sqrt{2}} \:  \bar{\Psi}_{\widetilde{H}^0} (v_d \: P_L - e^{i\phi_\mu^{M_1}} v_u \: P_R ) \Psi_{\widetilde{B}}  - \frac{g_2}{\sqrt{2}} \:  \bar{\Psi}_{\widetilde{H}^0} (v_d \: P_L + e^{i\phi_\mu^{M_2}} v_u \: P_R ) \Psi_{\widetilde{W}^0} \notag \\
& - g_2 \:  \bar{\Psi}_{\widetilde{H}^+} (v_d \: P_L + e^{i\phi_\mu^{M_2}} v_u \: P_R ) \Psi_{\widetilde{W}^+} +  \:\textrm{h.c.} \notag
\end{align}
These terms, which result in squarks, quarks, sleptons, leptons and
Higgsinos scattering due to
the spacetime-dependent Higgs vevs $v_u(x)$ and $v_d(x)$, contribute
to $CP$-violating sources $S^{\;\cancel{CP}}$ and $CP$-conserving relaxation
rates $\Gamma_M$ and $\Gamma_H$. The calculation of $S^{\: \cancel{CP}}$
has received the most attention, both in the CTP approach and in other
frameworks; however, there remains a significant dispersion in the recent
calculations of separate
groups~\cite{Lee:2004we, Carena:2002ss, Carena:2000id, Cline:2001rk,Konstandin:2005cd}.
The $CP$-conserving relaxation rates have been estimated in Ref.~\cite{Huet:1995sh},
and rigorously calculated and studied with CTP methods  in Ref.~\cite{Lee:2004we} in a manner consistent with the computation of the $CP$-violating sources.

Trilinear interactions proportional to the top Yukawa coupling $y_t$ are
\begin{align}
\mathcal{L}_{y_t} = & -y_t \: \widetilde{t}_R^* \:  \widetilde{t}_L \: \left(A_t \: H_u^0 + \mu^* H_d^{0*} \right) + y_t \: \widetilde{t}_R^* \:  \widetilde{b}_L \: \left( A_t \: H_u^+ - \mu^* H_d^{-*} \right) \label{eq:LintYuk}\\
&  + y_t \: \left( H_u^+ \: \bar{t}_R \: P_L \: b_L - H_u^0 \: \bar{t}_R \: P_L \: t_L \right) \notag \\
& + y_t \: e^{i \phi_\mu} \: \left(\widetilde{t}_R \: \bar{t}_L \: P_R \: \Psi_{\widetilde{H}^0}^C + \widetilde{t}_R \: \bar{b}_L \: P_R \: \Psi_{\widetilde{H}^+}^C \right) \notag \\ 
& + y_t \: e^{-i \phi_\mu} \: \left(\widetilde{t}_L \: \bar{t}_R \: P_L \: \Psi_{\widetilde{H}^0} + \widetilde{b}_L \: \bar{t}_R \: P_L \: \Psi_{\widetilde{H}^+} \right) \; + \; \textrm{h.c.} \; . \notag
\end{align}
For the bottom Yukawa coupling $y_b$ the corresponding interactions are
\begin{align}
\mathcal{L}_{y_{b}}=&
-y_b\:\widetilde{b}_R^*\: \widetilde{b}_L\:
\left(
A_b\: H_d^0 + \mu^* H_u^{0*}
\right)
+y_b\:\widetilde{b}_R^*\: \widetilde{t}_L\:
\left(
A_b\: H_d^- - \mu^* H_u^{+*}
\right)
\\
&+ y_b\:\left(
H_d^-\:\bar{b}_R\: P_L \: t_L  -  H_d^0\:\bar{b}_R\: P_L\: b_L
\right)
\notag\\
&+ y_b\:\left(
-\widetilde{b}_R \: \bar{b}_L \: P_R \: \Psi_{\widetilde{H}^0}
+\widetilde{b}_R \: \bar{t}_L \: P_R \: \Psi_{\widetilde{H}^+}
\right)
\notag\\
&+ y_b\:\left(
\widetilde{b}_L \: \bar{b}_R \: P_L \: \Psi_{\widetilde{H}^0}^C
-\widetilde{t}_L \: \bar{b}_R \: P_L \: \Psi_{\widetilde{H}^+}^C
\right)
\; + \; \textrm{h.c.} \; , \notag
\end{align}
where we have employed notations and rephasing conventions according to
Ref.~\cite{Lee:2004we}.
The corresponding interactions $\mathcal{L}_{y_{\tau}}$ for third generation
(s)leptons follow when replacing $b_R\to\tau_R$ and $t_L\to \nu_{\tau}$ in
$\mathcal{L}_{y_{b}}$.

In general, all interactions mediated by third generation Yukawa and triscalar
couplings may have a sizeable impact on the result of EWB, such that we consider
\begin{equation}
\mathcal{L}_Y=\mathcal{L}_{y_{t}}+\mathcal{L}_{y_{b}}+\mathcal{L}_{y_{\tau}}\,.
\end{equation}
These terms, which are Yukawa interactions and their supersymmetric counterparts,
lead to transport coefficients generically denoted $\Gamma_Y$.  These equilibration rates are
potentially important in that they communicate $CP$-asymmetries from the Higgs
sector to the quark sector, biasing sphalerons to produce a baryon asymmetry.
The dominant absorption/emission  contribution to $\Gamma_Y$ has been studied
within the CTP framework in Ref.~\cite{Cirigliano:2006wh}; the sub-dominant
scattering contribution to $\Gamma_Y$ has been partially calculated in
Refs.~\cite{Joyce:1994zn, Huet:1995sh}.

Supergauge interactions are
\begin{align}
\mathcal{L}_{\widetilde{V}} = & - \frac{g_1}{\sqrt{2}} \: \left[ \bar{\Psi}_{\widetilde{H}^+}  (H_d^{-*} \: P_L + e^{i\phi_\mu^{M_1}} H_u^+ \: P_R ) \Psi_{\widetilde{B}} +  \bar{\Psi}_{\widetilde{H}^0} (H_d^{0*} \: P_L - e^{i\phi_\mu^{M_1}} H_u^0 \: P_R ) \Psi_{\widetilde{B}} \right] \label{eq:LintVtilde} \\
& - \frac{g_2}{\sqrt{2}} \: \left[ \bar{\Psi}_{\widetilde{H}^+}  (- H_d^{-*} \: P_L + e^{i\phi_\mu^{M_2}} H_u^+ \: P_R ) \Psi_{\widetilde{W}^0} +  \bar{\Psi}_{\widetilde{H}^0} (H_d^{0*} \: P_L + e^{i\phi_\mu^{M_2}} H_u^0 \: P_R ) \Psi_{\widetilde{W}^0} \right]\notag \\
&- g_2 \: \left[ \bar{\Psi}_{\widetilde{H}^+}  (H_d^{0*} \: P_L + e^{i\phi_\mu^{M_2}} H_u^0 \: P_R ) \Psi_{\widetilde{W}^+} +  \bar{\Psi}_{\widetilde{W}^+} (H_d^{-*} \: P_L - e^{i\phi_\mu^{M_2}} H_u^+ \: P_R ) \Psi_{\widetilde{H}^0}^C \right]\notag \\
&- \frac{g_2}{\sqrt{2}} \: \left[ \widetilde{u}_L^{i*} \: \bar{\Psi}_{\widetilde{W}^0} \: P_L \: u_L^i - \; \widetilde{d}_L^{i*} \: \bar{\Psi}_{\widetilde{W}^0} \: P_L \: d_L^i
+\widetilde{\nu}_L^{i^*}\: \bar{\Psi}_{\widetilde{W}^0} P_L \: \nu_L^i
-\widetilde{e}_L^{i^*}\: \bar{\Psi}_{\widetilde{W}^0} P_L \: e_L^i
\right]
\notag \\
&- \frac{g_1}{3\sqrt{2}} \: \left[ \widetilde{u}_L^{i*} \: \bar{\Psi}_{\widetilde{B}} \: P_L \: u_L^i + \widetilde{d}_L^{i*} \: \bar{\Psi}_{\widetilde{B}} \: P_L \: d_L^i \right]
+\frac{g_1}{\sqrt{2}}\:\left[
\widetilde{\nu}_L^{i*}\bar{\Psi}_{\widetilde{B}} \: P_L \nu_L^i
+\widetilde{e}_L^{i*}\bar{\Psi}_{\widetilde{B}} \: P_L e_L^i
\right]
\notag \\
&- g_3 \sqrt{2} \: \left[ \widetilde{u}_L^{i*} \: \lambda^a \bar{\Psi}_{\widetilde{G}}^a \: P_L \: u_L^i + \widetilde{d}_L^{i*} \: \lambda^a \: \bar{\Psi}_{\widetilde{G}}^a \: P_L \: d_L^i \right]
\notag  \\
& - g_2 \: V_{ij}^* \: \widetilde{d}_L^{j*} \: \bar{\Psi}_{\widetilde{W}^+} \: P_L \: u_L^i - g_2 \: V_{ij} \: \widetilde{u}_L^{i*} \: \bar{\Psi}_{\widetilde{W}^+}^C \: P_L \: d_L^j
-g_2\:\widetilde{e}_L^{i*}\bar{\Psi}_{\widetilde{W}^+} \: P_L \: \nu_L^i
-g_2\:\widetilde{\nu}_L^{i*}\bar{\Psi}_{\widetilde{W}^+}^C \: P_L \: e_L^i
\notag \\
& - g_3 \sqrt{2} \: \left[ \widetilde{u}_R^{i} \: \bar{\lambda}^a \bar{u}_R^i P_L \Psi_{\widetilde{G}}^a + \widetilde{d}_R^{i} \: \bar{\lambda}^a \bar{d}_R^i P_L \Psi_{\widetilde{G}}^a \right]
\notag \\
& + \; \frac{2\sqrt{2}}{3} \: g_1 \: \left[ \widetilde{u}^i_R \: \bar{u}^i_R \: P_L \: \Psi_{\widetilde{B}} \right] - \frac{\sqrt{2}}{3} \: g_1 \: \left[ \widetilde{d}^i_R \: \bar{d}^i_R \: P_L \: \Psi_{\widetilde{B}} \right]
+\sqrt2\: g_1 \:\left[
\widetilde{e}_R^i\:\bar{e}_R^i\:P_L\:\Psi_{\widetilde{B}}
\right]
\: + \: \textrm{h.c.} \;\notag
,
\end{align}
where $\lambda^a$ and $\bar \lambda^a$ are the generators of $\mathbf{3}$
and $\mathbf{\bar 3}$ of ${\rm SU}(3)$.
These supergauge interactions -- the supersymmetric version of gauge interactions in the SM -- lead to transport coefficients, generically denoted by $\Gamma_{\widetilde{V}}$, which tend to equilibrate the chemical potentials for particles and their superpartners.  All previous studies have assumed the limit $\Gamma_{\tilde{V}} \to \infty$, which leads to superequilibrium.  One of the purposes of the present work is to calculate $\Gamma_{\widetilde{V}}$ and to solve the Boltzmann equations without the assumption of superequilibrium.

In employing these interactions to compute the supergauge equilibration rates, we work with the mass eigenstates of the unbroken phase: gauginos and Higgsinos (rather than charginos and neutralinos), left- and right-handed quarks and squarks, and Higgs scalars. Deep inside the bubble, this choice is clearly not appropriate, owing to large flavor mixing induced by the non-zero Higgs vevs. A proper treatment of this flavor mixing requires a modification of the transport equations that allows for an all-orders summation of the spacetime varying Higgs vevs~\cite{Konstandin:2005cd,resuminprogress}. In the absence of such a treatment, we will work in the \lq\lq vev insertion" approximation, defined by assuming that the dynamics of chiral charge production are dominated by the region near the phase boundary and that the Higgs vevs in this region are small compared to the temperature and slowly varying ({\em i.e.}, the wall is relatively thick). Under these assumptions one may treat the flavor mixing perturbatively. We find below that the particle densities are generally largest in magnitude near the phase boundary as one would expect if the vev insertion approximation is valid. Nonetheless, we emphasize that our specific numerical conclusions are provisional and await a more complete treatment of the flavor mixing dynamics in the broken phase within the bubble. 

Within the spectrum of unbroken phase eigenstates, the Higgs scalars
provide additional complications.
%Clearly, if we diagonalize the
%finite-temperature Higgs mass matrix, some degrees of freedom may be naively
%tachyonic, due to the fact that
%electroweak symmetry is broken at $T=0$.
To appreciate the difficulties more clearly, suppose for simplicity
the lightest Higgs mass eigenvector field $\phi$ is a fixed linear
combination of $H_u$ and $H_d$ as we pass through the bubble wall.
Then the naive dispersion relationship for $\phi$ modes in the WKB
approximation will be
\begin{equation}
\omega_{\bf k}= \sqrt{|{\bf k}|^2 + V^{\prime\prime}(\phi)}
\label{eq:naivedispersion}
\end{equation}
where $V(\phi)$ is the thermal effective potential and $v(x)=\langle
\phi \rangle$ solves the Euler-Lagrange equations involving
$V^{\prime}(\phi)$ leading to the cancellation of tadpoles.  For
states with sufficiently large momentum, the fact that
$V^{\prime\prime}(\phi)<0$ in a particular $\phi$ range will be
unimportant.  However, sufficiently soft modes will become unstable if
$V^{\prime\prime}(\phi)<0$ and may lead to significant backreaction
corrections to $v(x)$ on time scales of order $\textrm{Im} \,
{\omega_{\bf k}^{-1}}$ and can also lead to particle production
effects.  Note that this statement of the problem is a bit more subtle
than it seems because $v(x)$ is not spatially homogeneous while the
thermal effective potential $V(v)$ obtained by the usual construction
methods corresponds to the energy of the system with $v$ fixed to be
spatially homogeneous (e.g. see \cite{Weinberg:1987vp}).
%  Hence, any truly
%stable $v_{\mbox{stable}}(x)$ satisfying the Euler-Lagrange classical
%equation of the system can lead to a non-tachyonic dispersion
%relationship for the perturbations, and there can be a corresponding
%effective action containing an effective $V_{\mbox{stable}}(\phi)$
%with $V_{\mbox{stable}}^{\prime\prime}(\phi)>0$ after an appropriate field
%redefinition.  Another way to view this is that if we use an iterative
%approximation scheme to compute $v(x)$, we would start with an
%homogeneous effective potential $V(\phi)$, compute $v(x)$ using the
%Euler-Lagrange equations, compute thermal and quantum perturbations
%about $v(x)$, and then compute a back-reaction correction to the
%effective action -- not just the potential -- which presumably
%can lead to a $V_{\mbox{stable}}$.

In this paper, we will simply settle with an estimate of the error
incurred by neglecting this inhomogeneous bubble profile effect on the
Higgs transport equations, and leave a more detailed analysis to a
future work.\footnote{Any large effects coming from this is unlikely
to be computable analytically.}  If we denote the largest magnitude of
$V^{\prime\prime}(\phi)<0$ region when the temperature is near the critical
temperature $T_c$ as $k_c \equiv \max[|V^{\prime\prime}(\phi)|^{1/2}]$ , the
effect of neglecting of the $V^{\prime\prime}(\phi)<0$ region on the Higgs
transport equations can be estimated by the following fractional thermal
distribution number density:
\begin{equation}
\frac{n_{c}}{n}\approx\frac{H(k_{c}/T_c,m/T_{c})}{H(\infty,m/T_{c})}\,,
\end{equation}
where
\begin{equation}
H(k_{c}/T,m/T)\equiv\int_{m/T}^{\sqrt{k_{c}^{2}/T^{2}+m^{2}/T^{2}}}dxx\sqrt{x^{2}-m^{2}/T^{2}}\left[\exp\left(x\right)-1\right]^{-1}
\end{equation}
and where $m$ is the unbroken phase mass.  The function $n_{c}/n$ is a
monotonically increasing function with $k_{c}/T$ and a monotonically
decreasing function with increasing $m$. Although the exact value of
$k_{c}/T_{c}$ is model dependent, if we take $1\lesssim
k_{c}/T_{c}\lesssim2$ and take $m/T_c=0.7$, we arrive at fractional
correction estimate of\begin{equation}
0.09\lesssim\frac{n_{c}}{n}\lesssim0.35.\end{equation} Hence, we can
expect order 10\% of the Higgs scalar density will have significantly
distorted phase space distribution due to the instability of these
modes.  Although not completely negligible numerically, these
effects can be seen as refinements on the order unity effects
presented in this paper.

%In the absence of a complete solution to this issue, 
In the remainder of the paper, we will make the simplifying assumption
that quadratic fluctuations about the classical solution to the field
equations can be approximated by a set of scalar fields $(H_u^+,
H_u^0, H_d^-, H_d^0)$ with mass terms 
\be \mathcal{L} \supset -
\left(H_u^{+\dagger}, H_d^- \right) \left( \ba{cc} m_{H_u}^2 + |\mu|^2
+ \delta_u & b + \delta_b\\ b + \delta_b & m_{H_d}^2 + |\mu|^2 +
\delta_d \ea \right) \left( \ba{c} H_u^+ \\ H_d^{-\dagger} \ea \right)
\;, 
\ee 
and the same for $(H_u^0, H_d^{0\dagger})$ but with $b +
\delta_b \to - (b + \delta_b)$.  The terms $\delta_{u,d,b}$ denote
finite temperature corrections which lead to a minimum in the Higgs scalar potential at $v_u = v_d = 0$.  We
can re-express this mass matrix using the minimization conditions for
electroweak symmetry breaking at $T=0$~\cite{Martin:1997ns}:
\bea
m_{H_u}^2 + |\mu|^2 &=& m_A^2 \cos^2 \beta_0 + \frac{1}{2} m_Z^2 \cos
2\beta_0\,,
\\
m_{H_d}^2 + |\mu|^2 &=& m_A^2 \sin^2 \beta_0 - \frac{1}{2}
m_Z^2 \cos 2\beta_0\,,
\\
b &=& m_A^2 \sin\beta_0 \cos\beta_0 \,,
\eea
where $m_Z$ and $m_A$ are the $Z$ and pseudoscalar Higgs boson masses
at $T=0$ and
\be \tan\beta_0 \equiv \left. \frac{v_u}{v_d}
\right|_{T=0} \;.
\ee
Therefore, the mass term becomes
\be
\mathcal{L} \supset - \left(H_u^{+\dagger}, H_d^- \right) \left(
\ba{cc} m_A^2 \cos^2\beta_0 + \frac{1}{2} m_z^2 \cos 2\beta_0 +
\delta_u & \frac{1}{2} m_A^2 \sin 2\beta_0 + \delta_b \\ \frac{1}{2}
m_A^2 \sin 2\beta_0 + \delta_b & m_A^2 \sin^2\beta_0 - \frac{1}{2}
m_Z^2 \cos 2\beta_0 + \delta_d \ea \right) \left( \ba{c} H_u^+ \\
H_d^{-\dagger} \ea \right) \;.
\ee
The eigenvalues, corresponding to
the charged Higgs scalar masses in the unbroken phase, are
\be
m_{H_{1,2}}^2 = \frac{1}{2} \left[ m_A^2 + (\delta_u + \delta_d) \mp
\sqrt{ \left((m_A^2 + m_Z^2) \cos 2\beta_0 + (\delta_u - \delta_d)
\right)^2 + ( m_A^2 \sin 2\beta_0 + 2 \delta_b)^2 } \; \right] \;.
\ee
We see that if $\delta_{u,d,b} \to 0$, then $m_{H_1}^2 < 0$ --- a
consequence of the fact that electroweak symmetry is broken at low
temperatures.  Furthermore, the mixing angle $\alpha$, defined by
\be
\left( \ba{c} H_u^+ \\ H_d^{-\dagger} \ea \right) = \left( \ba{cc}
\cos\alpha & \sin\alpha \\ -\sin\alpha & \cos\alpha \ea \right) \:
\left( \ba{c} H_1^+ \\ H_2^+ \ea \right) \;,
\ee
is given by
\be
\label{HiggsMix}
\tan 2\alpha = \frac{m_A^2 \sin 2\beta_0 + 2 \delta_b}{m_A^2 \cos 2\beta_0
+ m_Z^2 \cos 2\beta_0 + \delta_u - \delta_d } \;.
\ee
(The
neutral Higgs mass matrix differs from the charged Higgs mass matrix
only by $b + \delta_b \to - (b+\delta_b)$; the mass eigenvalues are
the same and the mixing angle differs by an overall sign). We
emphasize that we are diagonalizing the Higgs potential about its
minimum at $\langle H_u \rangle = \langle H_d \rangle = 0$ in the
unbroken phase, as opposed to the usual zero-temperature
treatment~\cite{Martin:1997ns}; our results for $m_{H_{1,2}}^2$ and
$\alpha$ do not simplify to the zero-temperature Higgs masses and
mixing angles in the limit $\delta_{u,d,b} \to 0$.

If we assume that $m_A \gg m_Z, \: \delta_{u,d,b}$, and $\tan\beta_0 \gg 1$, then
\bea
m_{H_1}^2 &\simeq& \delta_u - \frac{1}{2} m_Z^2 \\
m_{H_2}^2 &\simeq& m_A^2 + \frac{1}{2} m_Z^2 + \delta_d
\eea 
and $\alpha \simeq 1/\tan\beta_0$.  In our analysis, we will
use Eq.~(\ref{HiggsMix}).
In general, this mixing angle will be spacetime-dependent due to the appearance of terms in the mass matrix proportional to $v_u(x), \; v_d(x)$. However, it has been found that
\be
\Delta\beta \equiv \left. \frac{}{} \beta(T) \right|_{z\to\infty} \; - \; \left. \frac{}{} \beta(T) \right|_{z\to -\infty}
\ee
is numerically small: $\Delta \beta \lesssim 10^{-2}$~\cite{Moreno:1998bq}.  Therefore, we neglect the spacetime-dependence of the Higgs mixing angle.  (In extended supersymmetric EWB scenarios, it is conceivable that $\Delta \beta$ might be larger, necessitating a proper treatment of spacetime-dependent Higgs mixing.)

The purpose of the preceding analysis is to motivate realistic masses and
mixing angles in the Higgs sector. We defer a rigorous
determination of Higgs boson masses and mixing during EWB to a future study.

% \begin{itemize}
% 
% \item We take the Higgs mixing angle to be $\alpha = \beta(T)$, spacetime-independent due to the smallness of $\Delta \beta$.
% 
% \item We expect the lighter Higgs boson has mass $m_{H_1} \lesssim 100$ GeV, since $\delta_u \lesssim T^2$.  We take $m_{H_1} = 70$ GeV.  We note that $m_{H_1}$ corresponds to the curvature of the Higgs potential at finite temperature about the minimum at $v_u = v_d = 0$, not the mass of lightest Higgs boson at $T=0$; hence, there is no inconsistency with the LEP II bound.
% 
% \item Since $\delta_d \lesssim T^2$, if we take $m_A = 150$ GeV, then we have $m_{H_2} \lesssim 200$ GeV.  If $m_A$ becomes larger, then $\Delta \beta$, and thus $Y_B$, become suppressed~\cite{Moreno:1998bq}.  We take $m_{H_2} = 200$ GeV.
% 
% \end{itemize}

\subsection{Supergauge equilibration rates}
\label{section:Supergauge}
\begin{figure}[ht]
\includegraphics[scale=1]{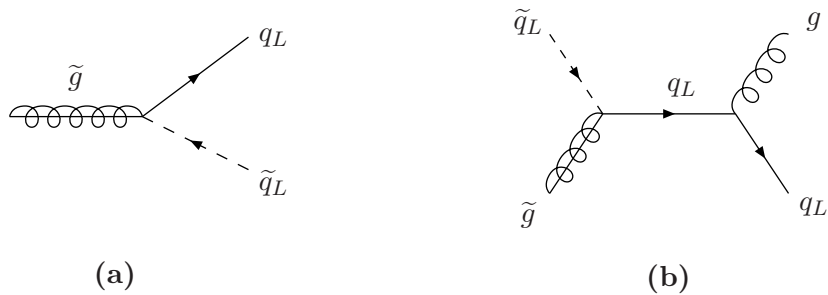}
\caption{Examples of absorption/decay (a) and scattering (b) processes which lead to superequilibrium.}
\label{fig:feyn_figs}
\end{figure}

Supergauge interactions generate three-body absorption/decay processes as
illustrated in FIG.~\ref{fig:feyn_figs}(a).  These processes drive the
plasma toward superequilibrium, the condition where the
chemical potentials for a
particle and its superpartner are equal.  Following
closely the derivation of $\Gamma_Y$ that is
presented in Ref.~\cite{Cirigliano:2006wh}, we calculate the fully
thermally averaged rates $\Gamma_{\widetilde V}$ in the on-shell limit.

We compute the supergauge interaction rates $\Gamma_{\widetilde{V}}$ arising from emission/absorption processes in the thermal
plasma [FIG.~\ref{fig:feyn_figs}(a)].  Each supergauge interaction term in Eq.~(\ref{eq:LintVtilde}) can be cast in the general form
\be
\mathcal{L}_{int} = \phi \: \bar{\psi} \left( g_L \, P_L + g_R \, P_R \right) \widetilde{V} + \; \textrm{h.c.}
\ee
for gaugino $\widetilde{V}$, and (bosonic, fermionic) superpartners $(\phi,\psi)$.  These interactions give contributions to the RHS of Eqns.~(\ref{eq:scalar1},\ref{eq:fermion1}) of the form
\be
\partial_\mu \, j_\phi^\mu(X) = - \partial_\mu \, j_\psi^\mu(X) = S_{\widetilde V}\,,
\ee
where
\be
S_{\widetilde V} \equiv \left[ \left( \, \left|g_L^2\right| + \left|g_R^2\right|\,  \right) \: \mathcal{I}_F \left(m_\psi, m_\phi, m_{\widetilde V} \right) + 2 \, \textrm{Re} \left( g_L g_R^* \right) \: \widetilde{\mathcal{I}}_F \left(m_\psi, m_\phi, m_{\widetilde V} \right) \right] \, \left( \mu_\psi - \mu_\phi - \mu_{\widetilde V} \right) \;. \label{eq:Vtildesource}
\ee
The functions $\mathcal{I}_F$ and $\widetilde{\mathcal{I}}_F$ are defined to be \cite{Cirigliano:2006wh}
\begin{align}
\mathcal{I}_F(m_1,m_\phi,m_2) &= \frac{1}{16\pi^3 T} \, 
\left(m_1^2 + m_2^2 - m_\phi^2 \right) \, 
\int_{m_1}^\infty d\omega_1 \int_{\omega_\phi^-}^{\omega_\phi^+}d\omega_\phi \\
\times \biggl\{n_B(\omega_\phi)&\bigl[1-n_F(\omega_1)\bigr]n_F(\omega_1-\omega_\phi)\bigl[\theta(m_1-m_2-m_\phi) - \theta(m_\phi - m_1 - m_2)\bigr] \notag \\
&\quad \qquad + n_B(\omega_\phi)n_F(\omega_1)\bigl[1-n_F(\omega_1+\omega_\phi)\bigr]\theta(m_2-m_1-m_\phi)\biggr\}\,, \notag
\end{align}
with integration limits on $\omega_\phi$ given by
\begin{equation}
\begin{split}
\omega_\phi^\pm = \frac{1}{2m_1^2}\Bigl\{&\omega_1\abs{m_\phi^2 + m_1^2 - m_2^2} \\
& \pm \bigl[(\omega_1^2 - m_1^2)\bigl(m_1^2-(m_2 + m_\phi)^2\bigr)\bigl(m_1^2-(m_2 - m_\phi)^2\bigr)\bigr]^{1/2}\Bigl\};
\end{split}
\end{equation}
and lastly
\be
\widetilde{\mathcal{I}}_F(m_1, m_\phi, m_2) = \frac{2 \, m_1 m_2}{m_1^2 + m^2_2 - m_\phi^2} \: \mathcal{I}_F(m_1,m_\phi,m_2) \;.
\ee

The gaugino chemical potential $\mu_{\widetilde{V}}$ only appears on the RHS of Eq.~\eqref{eq:Vtildesource} when the gaugino is a Dirac fermion (i.e., for $\widetilde{W}^\pm$).  For Majorana gauginos, no gaugino chemical potential appears.  Although a Majorana chemical potential does not correspond to a conserved Noether current, it can arise as a deviation from a pure Fermi-Dirac distribution when annihilation processes are out of equilibrium.  However, such a deviation does not contribute to Eq.~\eqref{eq:Vtildesource} due to CP-symmetry.  The charge current densities and corresponding chemical potentials for Dirac fermions and complex scalars are all odd under CP.  However, a Majorana chemical potential is even under CP.  Therefore, in the limit that we can neglect CP-violating phases in our interaction rates, Majorana chemical potentials do not contribute to the Boltzmann equations for charge current densities.  We have explicitly verified that  a non-vanishing Majorana chemical potential ultimately cancels from $S_{\widetilde V}$ to linear order in $\epsilon_\mu$ and at zeroth order in $CP$-violating phases $\phi_{CP}$.  Physically speaking, an excess of Majorana gauginos $\widetilde{V}$ does not bias a charge asymmetry in $\psi$ and $\phi$ since the rates for $\widetilde V \leftrightarrow \psi \phi$ and $\widetilde V \leftrightarrow \bar\psi \phi^\dagger$ are equal\footnote{In leptogenesis scenarios, a heavy Majorana neutrino can bias a chiral lepton asymmetry, as long as the relevant $CP$-violating phases are non-zero.  This is consistent with the statement that, in the present discussion, transport coefficients which couple Majorana chemical potential $\mu_{\widetilde V}$ to $\mu_\phi$, $\mu_\psi$ arise at order $\mathcal{O}(\sin\phi_{CP})$.}.

At present, we work exclusively within the MSSM, where existing measurements constrain $\phi_{CP} \ll 1$; consequently, we neglect these contributions.  In extensions of the MSSM where additional $CP$-violating phases are less constrained and may be large, the non-equilibrium dynamics of Majorana fermions may be important. We also emphasize that these arguments apply to the degrees of freedom
that are present in the symmetric phase. This is consistent within
the framework of the present paper, as our main focus is the diffusion
process in the symmetric phase and as we calculate the source and
relaxation terms in a pertubative mass-insertion scheme. Note that in
the broken phase, the Higgsinos, which are treated as charged
particles in the symmetric
phase, mix {\it e.g.} with the Binos, which are Majorana
particles. The resulting neutralinos are Majorana particles, but due to
their Higgsino component, their out-of-equilibrium dynamics is crucial 
for EWB. A non-pertubative treatment of the mixing in the broken
phase will be subject of future investigations.

We can relate the chemical potentials $\mu$ to charge number densities $n$ via
\be
n \equiv \int \frac{d^3p}{(2\pi)^3} \left( \frac{1}{e^{(\omega_p - \mu)/T} \pm 1} - \frac{1}{e^{(\omega_p + \mu)/T} \pm 1} \right) = \frac{T^2}{6}\: k(m/T)\: \mu \, + \, \mathcal{O}(\epsilon_\mu^3)\,,
\ee
where $k(m/T)$ is a statistical factor \cite{Lee:2004we}
\be
k(m/T) = k(0)\, \frac{c_{F,B}}{\pi^2}\, \int_{m/T}^\infty\, dx \, x\ \frac{e^x}{(e^x\pm1)^2}\, \sqrt{x^2-m^2/T^2}\ \ \ 
\ee
with $k(0)=1$ for chiral fermions, $k(0)=2$ for Dirac fermions and complex scalars, $c_{F\, (B)}=6(3)$ for fermions (bosons), and the $+$ ($-$) sign for fermions (bosons).  Therefore, we can write
\be
S_{\widetilde V} = \Gamma_{\widetilde V}^{(\psi, \phi)} \: \left( \frac{n_{\psi}}{k_\psi} - \frac{n_{\phi}}{k_\phi} \right) \label{eq:SVtilde}
\ee
for Majorana $\widetilde V$, where
\be
\label{Gamma:Vtilde}
\Gamma_{\widetilde V}^{(\psi,\phi)} \equiv \frac{6}{T^2} \left[ \left( \, \left|g_L^2\right| + \left|g_R^2\right|\,  \right) \: \mathcal{I}_F \left(m_\psi, m_\phi, m_{\widetilde V} \right) + 2 \, \textrm{Re} \left( g_L g_R^* \right) \: \widetilde{\mathcal{I}}_F \left(m_\psi, m_\phi, m_{\widetilde V} \right) \right] \;.
\ee

If $\widetilde V$ is Dirac, as is the case for $\widetilde W^\pm$,
the situation is more subtle. Instead of Eq.~(\ref{eq:SVtilde}),
when inserting Eq.~(\ref{Gamma:Vtilde}) into Eq.~(\ref{eq:Vtildesource}), we have
\be
S_{\widetilde V} = \Gamma_{\widetilde V}^{(\psi, \phi)} \: \left( \frac{n_{\psi}}{k_\psi} - \frac{n_{\phi}}{k_\phi} - \frac{n_{\widetilde V}}{k_{\widetilde V}} \right) \;,
\ee
meaning that that non-equilibrium dynamics of ${\widetilde W^\pm}$ does not in general decouple from the dynamics that produces $Y_B$.  Let us consider the contributions from $S_{\widetilde W^\pm}$ to the Boltzmann equations (\ref{eq:fermion1}) for the third generation LH quarks:
\begin{subequations}
\label{eq:Wtildepm}
\bea
\partial_\mu j^\mu_{t_L} &=& - \, g_2^2 \, \left| V_{tb} \right|^2 \, \mathcal{I}_F(m_{t_L}, m_{\widetilde b_L}, m_{\widetilde W^\pm} ) \: \left(\mu_{t_L} - \mu_{\widetilde{b}_L} - \mu_{\widetilde W^\pm} \right)\,, \\
\partial_\mu j^\mu_{b_L} &=& - \, g_2^2 \, \left| V_{tb} \right|^2 \, \mathcal{I}_F(m_{b_L}, m_{\widetilde t_L}, m_{\widetilde W^\pm} ) \: \left(\mu_{b_L} - \mu_{\widetilde{t}_L} + \mu_{\widetilde W^\pm} \right) \;.
\eea
\end{subequations}
Now we define
\begin{subequations}
\bea
\mu_q &\equiv& \frac{1}{2} \left(\mu_{t_L} + \mu_{b_L} \right)\,, \\
\mu_{\widetilde q} &\equiv& \frac{1}{2} \left(\mu_{\widetilde t_L} + \mu_{\widetilde b_L} \right)\,, \\
\Delta \mu_{ q} &\equiv& \frac{1}{2} \left(\mu_{t_L} - \mu_{b_L} \right)\,, \\
\Delta \mu_{\widetilde q} &\equiv& \frac{1}{2} \left(\mu_{\widetilde t_L} - \mu_{\widetilde b_L} \right)  \;.
\eea
\end{subequations}
With these definitions, we obtain from Eqns.~(\ref{eq:Wtildepm})
\bea
\partial_{\mu} \left( j^\mu_{u_L} + j^\mu_{d_L} \right) &=& -  \, N_C \,  g_2^2 \, \left| V_{tb} \right|^2  \left[ \mathcal{I}_F(m_{t_L}, m_{\widetilde b_L}, m_{\widetilde W^\pm} ) + \mathcal{I}_F(m_{b_L}, m_{\widetilde t_L}, m_{\widetilde W^\pm} ) \right] \left( \mu_q - \mu_{\widetilde q} \right) \notag \\
&\;& - \,  N_C \,  g_2^2 \, \left| V_{tb} \right|^2  \left[  \mathcal{I}_F(m_{t_L}, m_{\widetilde b_L}, m_{\widetilde W^\pm} ) - \mathcal{I}_F(m_{b_L}, m_{\widetilde t_L}, m_{\widetilde W^\pm} ) \right] \,  \\
 &\;& \qquad \qquad \qquad \times \left( \Delta \mu_q + \Delta \mu_{\widetilde q} - 2 \mu_{\widetilde W^\pm}\right) \notag
\eea
and 
\bea
\partial_{\mu} \left( j^\mu_{u_L} - j^\mu_{d_L} \right) &=&  - \, N_C \,  g_2^2 \, \left| V_{tb} \right|^2 \left[  \mathcal{I}_F(m_{t_L}, m_{\widetilde b_L}, m_{\widetilde W^\pm} ) - \mathcal{I}_F(m_{b_L}, m_{\widetilde t_L}, m_{\widetilde W^\pm} ) \right]  \left( \mu_q - \mu_{\widetilde q} \right) \notag \\
&\;& - \, N_C \,  g_2^2 \, \left| V_{tb} \right|^2  \left[  \mathcal{I}_F(m_{t_L}, m_{\widetilde b_L}, m_{\widetilde W^\pm} ) + \mathcal{I}_F(m_{b_L}, m_{\widetilde t_L}, m_{\widetilde W^\pm} ) \right] \,  \\
 &\;& \qquad \qquad \qquad \times \left( \Delta \mu_q + \Delta \mu_{\widetilde q} - 2 \mu_{\widetilde W^\pm} \right) \notag
\eea
In the unbroken phase, where $v_u = v_d = 0$, we have $m_{t_L}(T) = m_{b_L}(T) \equiv m_q$ and $m_{\widetilde t_L}(T) = m_{\widetilde b_L}(T) \equiv m_{\widetilde q}$, so that the differences
\be
\left[  \mathcal{I}_F(m_{t_L}, m_{\widetilde b_L}, m_{\widetilde W^\pm} ) - \mathcal{I}_F(m_{b_L}, m_{\widetilde t_L}, m_{\widetilde W^\pm} ) \right] \longrightarrow 0 \;,
\ee
while the sums
\be
\left[  \mathcal{I}_F(m_{t_L}, m_{\widetilde b_L}, m_{\widetilde W^\pm} ) + \mathcal{I}_F(m_{b_L}, m_{\widetilde t_L}, m_{\widetilde W^\pm} ) \right] \longrightarrow 2 \, \mathcal{I}_F(m_{q}, m_{\widetilde q}, m_{\widetilde W^\pm} ) \;.
\ee
Therefore, to the extent that we can neglect isospin-violating mass differences within the bubble, isovector asymmetries such as $\Delta \mu_{q}$ decouple from isoscalar densities such as $\mu_q$.  Although we have chosen only only one interaction as an illustration, we have verified that the decoupling of isoscalar and isovector densities occurs for all Yukawa and supergauge interactions (\ref{eq:LintYuk}, \ref{eq:LintVtilde}).  Since $Y_B$ is produced by weak sphalerons sourced by the chiral asymmetry
\be
n_{\rm left} \equiv \sum_{i=1}^3 \left( n_{u_L^i} + n_{d_L^i} \right)\;,
\ee
itself an iso-scalar density, isospin-violating asymmetries decouple from the determination $Y_B$.  In particular, $n_{\widetilde W^\pm}$ only couples to isospin-violating asymmetries and will decouple from the dynamics of $n_{\rm left}$.  Therefore, we consider the closed set of Boltzmann equations for the following twenty-five densities:
\begin{subequations}
\label{eq:densities}
\begin{align}
H_{1,2} &\equiv n_{H_{1,2}^+} + n_{H_{1,2}^0}\,,
&\widetilde{H} &\equiv n_{\widetilde H^+} + n_{\widetilde H^0}\,,
\\
q_{1,2} &\equiv n_{u^{1,2}_L} + n_{d^{1,2}_L}\,,
&\widetilde q_{1,2} &\equiv n_{\widetilde u^{1,2}_L} + n_{\widetilde d^{1,2}_L}\,,
\\
q &\equiv q_3 \equiv n_{t_L} + n_{b_L}\,,
&\widetilde q &\equiv \widetilde q_3 \equiv n_{\widetilde t_L} + n_{\widetilde b_L}\,,
\\
u_{1,2} &\equiv n_{u^{1,2}_R}\,,
&\widetilde u_{1,2} &\equiv n_{\widetilde u^{1,2}_R}\,,
\\
t&\equiv u_3 \equiv n_{t_R}\,,
&\widetilde t &\equiv \widetilde u_3 \equiv n_{\widetilde t_R}\,,
\\
d_{1,2} &\equiv n_{d^{1,2}_R}\,,
&\widetilde d_{1,2} &\equiv n_{\widetilde d^{1,2}_R}\,,
\\
b &\equiv  d_3 \equiv n_{b_R}\,,
&\widetilde b &\equiv \widetilde d_2 \equiv n_{\widetilde b_R}\,.
\\
\ell &\equiv \ell_3 \equiv n_{\tau_L}+n_{\nu^\tau_L}\,,
&\widetilde\ell &\equiv \widetilde\ell_3 \equiv n_{\widetilde\tau_L}+n_{\widetilde\nu^\tau_L}\,,
\\
\tau &\equiv n_{\tau_R}\,,
&\widetilde\tau &\equiv n_{\widetilde\tau_R}\,.
\end{align}
\end{subequations}
We defer to future work the study of isospin breaking effects within the bubble wall which couple isospin-violating asymmetries to those listed above.
Note that within the context of non-supersymmetric models, a study of deviation from isospin 
equilibrium has been performed and the effect has been
found to be quantitatively
relevant~\cite{Ref:IsoEquilibrium}.

The supergauge equilibration rates that enter the Boltzmann equations for these densities (\ref{eq:densities}) are
\begin{subequations}
\label{eq:GammaVtilde}
\begin{eqnarray}
\Gamma_{\widetilde V}^{(H_1,\widetilde H)} &=&
\frac{6 \, g_1^2}{T^2} \,
\left[
 \mathcal{I}_F (m_{\widetilde H},m_{H_1},m_{\widetilde B})
- \sin 2\alpha \, \widetilde{\mathcal{I}}_F(m_{\widetilde H},m_{H_1},m_{\widetilde B}) \right] \\
&\;& + \; \frac{18 \, g^2_2}{T^2} \, \left[
\mathcal{I}_F(m_{\widetilde H},m_{H_1},m_{\widetilde W}) + \sin 2\alpha \, \widetilde{\mathcal I}_F(m_{\widetilde H},m_{H_1},m_{\widetilde W}) \right]\,,  \notag \\
\Gamma_{\widetilde V}^{(H_2,\widetilde H)} &=&
\frac{6 \, g_1^2}{T^2} \,
\left[
 \mathcal{I}_F (m_{\widetilde H},m_{H_2},m_{\widetilde B})
+ \sin 2\alpha \, \widetilde{\mathcal{I}}_F(m_{\widetilde H},m_{H_2},m_{\widetilde B}) \right] \\
&\;& + \; \frac{18 \, g^2_2}{T^2} \, \left[
\mathcal{I}_F(m_{\widetilde H},m_{H_2},m_{\widetilde W}) - \sin 2\alpha \, \widetilde{\mathcal I}_F(m_{\widetilde H},m_{H_2},m_{\widetilde W}) \right]\,,  \notag \\
\Gamma_{\widetilde V}^{(q,\widetilde q)} &=&
\frac {2\, N_C \, g_1^2}{3 \, T^2} \, \mathcal{I}_F (m_q,m_{\widetilde q},m_{\widetilde B})
+\frac{ 18 \, N_C \, g_2^2}{T^2} \, \mathcal{I}_F\left(m_q,m_{\widetilde q},m_{\widetilde W}\right)\\
&\;& 
+\frac {12 (N_C^2 -1) \, g^2_3}{T^2} \, \mathcal{I}_F\left(m_q,m_{\widetilde q},m_{\widetilde G}\right)\,, \notag \\
\Gamma_{\widetilde V}^{(u,\widetilde u)} &=&
\frac {16 \, N_C \, g^2_1}{3 T^2} \mathcal{I}_F\left(m_{u},m_{\widetilde u},m_{\widetilde B}\right)
+ \frac{6(N_C^2-1) \, g_3^2}{T^2} \, \mathcal{I}_F\left(m_{u},m_{\widetilde u},m_{\widetilde G}\right)\,, \\
\Gamma_{\widetilde V}^{(d,\widetilde d)}&=&
\frac {4\, N_C \, g_1^2}{3T^2} \, \mathcal{I}_F\left(m_{d},m_{\widetilde d},m_{\widetilde B}\right)
+ \frac{6(N_C^2-1) \, g_3^2}{T^2} \, \mathcal{I}_F\left(m_{d},m_{\widetilde d},m_{\widetilde G}\right)\,,
\\
\Gamma_{\widetilde V}^{(\ell,\widetilde \ell)}&=&
\frac{6 g_1^2}{T^2} \mathcal{I}_F\left(m_{\widetilde\ell},m_\ell,m_{\widetilde B}\right)
+\frac{18 g_2^2}{T^2} \mathcal{I}_F\left(m_{\widetilde\ell},m_\ell,m_{\widetilde W}\right)\,,
\\
\Gamma_{\widetilde V}^{(\tau,\widetilde \tau)}&=&
\frac{12 g_1^2}{T^2} \mathcal{I}_F\left(m_{\widetilde\tau},m_\tau,m_{\widetilde B}\right)\,,
\end{eqnarray}
\end{subequations}
where in Eqns.~(\ref{eq:GammaVtilde}c-e) we have omitted a generational index since these expressions are identical for all generations.

\subsection{Yukawa, tri-scalar, and Higgs vev induced interactions}
\label{section:YukTri}

Presently, we summarize contributions to the Boltzmann equations that arise from interactions in Eqns.~(\ref{eq:LintM}, \ref{eq:LintYuk}).  Yukawa and SUSY-breaking tri-scalar interactions (\ref{eq:LintYuk}) lead to transport coefficients in the Boltzmann equations which couple Higgs, RH quark, and LH quark supermultiplets.  We assume that the tri-scalar $A$-terms are proportional to the corresponding Yukawa coupling~\cite{Martin:1997ns}. For example, the term
\be
\mathcal{L}_Y \supset y_t \: e^{i \phi_\mu} \: \left(\widetilde{t}_R \: \bar{t}_L \: P_R \: \Psi_{\widetilde{H}^0}^C + \widetilde{t}_R \: \bar{b}_L \: P_R \: \Psi_{\widetilde{H}^+}^C \right)
\ee
in Eq.~(\ref{eq:LintYuk}) leads to a contribution to the Boltzmann equations for densities $\widetilde t, q, \widetilde{H}$ of the form
\be
\partial_\mu \, j^\mu_{\widetilde H} =  \partial_\mu \, j^\mu_{q} = - \partial_\mu \, j^\mu_{\widetilde t} = \Gamma_Y^{(q, \widetilde t, \widetilde H)} \: \left( \frac{\widetilde t}{k_{\widetilde t}} - \frac{ q}{k_q} - \frac{\widetilde H}{k_{\widetilde H}} \right) \;.
\ee

We now list the complete set of equilibration rates \cite{Cirigliano:2006wh}
arising from $\mathcal{L}_Y$:
\begin{subequations}
\label{eq:gammayrates}
\begin{eqnarray}
\Gamma_Y^{(\widetilde t,\widetilde q ,H_1)} &=&
\frac {12 \, N_C \, y_t^2}{T^2}
\left| \, \sin \alpha \, \mu^* + \cos \alpha \, A_t \, \right|^2 \, \mathcal{I}_B\left(m_{\widetilde t},m_{\widetilde 1},m_{H_1} \right)\,, \\
\Gamma_Y^{(\widetilde t,\widetilde q ,H_2)} &=&
\frac {12 \, N_C \, y_t^2}{T^2}
\left| \, \cos \alpha \, \mu^* - \sin \alpha \, A_t \, \right|^2 \, \mathcal{I}_B\left(m_{\widetilde t},m_{\widetilde q},m_{H_2} \right)\,, \\
\Gamma_Y^{(\widetilde t,q ,\widetilde H)} \label{eq:GammaYHtildeQ}
&=&
\frac {12 \, N_C \, y_t^2}{T^2} \: \mathcal{I}_F \left(m_{\widetilde H},m_{\widetilde t},m_q \right)\,, \\
\Gamma_Y^{(t,q ,H_{1})} &=&
\frac {12 \, N_C \, y_t^2}{T^2} \, \cos^2 \alpha \, \mathcal{I}_F \left(m_t,m_q,m_{H_{1}} \right)\,, \\
\Gamma_Y^{(t,q ,H_{2})} &=&
\frac {12 \, N_C \, y_t^2}{T^2} \, \sin^2 \alpha \, \mathcal{I}_F \left(m_t,m_q,m_{H_{2}} \right)\,, \\
\Gamma_Y^{(t,\widetilde q ,\widetilde H)} &=& \frac {12 \, N_C \, y_t^2}{T^2} \, \mathcal{I}_F \left(m_t,m_{\widetilde H},m_{\widetilde q} \right)\,; \label{eq:GammaYHtildeQtilde}
\end{eqnarray}

\begin{eqnarray}
\Gamma_Y^{(\widetilde b,\widetilde q ,H_1)} &=&
\frac {12 \, N_C \, y_b^2}{T^2}
\left| \, \cos \alpha \, \mu^* - \sin \alpha \, A_b \, \right|^2 \,
\mathcal{I}_B\left(m_{\widetilde b},m_{\widetilde q},m_{H_1} \right)\,,
\\
\Gamma_Y^{(\widetilde b,\widetilde q ,H_2)} &=&
\frac {12 \, N_C \, y_b^2}{T^2}
\left| \, \sin \alpha \, \mu^* + \cos \alpha \, A_b \, \right|^2 \,
\mathcal{I}_B\left(m_{\widetilde b},m_{\widetilde q},m_{H_2} \right)\,,
\\
\Gamma_Y^{(\widetilde b,q ,\widetilde H)} \label{eq:GammaYHtildeQbottom}
&=&
\frac {12 \, N_C \, y_b^2}{T^2} \: \mathcal{I}_F \left(m_{\widetilde H},m_{\widetilde b},m_q \right)\,,
\\
\Gamma_Y^{(b,q ,H_{1})} &=&
\frac {12 \, N_C \, y_b^2}{T^2} \, \sin^2 \alpha \,
\mathcal{I}_F \left(m_b,m_q,m_{H_{1}} \right)\,, \\
\Gamma_Y^{(b,q ,H_{2})} &=&
\frac {12 \, N_C \, y_b^2}{T^2} \, \cos^2 \alpha \, \mathcal{I}_F \left(m_b,m_q,m_{H_{2}} \right)\,, \\
\Gamma_Y^{(b,\widetilde q ,\widetilde H)} &=&
\frac {12 \, N_C \, y_b^2}{T^2} \,
\mathcal{I}_F \left(m_b,m_{\widetilde H},m_{\widetilde q} \right)\,; \label{eq:GammaYHtildeQtildebottom}
\end{eqnarray}

\begin{eqnarray}
\Gamma_Y^{(\widetilde \tau,\widetilde \ell ,H_1)} &=&
\frac {12 \, N_C \, y_\tau^2}{T^2}
\left| \, \cos \alpha \, \mu^* - \sin \alpha \, A_\tau \, \right|^2 \,
\mathcal{I}_B\left(m_{\widetilde \tau},m_{\widetilde \ell},m_{H_1} \right)\,,
\\
\Gamma_Y^{(\widetilde \tau,\widetilde \ell ,H_2)} &=&
\frac {12 \, N_C \, y_\tau^2}{T^2}
\left| \, \sin \alpha \, \mu^* + \cos \alpha \, A_\tau \, \right|^2 \,
\mathcal{I}_B\left(m_{\widetilde \tau},m_{\widetilde \ell},m_{H_2} \right)\,,
\\
\Gamma_Y^{(\widetilde \tau,\ell ,\widetilde H)} \label{eq:GammaYHtildeQtau}
&=&
\frac {12 \, N_C \, y_\tau^2}{T^2} \: \mathcal{I}_F \left(m_{\widetilde H},m_{\widetilde \tau},m_\ell \right)\,,
\\
\Gamma_Y^{(\tau,\ell ,H_{1})} &=&
\frac {12 \, N_C \, y_\tau^2}{T^2} \, \sin^2 \alpha \,
\mathcal{I}_F \left(m_\tau,m_\ell,m_{H_{1}} \right)\,, \\
\Gamma_Y^{(\tau,\ell ,H_{2})} &=&
\frac {12 \, N_C \, y_\tau^2}{T^2} \, \cos^2 \alpha \, \mathcal{I}_F \left(m_\tau,m_\ell,m_{H_{2}} \right)\,, \\
\Gamma_Y^{(\tau,\widetilde \ell ,\widetilde H)} &=&
\frac {12 \, N_C \, y_\tau^2}{T^2} \,
\mathcal{I}_F \left(m_\tau,m_{\widetilde H},m_{\widetilde \ell} \right)\,, \label{eq:GammaYHtildeQtildetau}
\end{eqnarray}

\end{subequations}
where
\begin{equation}
\begin{split}
& \mathcal{I}_B (m_R,m_L,m_H) = - \frac{1}{16\pi^3 T}
 \,
\int_{m_R}^\infty d\omega_R \int_{\omega_L^-}^{\omega_L^+} d\omega_L  \\
& \ \ \ \times \biggl\{n_B(\omega_R)\bigl[1+n_B(\omega_L)\bigr] 
n_B(\omega_L-\omega_R)\bigl[\theta(m_R-m_L-m_H) - \theta(m_L-m_R-m_H)\bigr] \\
& \qquad  -n_B(\omega_R)n_B(\omega_L)\bigl[
1+n_B(\omega_L+\omega_R)\bigr]\theta(m_H-m_R-m_L)\biggr\},
\end{split}
\label{eq:IB}
\end{equation}
with integration limits given by:
%
%\begin{equation}
%\omega_L^\pm = \frac{1}{2m_R^2}\left[\omega_R\abs{m_R^2 + m_L^2 - m_H^2} \pm \sqrt{\omega_R^2 - m_R^2}\sqrt{(m_R^2 + m_L^2 - m_H^2)^2 - 4m_R^2 m_L^2}\right].
%\end{equation}
%
\begin{equation}
\begin{split}
\omega_L^\pm = \frac{1}{2m_R^2}\Bigl\{&\omega_R\abs{m_R^2 + m_L^2 - m_H^2}  \\
&\pm \bigl[(\omega_R^2 - m_R^2)\bigl(m_R^2 - (m_L + m_H)^2\bigr)\bigl(m_R^2-(m_L - m_H)^2\bigr)\bigr]^{1/2}\Bigl\}.
\end{split}
\end{equation}
These interactions communicate the effects of $CP$-violation from the Higgs(ino) sector to the (s)quark sector.  If supergauge interactions are assumed to be in equilibrium, only the sum of $\Gamma_Y$ rates listed above enters the Boltzmann equations that determine $Y_B$.
This procedure has been applied in our recent
publications~\cite{Chung:2008aya,Chung:2009cb}.
Without this assumption, we must distinguish between each rate.  For example, the rate for the transfer of $CP$-violating effects  from $\widetilde{H} \to t$ (\ref{eq:GammaYHtildeQ}) will be different from the rate of transfer from $\widetilde{H} \to \widetilde{t}$ (\ref{eq:GammaYHtildeQtilde}); this difference may impact $Y_B$ if transfer
between $t \leftrightarrow \widetilde t$ is inefficient, {\it cf.}
Section~\ref{sec:NumNoSG}.

% We also include $2 \to 2$ scattering contributions to the rates in Eqns.~(\ref{eq:gammayrates}d,e), as estimated in Ref.~\cite{Huet:1995sh}.  Suitably generalized to the case of Higgs mixing, these contributions are
% \bea
% \Gamma_Y^{(t,q ,H_{1})} &=&
% \frac {27}{2} \, \cos^2 \alpha \, \alpha_s \, y�_t^2 \, \left( \frac{\zeta(3) }{\pi^2} \right)^2  \\
% \Gamma_Y^{(t,q ,H_{2})} &=&
% \frac {27}{2} \, \sin^2 \alpha \, \alpha_s \, y�_t^2 \, \left( \frac{\zeta(3) }{\pi^2} \right)^2 \notag \;.
% \eea
% We do not include any other scattering contributions, deferring their inclusion to a future study.

\subsection{Four-body and off-shell contributions}
\label{sec:fourbodyandoffshell}

In addition, four-body scattering interactions,
such as those illustrated in FIG.~\ref{fig:feyn_figs}(b), will also
contribute to the equilibration process~\cite{Joyce:1994zn}. Although 
the associated rates are phase space suppressed and
are higher order
in the gauge couplings than the three-body rates, they can become leading
order when the three-body processes are kinematically forbidden. At
the same order in gauge coupling constants, also off-shell processes
contribute~\cite{Arnold:2001ms,Arnold:2003zc,Garbrecht:2008cb}, which are
loop instead of phase space suppressed.

A calculation of these contributions to the interaction rates may hence be important
for EWB in regions of parameter space where three body interactions
are kinematically forbidden. We anticipate a full calculation
that takes into account the masses in the Higgs sector and of the sfermions and the gauginos,
which typically can be of the same order as the temperature, to be rather involved
and do not perform it within the present work. For our parametric examples, we therefore avoid 
here parametric regions where three-body rates are
kinematically forbidden as much as it is possible. The only exception
that we make concerns the interaction between $H_1$, $t$ and $q$, where
it is hard to avoid kinematic suppression for three-body interactions
and to allow for electroweak symmetry breaking at the same time. We
specify the estimate  for the four-body interaction
that we adopt for that case in Section~\ref{sec:NumYukSGEq}.

\subsection{Sources and relaxation terms}
\label{sec:sourcerelax}

$CP$-violating sources can arise from the Lagrangian
terms~(\ref{eq:LintM}) involving squarks and Higgsinos.  Physically, they correspond to $CP$-asymmetric
reflection and transmission rates for squarks and Higgsinos scattering from the
bubble wall. 
In Ref.~\cite{Lee:2004we}, only
contributions to source terms that become resonant are calculated; off-resonance these
contributions become comparable to other terms that have not been calculated.
Therefore, we do not include in the Boltzmann equations the chiral relaxation rates
for squarks and Higgs bosons which are non-resonant for the present choices of
parameters, deferring their inclusion until a complete calculation of these rates
has been accomplished. We comment on the validity of this approximation in
Sec.~\ref{sec:numerical}.
%The squark CP-violating source possesses a resonance for
%$m_{\widetilde q} \sim m_{\widetilde t,b}$. However, within the MSSM
%with a strong first order phase transition, bounds on 
%the lightest Higgs mass and $\rho$ parameter require
%$m_{\widetilde q} \gg m_{\widetilde t}$. Besides, we also
%assume here that $m_{\widetilde q} \gg m_{\widetilde b}$.
%In this regime, squark CP-violation
%is insufficient to produce the observed BAU, and the chiral
%squark relaxation terms are negligible when compared to
%the quark ones.  The Higgsino CP-violating source
%also possesses a resonance when $m_{\widetilde H}$ approaches the mass of either
%electroweak gaugino. 
In the present work, we assume a resonant Higgsino source
with $|\mu| = M_1 = 200$ GeV, which is sufficient to produce the observed BAU.
The precise formula for $S^{\, \cancel{CP}}_{\widetilde H}$ is given in
Ref.~\cite{Lee:2004we}.

The choice of Higgsino and Bino as the source of resonant $CP$-violation
rather than of Higgsino and Wino is motivated by the recent evaluation
of the two loop electric dipole moments (EDMs)
of the electron and the neutron in the
limit of heavy first two generation squarks and sleptons~\cite{Li:2008kz}. The results indicate that
the size of the $CP$-violating phase between $\mu$ and $M_1$ is much less
constrained by the non-observation of permanent EDMs than the phase between
$\mu$ and $M_2$. Therefore,
Bino-driven EWB~\cite{Chung:2008aya,Li:2008ez,Chung:2009cb}
may prove to be a more
viable option, particularly as the sensitivity of EDM searches improve.

In addition to the source terms, there exist chiral relaxation rates that also arise
from particles scattering with the Higgs background field (\ref{eq:LintM})
and that tend to wash out $CP$-violating asymmetries.  These processes exist in both
the quark and Higgs sectors and are generically denoted as $\Gamma_M$ and $\Gamma_H$,
respectively.  In the present work, since we must distinguish between the densities
for superpartners, we must likewise distinguish between the quark and squark
contributions to $\Gamma_M$, and between the Higgs boson and Higgsino contributions
to $\Gamma_H$.  In the Boltzmann equations to follow, we
write down terms for all relaxation rates that are possible
within the broken phase, both resonant and non-resonant.
For the numerical examples in Section~\ref{sec:numerical} however,
we only include the contributions that may become resonantly enhanced:
the Higgsino-Bino (Higgsino-Wino) contribution
to $\Gamma_H$, resonantly enhanced for $|\mu| = M_1 = 200\, {\rm GeV}$,
($|\mu| = M_2$); and 
the matter fermion contributions to $\Gamma_M$, which are
resonantly enhanced since the thermal masses satisfy the relations
$|m_q-m_t|\ll T$, $|m_q-m_b|\ll T$ and $|m_\ell-m_\tau|\ll T$.

To facilitate comparison with Ref.~\cite{Lee:2004we}, we note
that we apply the slight changes of notation
$\Gamma_H^{\widetilde H,\widetilde V}=\Gamma_h$,
$\Gamma_M(t,q)=(6/T^2)\Gamma_{t}^-$,
$\Gamma_M(\widetilde t,\widetilde q)=(6/T^2)\Gamma_{\widetilde t}^-$.

\subsection{Thermal masses}
\label{sec:ThMass}

The masses that appear in the expressions for the equilibration rates
$\Gamma_Y$, $\Gamma_{\widetilde V}$, in the relaxation terms
$\Gamma_H$ and $\Gamma_M$ and in the
source terms $S^{\, \cancel{CP}}_{\widetilde H}$,
are understood to include thermal corrections, which we take from
Ref.~\cite{Chung:2009cb}, assuming light right-handed
stop, sbottom and stau particles.
While we treat the thermal masses as Dirac mass terms for fermions,
strictly speaking, the thermal masses characterize a modification of the dispersion
relation only and imply no chiral symmetry breaking in the symmetric phase. We provide a justification
of this procedure through a discussion of the numerical
inaccuracies that are incurred in Section~\ref{sec:particlehole}.

\subsection{Diffusion constants}
\label{sec:DiffCon}

When generalizing the distribution functions $f_B$ and $f_F$ to be dependent
also on the spatial momentum $\mathbf p$ in an anisotropic way, it is possible
to derive the Fick diffusion law
\begin{equation}
\label{FickLaw}
\mathbf X=-D_X \nabla X
\end{equation}
from kinetic theory, where $X^\mu=(X, {\mathbf X})$ denotes the number density current for any of the particle
species considered here. The quantity $D_X$ is the diffusion constant for
the species $X$ and can be calculated from the moments of scattering
matrix elements of $X$.
For our purposes, it is useful to recast the diffusion law~(\ref{FickLaw})
by boosting (non-relativistically) to the frame where the bubble wall is at rest as
\begin{equation}
\label{DiffWallFrame}
\partial_\mu X^\mu=v_w\partial_z X - D_X \partial_z^2 X\,.
\end{equation}
The particular values that we use here for the diffusion constants are taken
from Ref.~\cite{Joyce:1994zn} and are summarized in
Table~\ref{table:input:fiducial} in Section~\ref{sec:numerical},
where we present results from our numerical studies. As a simplification,
common diffusion rates for particles and sparticles are taken, such that
{\it e.g.} $D_Q=D_q=D_{\widetilde q}$.
The physical interpretation of Eqs.~(\ref{FickLaw},~\ref{DiffWallFrame})
is simple: Within the plasma,
a particle gets rescattered and undergoes a random walk. These scatterings
are more frequent for colored particles such as quarks than for particles that
interact through electroweak and non-gauge couplings only, such as the $\tau$-lepton.
Therefore, $D_R$ is larger than $D_Q$, which implies that $\tau$-leptons
diffuse farther away from the bubble wall than quarks.

In Ref.~\cite{Joyce:1994zn}, the diffusion constants have been estimated using
an incomplete set of scattering matrix elements. In particular, it has been
pointed out in Ref.~\cite{Arnold:2001ms,Arnold:2003zc},
that infrared sensitive $t$-channel processes
have been neglected.
In general, one should therefore expect an order one inaccuracy within the
results of Ref.~\cite{Joyce:1994zn}.
The results in Ref.~\cite{Arnold:2001ms,Arnold:2003zc}
apply however to gauge theories with massless fundamental fermions only, which is why
they cannot be applied directly to sfermions and to Higgs bosons. In order to
quote a value for comparison, from Ref.~\cite{Arnold:2001ms,Arnold:2003zc}
we infer $D_q=4.7/T$ for six quark flavors, while from Ref.~\cite{Joyce:1994zn},
we adapt $D_Q=D_q=D_{\widetilde q}=6/T$.
In order to achieve predictions of accuracy better than order unity,
it will be necessary to reevaluate the diffusion constants in the future.

\subsection{Boltzmann equations}
\label{sec:Boltzmann}

We now present the full set of Boltzmann equations:
\begin{subequations}
\label{eq:system}
\begin{align}
\label{eq:tright}
\partial_\mu t^\mu  &=
-\Gamma_M^{(t,q)}\left( \frac{t}{k_t} - \frac{q}{k_q} \right) - \, \Gamma_Y^{(t,q,H_1)} \, \left(\frac{t}{k_t}-\frac{q}{k_q}-\frac{H_1}{k_{H_1}} \right)
-\Gamma_{\widetilde V}^{(t,\widetilde t)} \, \left(\frac{t}{k_t}-\frac{\widetilde t}{k_{\widetilde t}} \right)
\\
&\quad - \, \Gamma_Y^{(t,q,H_2)} \, \left(\frac{t}{k_t}-\frac{q}{k_q}-\frac{H_2}{k_{H_2}} \right)
-\Gamma_Y^{(t,\widetilde q,\widetilde H)} \, \left( \frac{t}{k_t} - \frac{\widetilde q}{k_{\widetilde q}}-\frac{\widetilde H}{k_{\widetilde H}}\right) \notag
+ \Gamma_{\rm ss} \, N_5\,,
\notag\\
\partial_\mu \widetilde t^\mu &= \label{eq:ttilderight}
- \Gamma_Y^{(\widetilde t,\widetilde q,H_1)} \left(\frac{\widetilde t}{k_{\widetilde t}}-\frac{\widetilde q}{k_{\widetilde q}}-\frac{H_1}{k_{H_1}} \right)
-\Gamma_Y^{(\widetilde t,\widetilde q,H_2)} \, \left(\frac{\widetilde t}{k_{\widetilde t}}-\frac{\widetilde q}{k_{\widetilde q}}-\frac{H_2}{k_{H_2}} \right)
+ S^{\: \cancel{CP}}_{\widetilde t}
\\
&\quad  -\Gamma_Y^{(\widetilde t,q,\widetilde H)} \, \left(\frac{\widetilde t}{k_{\widetilde t}}-\frac{q}{k_q}-\frac{\widetilde H}{k_{\widetilde H}}\right)
-\Gamma_{\widetilde V}^{(t,\widetilde t)} \, \left(\frac{\widetilde t}{k_{\widetilde t}}-\frac{t}{k_t} \right)
-\Gamma_M^{(\widetilde t,\widetilde q)}\left( \frac{\widetilde t}{k_{\widetilde t}} - \frac{\widetilde q}{k_{\widetilde q}} \right)\,,\notag \\
\label{eq:bright}
\partial_\mu b^\mu  &=
-\Gamma_M^{(b,q)}\left( \frac{b}{k_b} - \frac{q}{k_q} \right) - \, \Gamma_Y^{(b,q,H_1)} \, \left(\frac{b}{k_b}-\frac{q}{k_q}+\frac{H_1}{k_{H_1}} \right)
-\Gamma_{\widetilde V}^{(b,\widetilde b)} \, \left(\frac{b}{k_b}-\frac{\widetilde b}{k_{\widetilde b}} \right)
\\
&\quad - \, \Gamma_Y^{(b,q,H_2)} \, \left(\frac{b}{k_b}-\frac{q}{k_q}+\frac{H_2}{k_{H_2}} \right)
-\Gamma_Y^{(b,\widetilde q,\widetilde H)} \, \left( \frac{b}{k_b} - \frac{\widetilde q}{k_{\widetilde q}}+\frac{\widetilde H}{k_{\widetilde H}}\right) \notag
+ \Gamma_{\rm ss} \, N_5\,,
\notag\\
\partial_\mu \widetilde b^\mu &= \label{eq:btilderight}
- \Gamma_Y^{(\widetilde b,\widetilde q,H_1)} \left(\frac{\widetilde b}{k_{\widetilde b}}-\frac{\widetilde q}{k_{\widetilde q}}+\frac{H_1}{k_{H_1}} \right)
-\Gamma_Y^{(\widetilde b,\widetilde q,H_2)} \, \left(\frac{\widetilde b}{k_{\widetilde b}}-\frac{\widetilde q}{k_{\widetilde q}}+\frac{H_2}{k_{H_2}} \right)
+ S^{\: \cancel{CP}}_{\widetilde b}
\\
&\quad  -\Gamma_Y^{(\widetilde b,q,\widetilde H)} \, \left(\frac{\widetilde b}{k_{\widetilde b}}-\frac{q}{k_q}+\frac{\widetilde H}{k_{\widetilde H}}\right)
-\Gamma_{\widetilde V}^{(b,\widetilde b)} \, \left(\frac{\widetilde b}{k_{\widetilde b}}-\frac{b}{k_b} \right)
-\Gamma_M^{(\widetilde b,\widetilde q)}\left( \frac{\widetilde b}{k_{\widetilde b}} - \frac{\widetilde q}{k_{\widetilde q}} \right)\,,\notag \\
\partial_\mu q^\mu &= \label{eq:qleft}
- \Gamma_M^{(t,q)}\left(\frac{q}{k_q}-\frac{t}{k_t}\right)
- \Gamma_M^{(b,q)}\left(\frac{q}{k_q}-\frac{b}{k_b}\right)
-\Gamma_{\widetilde V}^{(q,\widetilde q)} \, \left(\frac{q}{k_q}-\frac{\widetilde q}{k_{\widetilde q}}\right)
-2 \, \Gamma_{\rm ss} \, N_5
\\
&\quad
-\Gamma_Y^{(t,q,H_1)} \, \left(\frac{q}{k_q}-\frac{t}{k_t}+\frac{H_1}{k_{H_1}}\right)
- \Gamma_Y^{(t,q,H_2)} \, \left(\frac{q}{k_q}-\frac{t}{k_t}+\frac{H_2}{k_{H_2}}\right)
-\Gamma_Y^{(\widetilde t,q,\widetilde H)} \left(\frac{q}{k_q}-\frac{\widetilde t}{k_{\widetilde t}}+\frac{\widetilde H}{k_{\widetilde H}}\right)
\notag\\
&\quad
-\Gamma_Y^{(b,q,H_1)} \, \left(\frac{q}{k_q}-\frac{b}{k_b}-\frac{H_1}{k_{H_1}}\right)
- \Gamma_Y^{(b,q,H_2)} \, \left(\frac{q}{k_q}-\frac{b}{k_b}-\frac{H_2}{k_{H_2}}\right)
-\Gamma_Y^{(\widetilde b,q,\widetilde H)}
 \left(\frac{q}{k_q}-\frac{\widetilde b}{k_{\widetilde b}}-\frac{\widetilde H}{k_{\widetilde H}}\right)\,,
\notag\\
\partial_\mu \widetilde q^\mu &=\label{eq:qtildetildeleft}
-\Gamma_M^{(\widetilde t,\widetilde q)}\left( \frac{\widetilde q}{k_{\widetilde q}} - \frac{\widetilde t}{k_{\widetilde t}} \right)
-\Gamma_M^{(\widetilde b,\widetilde q)}\left( \frac{\widetilde q}{k_{\widetilde q}} - \frac{\widetilde b}{k_{\widetilde b}} \right)
- \Gamma_{\widetilde V}^{(q,\widetilde q)} \, \left(\frac{\widetilde q}{k_{\widetilde q}}-\frac{q}{k_q}\right)
- S^{\; \cancel{CP}}_{\widetilde t}- S^{\; \cancel{CP}}_{\widetilde b}
\\
&\quad -\Gamma_Y^{(\widetilde t,\widetilde q,H_1)} \, \left(\frac{\widetilde q}{k_{\widetilde q}}-\frac{\widetilde t}{k_{\widetilde t}}+\frac{H_1}{k_{H_1}}\right)
-\Gamma_Y^{(\widetilde t,\widetilde q,H_2)} \, \left(\frac{\widetilde q}{k_{\widetilde q}}-\frac{\widetilde t}{k_{\widetilde t}}+\frac{H_2}{k_{H_2}}\right)
-\Gamma_Y^{(t,\widetilde q,\widetilde H)} \, \left(\frac{\widetilde q}{k_{\widetilde q}}-\frac{t}{k_t}+\frac{\widetilde H}{k_{\widetilde H}}\right)
\notag
\\
&\quad -\Gamma_Y^{(\widetilde b,\widetilde q,H_1)} \, \left(\frac{\widetilde q}{k_{\widetilde q}}-\frac{\widetilde b}{k_{\widetilde b}}-\frac{H_1}{k_{H_1}}\right)
-\Gamma_Y^{(\widetilde b,\widetilde q,H_2)} \, \left(\frac{\widetilde q}{k_{\widetilde q}}-\frac{\widetilde b}{k_{\widetilde b}}-\frac{H_2}{k_{H_2}}\right)
-\Gamma_Y^{(b,\widetilde q,\widetilde H)} \, \left(\frac{\widetilde q}{k_{\widetilde q}}-\frac{b}{k_b}-\frac{\widetilde H}{k_{\widetilde H}}\right)\,,
\notag
\\
\label{eq:tauright}
\partial_\mu \tau^\mu  &=
-\Gamma_M^{(\tau,\ell)}\left( \frac{\tau}{k_\tau} - \frac{\ell}{k_\ell} \right) - \, \Gamma_Y^{(\tau,\ell,H_1)} \, \left(\frac{\tau}{k_\tau}-\frac{\ell}{k_\ell}+\frac{H_1}{k_{H_1}} \right)
-\Gamma_{\widetilde V}^{(\tau,\widetilde \tau)} \, \left(\frac{\tau}{k_\tau}-\frac{\widetilde \tau}{k_{\widetilde \tau}} \right)
\\
&\quad - \, \Gamma_Y^{(\tau,\ell,H_2)} \, \left(\frac{\tau}{k_\tau}-\frac{\ell}{k_\ell}+\frac{H_2}{k_{H_2}} \right)
-\Gamma_Y^{(\tau,\widetilde \ell,\widetilde H)} \, \left( \frac{\tau}{k_\tau} - \frac{\widetilde \ell}{k_{\widetilde \ell}}+\frac{\widetilde H}{k_{\widetilde H}}\right)\,,
\notag\\
\partial_\mu \widetilde \tau^\mu &= \label{eq:tautilderight}
- \Gamma_Y^{(\widetilde \tau,\widetilde \ell,H_1)} \left(\frac{\widetilde \tau}{k_{\widetilde \tau}}-\frac{\widetilde \ell}{k_{\widetilde \ell}}+\frac{H_1}{k_{H_1}} \right)
-\Gamma_Y^{(\widetilde \tau,\widetilde \ell,H_2)} \, \left(\frac{\widetilde \tau}{k_{\widetilde \tau}}-\frac{\widetilde \ell}{k_{\widetilde \ell}}+\frac{H_2}{k_{H_2}} \right)
+ S^{\: \cancel{CP}}_{\widetilde \tau}
\\
&\quad  -\Gamma_Y^{(\widetilde \tau,\ell,\widetilde H)} \, \left(\frac{\widetilde \tau}{k_{\widetilde \ell}}-\frac{\ell}{k_\ell}+\frac{\widetilde H}{k_{\widetilde H}}\right)
-\Gamma_{\widetilde V}^{(\tau,\widetilde \tau)} \, \left(\frac{\widetilde \tau}{k_{\widetilde \tau}}-\frac{\tau}{k_\tau} \right)
-\Gamma_M^{(\widetilde \tau,\widetilde \ell)}\left( \frac{\widetilde \tau}{k_{\widetilde \tau}} - \frac{\widetilde \ell}{k_{\widetilde \ell}} \right)\,,\notag \\
\partial_\mu \ell^\mu &= \label{eq:ellleft}
- \Gamma_M^{(\tau,\ell)}\left(\frac{q}{k_q}-\frac{\ell}{k_\ell}\right)
-\Gamma_{\widetilde V}^{(\ell,\widetilde \ell)} \, \left(\frac{\ell}{k_\ell}-\frac{\widetilde \tau}{k_{\widetilde \tau}}\right)
\\
&\quad
-\Gamma_Y^{(\tau,\ell,H_1)} \, \left(\frac{\ell}{k_\ell}-\frac{\ell}{k_\ell}-\frac{H_1}{k_{H_1}}\right)
- \Gamma_Y^{(\tau,\ell,H_2)} \, \left(\frac{\ell}{k_\ell}-\frac{\tau}{k_\tau}-\frac{H_2}{k_{H_2}}\right)
-\Gamma_Y^{(\widetilde \tau,\ell,\widetilde H)}
 \left(\frac{\ell}{k_\ell}-\frac{\widetilde \tau}{k_{\widetilde \tau}}-\frac{\widetilde H}{k_{\widetilde H}}\right)\,,
\notag\\
\partial_\mu \widetilde \ell^\mu &=\label{eq:elltildetildeleft}
-\Gamma_M^{(\widetilde \tau,\widetilde \ell)}\left( \frac{\widetilde \ell}{k_{\widetilde \ell}} - \frac{\widetilde \tau}{k_{\widetilde \tau}} \right)
- \Gamma_{\widetilde V}^{(\ell,\widetilde \ell)} \, \left(\frac{\widetilde \ell}{k_{\widetilde \ell}}-\frac{\ell}{k_\ell}\right)
- S^{\; \cancel{CP}}_{\widetilde \tau}
\\
&\quad -\Gamma_Y^{(\widetilde \tau,\widetilde \ell,H_1)} \, \left(\frac{\widetilde \ell}{k_{\widetilde \ell}}-\frac{\widetilde \tau}{k_{\widetilde \tau}}-\frac{H_1}{k_{H_1}}\right)
-\Gamma_Y^{(\widetilde \tau,\widetilde \ell,H_2)} \, \left(\frac{\widetilde \ell}{k_{\widetilde \ell}}-\frac{\widetilde \tau}{k_{\widetilde \tau}}-\frac{H_2}{k_{H_2}}\right)
-\Gamma_Y^{(\tau,\widetilde \ell,\widetilde H)} \, \left(\frac{\widetilde \ell}{k_{\widetilde \ell}}-\frac{\tau}{k_\tau}-\frac{\widetilde H}{k_{\widetilde H}}\right)\,,
\notag
\\
\partial_\mu H_{i}^\mu &= \label{eq:hi}
-\Gamma_Y^{(t,q,H_{i})} \, \left(\frac{H_{i}}{k_{H_i}}-\frac{t}{k_t}+\frac{q}{k_q}\right) -\Gamma_Y^{(\widetilde t,\widetilde q,H_{1,2})} \, \left(\frac{H_i}{k_{H_i}}-\frac{\widetilde t}{k_{\widetilde t}}+\frac{\widetilde q}{k_{\widetilde q}}\right) \\
&\quad -\Gamma_Y^{(b,q,H_{i})} \, \left(\frac{H_{i}}{k_{H_i}}+\frac{b}{k_b}-\frac{q}{k_q}\right) -\Gamma_Y^{(\widetilde b,\widetilde q,H_{1,2})} \, \left(\frac{H_i}{k_{H_i}}+\frac{\widetilde b}{k_{\widetilde b}}-\frac{\widetilde q}{k_{\widetilde q}}\right)
\notag\\
&\quad -\Gamma_Y^{(\tau,\ell,H_{i})} \, \left(\frac{H_{i}}{k_{H_i}}+\frac{\tau}{k_\tau}-\frac{\ell}{k_\ell}\right) -\Gamma_Y^{(\widetilde \tau,\widetilde \ell,H_{1,2})} \, \left(\frac{H_i}{k_{H_i}}+\frac{\widetilde \tau}{k_{\widetilde \tau}}-\frac{\widetilde \ell}{k_{\widetilde \ell}}\right)
\notag\\
&\quad  -\Gamma_{\widetilde V}^{(H_{i},\widetilde H)} \, \left(\frac{H_i}{k_{H_i}}-\frac{\widetilde H}{k_{\widetilde H}}\right) - \Gamma_H^{(H_1,H_2)} \left( \frac{H_i}{k_{H_i}} \right) \notag \; , \qquad i=1,2\,,\\
\partial_\mu \widetilde H^\mu &= \label{eq:htilde}
-\Gamma_Y^{(t,\widetilde q,\widetilde H)} \, \left(\frac{\widetilde H}{k_{\widetilde H}}-\frac{t}{k_t}+\frac{\widetilde q}{k_{\widetilde q}}\right) -\Gamma_Y^{(\widetilde t,q,\widetilde H)} \, \left(\frac{\widetilde H}{k_{\widetilde H}}-\frac{\widetilde t}{k_{\widetilde t}}+\frac{q}{k_q}\right) \\
&\quad
-\Gamma_Y^{(b,\widetilde q,\widetilde H)} \, \left(\frac{\widetilde H}{k_{\widetilde H}}+\frac{b}{k_b}-\frac{\widetilde q}{k_{\widetilde q}}\right) -\Gamma_Y^{(\widetilde b,q,\widetilde H)} \, \left(\frac{\widetilde H}{k_{\widetilde H}}+\frac{\widetilde b}{k_{\widetilde b}}-\frac{q}{k_q}\right)
\notag\\
&\quad
-\Gamma_Y^{(\tau,\widetilde \ell,\widetilde H)} \, \left(\frac{\widetilde H}{k_{\widetilde H}}+\frac{\tau}{k_\tau}-\frac{\widetilde \ell}{k_{\widetilde \ell}}\right) -\Gamma_Y^{(\widetilde \tau,\ell,\widetilde H)} \, \left(\frac{\widetilde H}{k_{\widetilde H}}+\frac{\widetilde \tau}{k_{\widetilde \tau}}-\frac{\ell}{k_\ell}\right)
\notag\\
&\quad
-\Gamma_H^{(\widetilde H,\widetilde V)} \, \left(\frac{\widetilde H}{k_{\widetilde H}}\right)
-\Gamma_{\widetilde V}^{(H_1,\widetilde H)} \, \left(\frac{\widetilde H}{k_{\widetilde H}}-\frac{H_1}{k_{H_1}}\right)
-\Gamma_{\widetilde V}^{(H_2,\widetilde H)} \, \left(\frac{\widetilde H}{k_{\widetilde H}}-\frac{H_2}{k_{H_2}}\right)
+S^{\; \cancel{CP}}_{\widetilde H}\,, \notag \\
\partial_\mu q^{\mu}_i &=
-\Gamma_{\widetilde V}^{(q_i,\widetilde q_i)} \, \left(\frac{q_i}{k_{q_i}}-\frac{\widetilde q_i}{k_{\widetilde q_i}} \right)
-2 \, \Gamma_{\rm ss} \, N_5 \; , \qquad i=1,2\,, \\
\partial_\mu \widetilde q^{\mu}_i &= 
-\Gamma_{\widetilde V}^{(q_i,\widetilde q_i)} \, \left(\frac{\widetilde q_i}{k_{\widetilde q_i}}  - \frac{q_i}{k_{q_i}}\right) \; , \qquad i=1,2\,, \\
\partial_\mu  u^{\mu}_i &=
-\Gamma_{\widetilde V}^{(u_i,\widetilde u_i)} \, \left(\frac{u_i}{k_{u_i}}-\frac{\widetilde u_i}{k_{\widetilde u_i}} \right)+ \Gamma_{\rm ss}\, N_5 \; , \qquad i=1,2\,, \label{eq:uright} \\\
\partial_\mu  \widetilde u^{\mu}_i &=
-\Gamma_{\widetilde V}^{(u_i,\widetilde u_i)} \, \left(\frac{\widetilde u_i}{k_{\widetilde u_i}} - \frac{u_i}{k_{u_i}} \right) \; , \qquad i=1,2\,, \label{eq:uitilderight} \\
\partial_\mu  d_i^\mu &=
-\Gamma_{\widetilde V}^{(d_i,\widetilde d_i)} \, \left(\frac{d_i}{k_{d_i}}-\frac{\widetilde d_i}{k_{\widetilde d_i}} \right)+ \Gamma_{\rm ss} \, N_5 \; \qquad i=1,2\,, \\
\label{eq:ditilderight}
\partial_\mu \widetilde d^{\mu}_i &= 
-\Gamma_{\widetilde V}^{(d_i,\widetilde d_i)} \, \left(\frac{\widetilde d_i}{k_{\widetilde d_i}} - \frac{d_i}{k_{d_i}} \right) \; \qquad i=1,2 \,.
\end{align}
\end{subequations}
The rate for strong sphaleron transitions is given
by~\cite{Giudice:1993bb,Huet:1995sh,Moore:1997im}:
\be
\Gamma_{\rm ss} =  16 \, \kappa_s \, \alpha_s^4 \, T
\ee
with $\kappa_s \simeq 1$, and
\begin{equation}
N_5 \equiv \sum_{i=1}^3 \left(\frac{2\,q_i}{k_{q_i}} - \frac{u_i}{k_{u_i}} - \frac{d_i}{k_{d_i}} \right)\;.
\end{equation}

In order to facilitate a comparison with results appearing previously in the literature, we give
the our values for the statistical
factors in the massless limit: $k_q=k_{\widetilde q}/2=6$,
$k_{u_R}=k_{d_R}=k_{\widetilde u_R}/2=k_{\widetilde d_R}/2=3$,
$k_{H_1}=k_{H_2}=4$ and $k_{\widetilde H}=4$.

Before solving the system of Boltzmann equations in Secs.~\ref{sec:analytic},~\ref{sec:numerical}, we conclude the present section by considering two issues related to the equilibration rates $\Gamma_{Y, \widetilde V}$ derived above.  First, we consider the impact on these rates from a more rigorous treatment of massless fermion propagators in a thermal plasma.  Second, we present a numerical comparison of $\Gamma_Y$ and $\Gamma_{\widetilde V}$ rates to demonstrate that one in general does not have $\Gamma_{\widetilde V} \gg \Gamma_Y$.

\subsection{Thermal fermion propagators and particle/hole modes}
\label{sec:particlehole}

For three-body processes involving scalars and fermions with masses of order $T$, the forms for the Greens functions given in Eqns. (\ref{eq:gf>},\ref{eq:gf<}) (and the fermion analogs) suffice for the computation of the three-body rates. In the case of fermions that are massless (or nearly massless) at zero temperature, the structure of the thermal Greens functions becomes far more complicated. The renormalized fermionic spectral functions contain additional poles -- so called \lq\lq hole modes" -- generated by mixing between single and multiparticle states in the thermal bath~\cite{Klimov,Weldon:1989ys}. The resulting propagators are given by~\cite{Weldon:1999th,Lee:2004we}
\be
\label{eq:slambdaint}
S^\lambda(x,y;\mu)=\int {d^4k\over (2\pi)^4} e^{-ik\cdot(x-y)} 
g_F^\lambda(k_0, \mu) 
\left[\frac{\gamma_0-\boldgamma\mcdot\vect{\hat k}}{2}\rho_+(k_0, \vect{k}, \mu)
+ \frac{\gamma_0+\boldgamma\mcdot\vect{\hat k}}{2}\rho_-(k_0, \vect{k}, \mu)\right]
\,,
\ee
where $\vect{\hat k}$ is the unit vector in the $\vect{k}$ direction, and
\begin{equation}
\label{eq:rhoplus}
\begin{split}
\rho_+(k_0,\vect{k},\mu) = i\biggl[&\frac{Z_p(k,\mu)}{k_0-\mathcal{E}_p(k,\mu)}
-\frac{Z_p(k,\mu)^*}{k_0-\mathcal{E}_p(k,\mu)^*} \\
+ &\frac{Z_h(k,-\mu)^*}{k_0+\mathcal{E}_h(k,-\mu)^*}
- \frac{Z_h(k,-\mu)}{k_0+\mathcal{E}_h(k,-\mu)}+F(k_0^*,k,\mu)^*-F(k_0,k,\mu)\biggr]\,,
\end{split}
\end{equation}
and
\begin{equation}
\label{eq:rhominus}
\rho_-(k_0,\vect{k},\mu) = [\rho_+(-k^{0*},\vect{k},-\mu)]^*\,.
\end{equation}
%i\biggl[{Z_p(k,-\mu)^*\over k_0+E_p(k,-\mu)^*}
%-{Z_p(k,-\mu)\over k_0+E_p(k,-\mu)}\\
%\nonumber
%&&+{Z_h(k,\mu)\over k_0-E_h(k,\mu)}
%-{Z_h(k,\mu)^*\over k_0-E_h(k,\mu)^*}+F(-k_0^*,k,-\mu)^*-F(-k_0,k,-\mu)\biggr]
%\eea
%
Here, $\mathcal{E}_p(k,\mu)$ and $-\mathcal{E}_h(k,-\mu)^*$ are the two (complex) roots (in $k_0$) of
\be
k_0-k+D_+(k_0, k,\mu)+i\epsilon \label{poleeqn}
\ee
where $iD_{\pm}(k_0,k,\mu)$ are contributions to the inverse, retarded propagator proportional to $(\gamma_0\mp\boldgamma\mcdot\vect{\hat k})/2$ arising from interactions.  The function $F(k_0, k,\mu)$ gives the non-pole part of the propagator, and $k=\abs{\vect{k}}$.   
The mass of these particle/hole excitations is neither zero, nor equal to the standard thermal mass (denoted $m_T$); rather it is a $k$-dependent function given by
\be
m^2_{p,h}(k) = \textrm{Re}\left[\mathcal{E}_{p,h}(k)\right]^2 - k^2 \;.
\ee
For example, the mass of the particle mode obeys
\bea
m_p &\to& m_T  \qquad \textrm{for} \; k \to 0 \\
m_p &\to& 2 \, m_T \qquad \textrm{for} \; k \gg T  \;.
\eea

The residue functions $Z_p(k,\mu)$ and $Z_h(k,\mu)$ govern the relative importance of particle and hole contributions to the thermal propagators. In the absence of interactions, one has $Z_p(k,\mu)=1$ and $Z_h(k,\mu)=0$, in which case $S^\lambda(x,y;\mu)$ takes on the form given in Eqns.~(\ref{eq:spectral1}-\ref{eq:slambdafree}). However, the departure from these limits is not perturbative in the strength of the interaction ({\em e.g.}, the gauge coupling $g$), but rather depends strongly on the magnitude of the three-momentum, $k$. At $k=0$ one has $Z_p=Z_h=1/2$, while for $k$ or order the temperature or thermal mass, one has $Z_h/Z_p \ll 1$. In earlier work~\cite{Lee:2004we}, we studied the impact of the particle-hole structure of $ S^\lambda(x,y;\mu)$ on fermionic contributions to the CP-violating source terms and CP-conserving relaxation rates generated by interactions with the spacetime varying Higgs vevs. We found that the gaugino and Higgsino contributions were dominated by large momenta (of order the gaugino and Higgsino masses), leading to negligible effects associated with the hole modes. In contrast, the hole modes generate non-negligible contributions to the quark relaxation rates. 

Here, we analyze the impact of hole modes on the three-body rates since it is not {\em a priori} apparent that the loop integrals are dominated by momenta in the regime for which $Z_h/Z_p \ll 1$. To that end, we consider one particular three-body process involving the massless fermion $q$, the massive fermion $\widetilde{g}$, and the scalar $\widetilde{q}$. We compute the three-body rate $\Gamma_{\widetilde{V}}^{(q, \widetilde{q})}$ generated by the graph of FIG.~\ref{fig:feyn_figs}a but using Eq.~(\ref{eq:slambdaint}) for the $q$ Green's function.  (Although we are choosing a particular interaction, this discussion is generic to any process involving one massless fermion, one massive fermion, and one massive scalar.)  For illustrative purposes, we include only the gluino contribution, neglecting here those contributions from electroweak gauginos. 

\begin{figure}[ht]
\includegraphics[scale=1,width=8cm]{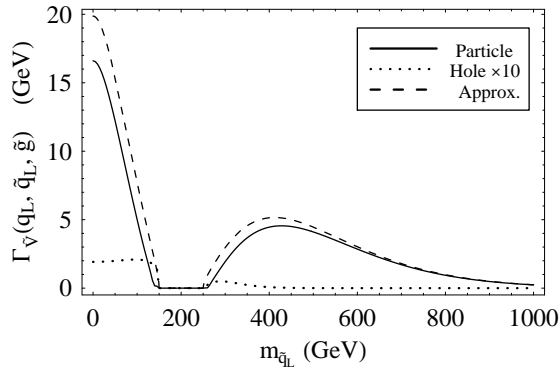}
\caption{We plot $\Gamma_{\widetilde{V}}^{(q, \widetilde{q})}$ (gluino-only) as a function of $m_{\widetilde{q}_L}$.  The solid and dotted lines denote, respectively, the particle and hole ($\times 10$) contributions to this rate.  The dashed line indicates $\Gamma_{\widetilde{V}}^{(q, \widetilde{q})}$ calculated using free thermal Green's functions with a thermal mass, corresponding to the last term in Eq.~(\ref{eq:GammaVtilde}c).}
\label{fig:quasi}
\end{figure}

In FIG.~\ref{fig:quasi}, we show the resulting $m_{\widetilde{q}}$-dependence of $\Gamma_{\widetilde{V}}^{(q, \widetilde{q})}$ for $m_{\widetilde{g}}=200$ GeV and $T=100$ GeV.  The solid and dotted curves show, respectively, the particle and hole ($\times 10$) contributions to $\Gamma_{\widetilde{V}}^{(q, \widetilde{q})}$, calculated using the full quasiparticle Green's function given in Eqns.~(\ref{eq:slambdaint}-\ref{eq:rhominus}), in the limit of zero thermal widths.  The hole contribution is much smaller than the particle contribution because the process is dominated by momenta $k \gtrsim T$; for these momenta, we always have $Z_h \lesssim 10^{-3}$.    

For comparison, the dashed curve in FIG.~\ref{fig:quasi} shows $\Gamma_{\widetilde{V}}^{(q, \widetilde{q})}$ calculated using free Green's functions (\ref{eq:slambdafree}) with the inclusion of thermal masses.  The agreement between solid and dashed curves in generally quite good; the majority of the discrepancy is due to the fact that $Z_p \ne 1$.  At $m_{\widetilde{q}_L} > 600$ GeV, the rate $\Gamma_{\widetilde{V}}^{(q, \widetilde{q})}$ is dominated by quasi-particles with momenta $k > (\textrm{few})\times 100$ GeV, for which $Z_p \gtrsim 0.95$; consequently, the agreement between solid and dashed curves is better than $\sim 95\%$. For $m_{\widetilde{q}_L} < 100$ GeV, the rate is dominated by quasi-particles with momenta $k \sim 100$ GeV, for which $Z_p \sim 0.8$; consequently, there is a $~20\%$ discrepancy between solid and dashed curved.  In the present work, we compute $\Gamma_{\widetilde{V}}^{(q, \widetilde{q})}$ and all other three-body rates using free Green's functions with thermal masses, rather than the full quasi-particle Green's functions.  Since these quasi-particle effects provide only small corrections to the three-body rate, we do not consider the more complicated case of two massless fermions and one massive scalar.

\section{Transport equations and $Y_B$: analytic study}
\label{sec:analytic}

In this section, we present semi-analytical approximate
solutions to the Boltzmann equations. In addition to the
discussion that has already been presented in Refs.~\cite{Chung:2008aya,Chung:2009cb},
we pay particular attention to the relaxation of particular charge
densities towards superequilibrium in the diffusion wake ahead of the
bubble wall. For this purpose, we first
identify the condition that determines whether or not a particular
interaction rate is sufficiently fast to lead to chemical equilibrium.  
We then show that fast supergauge interactions rates are a sufficient, but not 
necessary, condition for superequilibrium.  Indeed, it turns out that
the combined effect of Yukawa and triscalar interactions can also
lead to superequilibrium for particular species.

\subsection{Conditions for superequilibrium}
\label{section:superconditions}

Having specified the relevant interactions during EWB, we discuss here under
which circumstances a particular interaction maintains chemical equilibrium and when this equilibrium is physically relevant. As an example, we consider supergauge interaction rates in the transport
equations for $q$ and $\widetilde{q}$:
\bea
\partial_\nu q^\nu &=&  - \;  \Gamma_{\widetilde{V}}^{(q,\widetilde{q})} \, \left( \frac{q}{k_q} - \frac{\widetilde{q}}{k_{\widetilde{q}}} \right) + \; ... \\
\partial_\nu \widetilde{q}^\nu &=&  -  \; \Gamma_{\widetilde{V}}^{(q,\widetilde{q})} \, \left(  \frac{\widetilde{q}}{k_{\widetilde{q}}} - \frac{q}{k_q}  \right) + \; ... 
\eea
For the moment, we focus only on the gaugino interactions that maintain
chemical equilibrium between $q$ and $\widetilde{q}$.  Expressing these two
equations in terms of chemical potentials, rather than charge number densities, and
taking the difference, we have
\be
\label{eq:muequilib}
\left[ \frac{d}{dt} - D_q \nabla^2\right] \, (\mu_q - \mu_{\widetilde{q}}) = - \; \Gamma_{\widetilde{V}}^{(q,\widetilde{q})} \, \left( \frac{1}{k_q} + \frac{1}{k_{\widetilde{q}}} \right) \, (\mu_q - \mu_{\widetilde{q}}) + \; ...
\ee

Formally, chemical equilibrium corresponds to the equality of chemical potentials.  
Eq.~\eqref{eq:muequilib} implies that the difference of chemical potentials, $\mu_q-\mu_{\tilde q}$, will relax to zero with a characteristic time scale 
\be
\label{tau:quarksquark}
\tau_{\textrm{eq}} \equiv \left[ \Gamma_{\widetilde{V}}^{(q,\widetilde{q})} \,  \left( \frac{1}{k_q} + \frac{1}{k_{\widetilde{q}}} \right) \right]^{-1}  \ \ \ .
%\label{eq:gammaeff}
%\tau_{\textrm{eq}}^{(\widetilde q,q)}\equiv
%\left[ \Gamma_{\widetilde{V}}^{(q,\widetilde{q})}/{k_{\widetilde{q}}} \right]^{-1}
\ee

Consider now the presence of a density $q$ ({\it e.g.} induced by decays ${\tilde H}^\dagger \to q + {\tilde t}^\ast$). In general, it will equilibrate with $\widetilde{q}$ on the time scale $\tau_{\textrm{eq}}$, so long as the corresponding reaction $q+\widetilde{V} \leftrightarrow \widetilde{q}$ is kinematically allowed. In the limit that  gauginos become heavy compared to the temperature, we expect this equilibration process to be Boltzmann suppressed, that is:
\be
\label{massivegauginos}
\tau_\mathrm{eq}\to\infty\ ,\qquad m_{\widetilde{V}} \to \infty\ \ \ .
\ee
This suppression of the equilibration rate arises as one would expect because the gauginos decouple from the plasma. On the other hand, taking the limit $m_{\widetilde{q}} \to \infty$ leads to
\be
\tau_{\textrm{eq}}^{(\widetilde q,q)} \propto \frac{1}{m_{\widetilde{q}}} \longrightarrow 0\, .
\ee
This result is somewhat counter intuitive, since the $\widetilde{q}$ density becomes Boltzmann suppressed for large $m_{\widetilde{q}}$. However, the corresponding chemical potential is proportional to $\widetilde{q}/k_{\widetilde{q}}$, and $k_{\widetilde{q}}$ also decreases with larger $m_{\widetilde{q}}$. As a result, the chemical potentials of a particle
and its superpartner may equilibrate quickly in the presence of gauginos,
even when the sparticle
mass is much larger than the temperature. This is because the sparticle
chemical potential can adapt to the particle chemical potential by a small
change in the physical sparticle density.

%Suppose now the presence of a density $q$ ({\it e.g.} induced by decays ${\tilde H}^\dagger \to q + {\tilde t}^\ast$, and consider the rate at
%which $\widetilde q$ equilibrates with $q$.
%This is not $\left(\Gamma_{\widetilde{V}}^{(q,\widetilde{q})}\right)^{-1}$, but is instead
%\be
%\label{tau:quarksquark}
%\tau_{\textrm{eq}} \equiv \left[ \Gamma_{\widetilde{V}}^{(q,\widetilde{q})} \,  \left( \frac{1}{k_q} + \frac{1}{k_{\widetilde{q}}} \right) \right]^{-1}  \; \label{eq:gammaeff}
%\tau_{\textrm{eq}}^{(\widetilde q,q)}\equiv
%\left[ \Gamma_{\widetilde{V}}^{(q,\widetilde{q})}/{k_{\widetilde{q}}} \right]^{-1}
%\ee
%because of our conventions used in the definition of reaction rates.
%In the limit that $m_{\widetilde{V}} \to \infty$, we have $\tau_{\textrm{eq}} \to \infty$; but if $m_{\widetilde{q}} \to \infty$, we have
%\be
%\tau_{\textrm{eq}}^{(\widetilde q,q)} \propto \frac{1}{m_{\widetilde{q}}} \longrightarrow 0\,.
%\ee
% This is because in the large $m_{\widetilde{q}}$ limit, the relaxation rate for
% the chemical potential is large, since a small change in the number density of
% $\widetilde{q}$ induces a large change in $\mu_{\widetilde{q}}$.
% In other words, the supergauge interaction does not turn off if $\widetilde{q}$ is heavy; on the contrary, large $m_{\widetilde{q}}$ helps to induce superequilibrium.
%The only way to suppress supergauge interaction rates is therefore to have massive gauginos: 
%\be
%m_{\widetilde{V}} \gg T\;. \label{massivegauginos}
%\ee

From the standpoint of the generation of $n_\mathrm{left}$,  it is important to analyze the converse situation, namely,  how the presence of a heavy superpartner affects the chemical potential, and hence the number density, of the lighter SM particles through the network of transport equations. From our numerical studies, we find that when a (s)particle ({\em e.g.} $\widetilde{q}$), becomes heavy compared to the temperature, its transport equation effectively decouples from the system of transport equations. Consequently, even though a process of chemical equilibration involving this heavy (s)particle may take place quickly, its occurrence will be irrelevant for the generation of the lighter particle densities. As we will see below, this decoupling effect can lead to \lq\lq bottlenecks" in a chain of reactions that might otherwise lead to chemical equilibrium and significant effects on particle number densities.

With these considerations in mind, we will consider the equilibration
time scales as in Eq.~\eqref{tau:quarksquark} to be physically relevant only when the (s)particles involved are not too heavy compared to the temperature. In this regime, the corresponding $k$-factors are never too suppressed with respect to their values in the massless limit. When more than two particles are involved in the equilibrating reaction, the corresponding equilibration rate will go like
\be
\tau_\mathrm{eq}^{-1}\sim \Gamma \times \sum_j \frac{1}{k_j}\ \ \ .
\ee
For purposes of determining the criteria for various reactions to reach equilibrium on time scales short compared to other processes,  we will take the longest possible time scale for each reaction by retaining only the minimum value of the $1/k_j$ appearing above:
\begin{equation}
\tau_{\rm eq}^{-1}=\Gamma^{(x_1,x_2,...)}\min_{i}\left\{\frac{1}{k_{x_i}}\right\}\,.
\label{eq:gammaeff}
\end{equation}
A sufficient condition for determining whether chemical equilibrium is maintained during the process of diffusion ahead of the bubble wall is then 
\be
\tau_{\textrm{eq}} < \tau_{\textrm{diff}}
\ee
where
\be
\label{tau:diff}
\tau_{\textrm{diff}} \equiv \frac{\bar{D}}{v_w^2}
\ee
is the diffusion time scale, with effective diffusion constant $\bar{D}$ defined below.  If $\tau_{\textrm{eq}} > \tau_{\textrm{diff}}$, then an asymmetry between chemical potentials
%(e.g. $\mu_q - \mu_{\widetilde{q}}$)
will diffuse ahead of the moving bubble wall faster than it will equilibrate away; therefore, equilibrium will be broken (unless there is another interaction fast enough to maintain equilibrium).
When we evaluate all the interaction rates in Eqns.~(\ref{eq:tright}-\ref{eq:ditilderight}) numerically in Sec.~\ref{sec:numerical}, we will see which rates satisfy $\tau_{\textrm{diff}}/\tau_{\textrm{eq}} > 1$ so that they maintain chemical equilibrium.  (We note parenthetically that the Hubble rate $H \sim 10 \times T^2/M_{pl}$ is far too small to be relevant for EWB transport dynamics).
In our previous papers~\cite{Chung:2008aya,Chung:2009cb}, we have
taken $\tau_{\textrm{eq}}^{-1}=\Gamma^{(x_1,x_2,...)}$ for simplicity.
The defintition~(\ref{eq:gammaeff}) corresponds to a weaker
requirement in the sense that $\tau_{\textrm{eq}}$ goes to
zero in the case when all masses of the particles involved in the 
reaction are much larger than $T$,

Now, one might  expect that a sufficient condition  for breaking superequilibrium is simply Eq.~(\ref{massivegauginos}).   This expectation
is  false, as we  can see  from the  following argument.   Suppose all
gaugino  interaction  rates  vanished,  while the  Yukawa  rates  were
infinitely large.  Top quark
Yukawa interactions in chemical equilibrium lead to
the following  relations among  chemical potentials: 
\begin{subequations}
\label{Yuk1}
\bea
\mu_{q} + \mu_{H_1}  &=&  \mu_{t}  \\
\mu_{q} +  \mu_{H_2}  &=& \mu_{t} \label{Yuksinalpha}\\
\mu_{\widetilde{q}}     +     \mu_{H_1}     &=& \mu_{\widetilde{t}} \\ \mu_{\widetilde{q}}     +     \mu_{H_2}     &=& \mu_{\widetilde{t}}  \\
\mu_{q}   +   \mu_{\widetilde{H}}   &=& \mu_{\widetilde{t}}  \\
\mu_{\widetilde{q}}  + \mu_{\widetilde{H}} &=&    \mu_{t}
\eea
\end{subequations}
Using    these   relations, one has
\be
\label{supertop}
\mu_{q} = \mu_{\widetilde{q}}
\;, \qquad  \mu_{t} =  \mu_{\widetilde{t}} \;, \qquad  \mu_{H_1}   =  \mu_{H_2} = \mu_{\widetilde{H}}   \;.
\ee
Note that the interaction leading to~(\ref{Yuksinalpha}) is suppressed
by the Higgs mixing parameter $\sin\alpha$. However, Eqs.~(\ref{supertop})
follow from Eqs.~(\ref{Yuk1}) even when Eq.~(\ref{Yuksinalpha}) is excluded.
Analogously, when the interactions mediated by bottom and tau
Yukawa interactions are fast, it additionally follows that
\begin{equation}
\label{superdown}
 \mu_{b} =  \mu_{\widetilde{b}}\;, \qquad
\mu_\ell=\mu_{\widetilde \ell}\;, \qquad
 \mu_{\tau} = \mu_{\widetilde{\tau}}\,.
\end{equation}

This implies that  even   in the limit
$m_{\widetilde{V}} \to \infty$, where supergauge interactions are
suppressed,  superequilibrium can  be maintained  through Yukawa
interactions alone.
Fast supergauge interactions are a
sufficient condition for superequilibrium, but not a necessary one in
the presence  of other interactions.  Obviously, this  argument does not
apply to the first and  second generation (s)quark and (s)lepton sectors, where
the Yukawa rates are much  too small to enforce superequilibrium
in  the absence  of fast  supergauge  interactions.

%{\bf ** I understand the sense of the following argument: there is a bottleneck when one species decouples from the plasma. However, the earlier argument implies that when the sfermion goes to the decoupling limit, the equilibration time scale goes to zero. This has no physical relevance, it seems to me, because in the decoupling limit, the chemical potential of the heavy particle really plays no role and in the transport equations for the number density, its effect will go to zero by the 1/k factor. 

%Hence, I think we concluded that $\Gamma$ rather than $\Gamma/k$ is the relevant quantity. 

%To say it another way, the physical significance of superequilibrium is not simply that chemical potentials are equal, but that they are sufficiently large to lead to large $n_{\mathrm left}$. At the end of the day, we really solve a set of coupled equations for the densities since we ultimately need  the baryon number density, not the baryon chemical potential. 

%This whole discussion has to be clarified. **}

But also third generation (s)quarks and (s)leptons may not reach superequilibrium
in the case when one or more of the (s)quark and (s)lepton species are very heavy.
Then, particular Yukawa and triscalar rates can be small such that
$\tau_{\rm eq}$ defined according to Eq.~(\ref{eq:gammaeff}) fails the
criterion $\tau_{\rm eq}<\tau_{\rm diff}$. As discussed above, this occurs because the heavy
(s)quarks and (s)leptons can consitute a bottleneck for the transfer
of the Higgsino charge to the non-supersymmetric particles.
As a consequence,
not all of the relations~(\ref{Yuk1}) need to be satisfied at the same time
and superequilibrium~(\ref{supertop}) does not follow. We encounter an example for
such a situation in Section~\ref{sec:NumNoSG}.

To summarize, we identify two equilibration time-scales:
\begin{itemize}
\item
The time-scale $\tau_{\rm eq}^{q,\widetilde q}$ in Eq.~(\ref{tau:quarksquark}) is useful to show that superequilibrium
is maintained in the presence of light gauginos. In the case of a heavy
$\widetilde q$, the condition $\mu_q=\mu_{\widetilde q}$ is maintained
due to the smallness of $k_{\widetilde q}$, but it is of little physical significance since the density of $\widetilde q$ is small.
\item
The scale $\tau_{\rm eq}$ in Eq.~(\ref{eq:gammaeff}) is useful to
formulate sufficient conditions for equilibration. When
$\tau_{\rm eq}>\tau_{\rm diff}$, there may or may not be a bottleneck
preventing the establishment of chemical equilibrium on diffusion
time-scales. Whether chemical equilibrium is maintained depends in this
situation on the particular masses and interaction rates important
for the network of reactions.
\end{itemize}

\subsection{Analytical approximation}
\label{Sec:Analytical}
Before we explore numerical solutions to the diffusion equations, we briefly
recapitulate some salient points of
the analytical approximation that has been presented
in~\cite{Chung:2008aya,Chung:2009cb}.
First suppose that the relations~(\ref{supertop},\ref{superdown}) hold, because of
fast supergauge interactions
or fast Yukawa and triscalar interactions. A relaxation of this assumption, among
other things, is numerically studied in Section~\ref{sec:numerical}.

Then, it is convenient to introduce common particle and superparticle densities as
\begin{subequations}
\label{eq:supereq}
\label{eq:analyticsol}
\begin{align}
\frac{x}{k_x} =& \frac{\widetilde{x}}{k_{\widetilde{x}}} = \frac{X}{k_X}\,,\quad
X=x+\widetilde x\,,\quad k_X=k_x+k_{\widetilde x}\,,
\nonumber\\
&\textrm{for} \; x \in \{q_i, u_i, d_i, \ell,\tau\equiv r\}, \; i \in \{1,2,3\}\,;
\\
\frac{H_i}{k_{H_i}} =& \frac{\widetilde{H}}{k_{\widetilde{H}}} = \frac{H}{k_H}\,, \quad \textrm{for} \; i\in \{1,2\}\,,
\quad H=\widetilde H+H_1+H_2\,,
\quad k_H=k_{\widetilde H}+k_{H_1}+k_{H_2}
\;.
\end{align}
\end{subequations}
Adding the equations for particles and their superpartners then immediately reduces
the system of Boltzmann equations~(\ref{eq:system}) to the more compact system
of equations presented in Ref.~\cite{Chung:2009cb} that consists of
six equations only.

The next step towards an analytical solution is to realize
that the sums of chemical potentials
that multiply the Yukawa, triscalar\footnote{We assume in this section
that we are in a parametric region, where $t$-, $b$- and $\tau$-Yukawa and triscalar rates are fast.} and strong sphaleron
rates can simultaneously be set to zero, as these rates are typically faster than the diffusion rate and the corresponding reactions reach equilibrium before densities can diffuse away. Explicitly,
when imposing 
\begin{equation}
\frac{Q}{k_Q}+\frac{H}{k_H}-\frac{T}{k_T}=
\frac{Q}{k_Q}-\frac{H}{k_H}-\frac{B}{k_B}=
\frac{L}{k_L}-\frac{H}{k_H}-\frac{R}{k_R}=0\,,
\end{equation}
in conjunction with approximate
baryon number conservation (weak sphaleron transitions are out of equilibrium
on diffusion time-scales $\Gamma_{\rm ws}\ll\tau_{\rm diff}^{-1}$),
we can use these relations to
eliminate all number densities except for $H$ from the Boltzmann equations
through
\begin{subequations}
\label{kappa:all}
\begin{align}
Q&=\kappa_Q H = \frac{k_Q}{k_H}\frac{k_B-k_T}{k_B+k_Q+k_T} H\,,\label{kappa:Q}\\
T&=\kappa_T H = \frac{k_T}{k_H} \frac{2 k_B +k_Q}{k_B+k_Q+k_T} H\,,\label{kappa:T}\\
B&=\kappa_B H = -\frac{k_B}{k_H}\frac{2k_T+k_Q}{k_B+k_Q+k_T} H\,,\label{kappa:B}\\
L&=\kappa_L H = \vartheta_L \frac{k_L}{k_H} \frac{k_R D_R}{k_L D_L+k_R D_R} H\label{kappa:L}\,.
\end{align}
\end{subequations}
Here, we have defined
\begin{equation}
\vartheta_L=\left\{
\begin{array}{l}
1\quad\textnormal{if}\quad\Gamma_{y\tau}\gg\Gamma_{\rm diff}\\
0\quad\textnormal{if}\quad\Gamma_{y\tau}\ll\Gamma_{\rm diff}
\end{array}
\right.\,.
\end{equation}
If neither of these inequalities is amply fulfilled, the analytical approximation
will not be accurate. Similarly, for $D_R\not=D_L$, Eq.~(\ref{kappa:L})
is in conflict with the fact that the densities $R$ and $L$ diffuse at a
different rate ahead of the wall (unlike for the (s)quarks, which we assume here
to diffuse at the common rate $D_Q$, that is dominated by strong interactions).
However, since $D_R\gg D_L$, the the relation~(\ref{kappa:L}) is not a bad
approximation compared to other uncertainties in the analytical calculation.
This is because the right-handed (s)leptons diffuse on much larger distances
away from the bubble wall, such that their local density can be neglected. Note that global lepton number conservation
$\int dz (L+R)=0$ holds, when neglecting
weak sphaleron transitions. 
For more details on this point,
see the discussion in Ref.~\cite{Chung:2009cb}. 
%An analytical
%calculation that is also valid when there is no hierarchy in the diffusion
%constants, will appear in the near future~\cite{Chung:2009B} {\bf ** really ? **}.

Another feature of Eqs.~(\ref{kappa:all}) is that these relations
imply that the contribution from third generation quarks to the factor $N_5$ that multiplies the strong sphaleron rate,
$\Gamma_{\rm ss}$, vanishes. Since the $CP$-violating sources, triscalar couplings, and Yukawa interactions for the first two generation (s)fermions are highly suppressed by their Yukawa couplings, there exists no independent source for their densities apart from their coupling to the third generation via the strong sphalerons. 
Consequently, the vanishing third generation contribution to $N_5$ implies that no chiral charge densities within the first two generations are
generated~\cite{Chung:2008aya,Chung:2009cb}. This situation differs substantially from 
from that of earlier studies, where bottom Yukawa
couplings are neglected~\cite{Huet:1995sh,Lee:2004we}, and it leads to
significantly different results for $Y_B$ and its dependence on the MSSM parameters.

Applying the eliminations~(\ref{kappa:all}), the resulting diffusion equation is
\begin{equation}
\label{eq:diff}
v_w H^\prime-\bar D H^{\prime\prime}=-\bar\Gamma H + \bar S\,,
\end{equation}
and in the symmetric phase, ahead of the bubble wall, its relative
accuracy is ${\cal O}(\Gamma_{\rm diff}/{\Gamma_{Y}})$. In this equation,
the effective
diffusion constant, source terms  and damping rates are given
by~\cite{Chung:2009cb}
\begin{subequations}
\begin{align}
\bar D&=
\frac{D_H+D_Q(\kappa_T-\kappa_B)+D_L \vartheta_L\kappa_L}{1+\kappa_T-\kappa_B+
\vartheta_L \kappa_L}\,, \\
\label{eq:gammabar}
\bar \Gamma&=
\frac{\Gamma_h+\Gamma_{mt}+\Gamma_{mb}+\vartheta_L\Gamma_{m\tau}}
{k_H(1+\kappa_T-\kappa_B+\vartheta_L\kappa_L)}\,,\\
\bar S&=\frac{S^{\cancel{CP}}_{\widetilde H}+S^{\cancel{CP}}_{\widetilde H}
-S^{\cancel{CP}}_{\widetilde b}+\vartheta_L S^{\cancel{CP}}_{\widetilde \tau}
}{1+\kappa_T-\kappa_B+\vartheta_L \kappa_L}\,.
\end{align}
\end{subequations}
Using Eqs.~(\ref{kappa:all})
The left handed fermionic charge density, that couples to the weak sphaleron
is related to $H$ as
\begin{equation}
\label{n:left}
n_{\rm left}=\left(
\frac{k_q}{k_H}\frac{k_B-k_T}{k_B+k_Q+k_T}
+\vartheta_L \frac{k_\ell}{k_H}\frac{k_R D_R}{k_L D_L + k_R D_R}
\right)H\,.
\end{equation}

In the symmetric phase, where the sources and relaxation terms are vanishing,
the Higgs-Higgsino density is then given by
\begin{equation}
H={\cal A}{\rm e}^{v_w z/{\bar D}}\,.
\end{equation}
Assuming that the source terms are negligible for $z<-L_w/2$, and that the
relaxation terms have the particular form $\bar\Gamma(z)=\theta(z)\bar \Gamma$,
the normalization ${\cal A}$ can be found as
\begin{equation}
\label{eq:amplitude}
{\cal A}=\int\limits_0^\infty dy\,\bar S(y)
\frac{{\rm e}^{-\gamma_+ y}}{\bar D\gamma_+}
\,+\!\!\!\int\limits_{-L_w/2}^0 dy\,\bar S(y)\left(
\frac{\gamma_-}{v_w\gamma_+}+\frac{{\rm e}^{-v_wy/{\bar D}}}{v_w}
\right)\,,
\end{equation}
where
\begin{equation}
\label{eq:gammapm}
\gamma_\pm=\frac{1}{2\bar D}\left(
v_w\pm\sqrt{v_w^2+4\bar\Gamma \bar D}
\right)\,.
\end{equation}

These analytic results, which have been discussed extensively in Refs.~\cite{Chung:2008aya,Chung:2009cb}, suggest qualitative features one should expect from the full numerical solutions that we present in Section \ref{sec:numerical}. 

First, the presence of important bottom Yukawa interactions effectively quenches the contributions from the first and second generation quarks to $n_\mathrm{left}$. This quenching arises in this regime because Yukawa-induced equilibrium involving third generation quarks leads to vanishing third generation chiral charge. Non-vanishing first and second generation quark densities arise only as required to maintain strong sphaleron equilibrium and thus, in this limit, also vanish. The resulting expression for $n_\mathrm{left}$, given in Eq.~\eqref{n:left}), depends only on quantities arising from third generation left-handed fermions. 

Second, the efficiency with which the non-vanishing $H$ density -- induced by the corresponding CP-violating source terms for Higgsinos -- converts to $n_\mathrm{left}$ depends critically on the $k$-factors associated with the right-handed third generation sfermions. The combination of Yukawa-induced equilibrium, superequilibrium, and local baryon number conservation implies that the third generation LH quark density induced by $H$ depends on $k_B-k_T$. 
In the limit that the RH top and bottom squarks are degenerate, this contribution will vanish, while for a non-degenerate spectrum, the sign of this contribution will depend on which of the two squarks is heavier. The $k$-factor dependence of the third generation LH leptons is more complicated since the LH and RH sleptons may diffuse at different rates.  In the illustrative limit of equal diffusion constants for the two chiral species, the lepton contribution depends on the geometric mean of the two $k$-factors. If either the LH or RH sleptons become heavy compared to the temperature, the corresponding $k$-factor is suppressed, signaling a decoupling of the slepton from the plasma and quenching the lepton contribution to $n_\mathrm{left}$. Note that the sign of the lepton and bottom quark contributions are both opposite to that of the top quark contribution, implying that neglect of the former can lead to an overestimate of $n_\mathrm{left}$ -- and, thus, of $Y_B$ -- compared to the result when they are included. 

In the following discussion, we will see how these features emerge from the full numerical solutions to the coupled transport equations in regions of parameter space where the corresponding assumptions behind the analytic treatments are valid. We will also identify regions wherein these assumptions break down yet some of the qualitative features persist, as well as regions wherein we find significant departures.

%\subsection{Summary}
%\label{sec:AnSummary}

%{\tiny\it
%We summarize our results from this section as a guide for our expectations in the numerical analysis to follow.  
%\begin{enumerate}
%\item Superequilibrium will be maintained for $(t,\widetilde{t})$, $(q,\widetilde{q})$, and $(H_1,H_2,\widetilde{H})$, independent of the supergauge interaction rates, due to Yukawa processes.
%\item For fixed sign of the CP-violating sources, the baryon asymmetry can change sign, depending on the sign of $\kappa_q + \kappa_{12}$.
%\item If the gauginos are sufficiently heavy, superequilibrium can be broken for densities which do not participate in top Yukawa interactions: namely, $(u_i,\widetilde{u}_i), \; (q_i,\widetilde{q}_i )\; (i=1,2)$ and $(d_i,\widetilde{d}_i )\; (i=1,2,3)$.  If a particular squark density is not in chemical equilibrium with its superpartner, the baryon density will be independent of that squark's mass.
%\item If a squark is much heavier than its superpartner, the baryon asymmetry will be independent of whether or not the supermultiplet is in superequilibrium.
%\end{enumerate}
%}

\section{Transport equations and $Y_B$: numerical results}

\label{sec:numerical}

We numerically solve the full system of Boltzmann Eqs.~(\ref{eq:system})
in the presence of finite $\Gamma_{\tilde V}, \Gamma_{Y}, \Gamma_{\rm ss}$.  Initially, we focus on one particular benchmark point,
which is motivated from the requirement of a strong first order phase transition within the MSSM and for which our analytical approximations
are valid (Section~\ref{sec:NumYukSGEq}). Then, we explore regions of smaller $\tan\beta$, where the
analytical approximations break down because down-type Yukawa 
interactions do not equilibrate on diffusion time-scales (Section~\ref{sec:NumTanBeta}).
We also investigate how robust the assumption of supergauge equilibrium
is, in particular whether it can be maintained even when gaugino 
interactions are quenched (Section~\ref{sec:NumNoSG}).
Technical details on how we obtain our numerical solutions are given in
Appendix~\ref{appendix:numerical}.

\subsection{Supergauge and Yukawa interactions in equilibrium}
\label{sec:NumYukSGEq}

First, we present a numerical solution for
a particular point in parameter space. This point is chosen
based on the following criteria:
\begin{itemize}
\item
 it is motivated by existing studies of EWB within
the MSSM,
\item
there is a
reasonable agreement between analytical approximations and numerical
solutions,
\item
the results share some qualitative key features with the
scenarios recently presented in Refs.~\cite{Chung:2008aya,Chung:2009cb}.
\end{itemize}

\begin{table}
\begin{tabular}{|l|l||l|l||l|l||l|l|}
\hline
$T$	&$100\,{\rm Gev}$	&$\mu$			&$200\,{\rm GeV}$	&$g_1$		&$0.357$		&$\gamma_{t,\widetilde t}=\gamma_{b,\widetilde b}$	&$0.5 T$\\
$v_w$	&$0.05$			&$M_A$			&$200\,{\rm GeV}$	&$g_2$		&$0.640$		&$\gamma_{\tau,\widetilde \tau}$			&$0.003 T$\\
$L_w$	&$0.25/T$		&$M_Z$			&$91\,{\rm GeV}$	&$g_3$		&$1.243$		&$\gamma_{\widetilde H}$				&$0.025 T$\\
$v_H(T)$&$125\,{\rm Gev}$	&$M_1$			&$200\,{\rm GeV}$	&$A_{t,b,\tau}$	&$300\,{\rm GeV}$	&$\gamma_{\widetilde W}$				&$0.065 T$\\
$D_H$	&$110/T$		&$M_2$			&$550\,{\rm GeV}$	&$\tan\beta$	&$15$			&$\gamma_{\widetilde B}$				&$0.003 T$\\
$D_Q$	&$6/T$			&$M_{\widetilde t}^2$	&$-(70\,{\rm GeV})^2$	&&&&\\
$D_L$	&$100/T$		&$M_{\widetilde b}^2$	&$(500\,{\rm GeV})^2$	&&&&\\
$D_E$	&$380/T$		&$M_{\widetilde\tau}^2$	&$(100\,{\rm GeV})^2$	&&&&\\
\hline
\end{tabular}
\caption{
\label{table:input:fiducial}
Input parameters at the benchmark point. The masses for supersymmetric 
particles that do not occur in this table have been chosen to
be $2\,{\rm TeV}$, such that they effectively deccouple.
}
\end{table}

The set of parameters that we choose is given in
Table~\ref{table:input:fiducial}.
From these input parameters, we derive the parameters appearing in
the Boltzmann equations~(\ref{eq:system}) in the following way:
\begin{itemize}
\item
The squark and slepton mass parameters in Table~\ref{table:input:fiducial} are
understood to be evaluated in the symmetric phase
and without thermal corrections. We indicate this by the use of a
capital $M$. The corresponding quark and lepton masses in the symmetric phase
without thermal correction are zero. In addition,
we derive the the masses of the Higgs boson eigenstates from $M_Z$ and $M_A$ as explained
in Section~\ref{section:InteractionLagrangian}, where we take account of
the thermal corrections that are summarized
in Ref.~\cite{Chung:2009cb}.
\item
The source and damping terms are computed
following~\cite{Lee:2004we,Riotto:1998zb},
as explained in detail in Section~\ref{sec:sourcerelax}.
Additional input parameters here are the thermal widths of some particles
$X$, which we denote as $\gamma_X$. The particular values we adopt
are again given in Table~\ref{table:input:fiducial}. They are taken from
Refs.~\cite{Enqvist:1997ff,Elmfors:1998hh} or motivated by the discussion
therein.
%In order to give an impression of the size of these terms and to
%facilitate the reproducability of the results, we also present the numerical
%values for the various source and damping rates in Table~\ref{XXXX}.
\end{itemize}

As mentioned above, we have chosen the mass parameters such that
they are favorably disposed toward
the viability of EWB in the MSSM.  The light $\widetilde{t}_R$ provides a
strong first-order phase transition \cite{Laine:1998qk,Carena:1997,lightstopMSSM}.
The heavy $\widetilde{q}$ is needed to increase the mass of the lightest
Higgs boson beyond LEP~II constraints.  Making the first and second generation
squarks heavy suppresses one-loop contributions to EDMs and precision
electroweak observables.

As a consequence of the large mass of
$\widetilde q$, there is wide agreement that EWB in the MSSM is viable only for
$CP$-violation in the Higgsino/gaugino sector and only close the resonance
region of $M_1 \sim |\mu|$ (or $M_2 \sim |\mu|$), and not from the
quark/squark sector.  Consequently, we choose
$M_1 = |\mu|$ for our benchmark point to maximize the Higgsino $CP$-violating
source.  While we calculate our $CP$-violating sources as per
Ref.~\cite{Lee:2004we,Riotto:1998zb}, there is still some disagreement about
the magnitude and parametric dependence of $S^{\; \cancel{CP}}_{\widetilde{H}}$,
{\it cf.}~\cite{Carena:2000id,Carena:2002ss,Konstandin:2005cd}.
However, in all of the present analysis, we keep $S^{\; \cancel{CP}}_{\widetilde{H}}$
fixed, as our emphasis is on the effects $CP$-conserving  transport coefficients for a given $CP$-violating source.
Therefore, our present work can be adapted to other calculations of
$S^{\; \cancel{CP}}_{\widetilde{H}}$ simply by an appropriate rescaling,
since all particle densities and the BAU scale linearly with
$S^{\; \cancel{CP}}_{\widetilde{H}}$.

In addition, following Refs.~\cite{Carena:2000id,Carena:1997gx}, we take the
Higgs vev profiles to be
\begin{eqnarray}
v(z)&=&\frac12 v_H(T)\left(1+\tanh\left(2\alpha\frac{z}{L_w}\right)\right)\,,\\
\beta(z)&=&\beta_0(T)-\frac 12 \Delta\beta\left(1-\tanh\left(2\alpha\frac{z}{L_w}\right)\right)\,,
\end{eqnarray}
with $\alpha=3/2$, which provide an accurate analytic approximation to profiles
obtained numerically in Ref.~\cite{Moreno:1998bq}.  We follow
Refs.~\cite{Riotto:1998zb, Lee:2004we} in our calculation of $CP$-violating
sources and $CP$-conserving relaxation rates for quarks and Higgsinos.
To facilitate comparison to other work, we provide the explicit numerical
values for these results:
\begin{subequations}
\label{num:relax:rates}
\bea
S_{\widetilde{H}}^{\: \cancel{CP}} &\simeq& - 9.4 \: \textrm{GeV} \, \times \sin\phi_\mu \, v_w \, \beta^\prime(z) \, v(z)^2\,, \\
\Gamma_H^{(\widetilde H, \widetilde V)} &\simeq& 1.4 \times 10^{-2} \; \textrm{GeV}^{-1} \times v(z)^2\,, \\
\Gamma_M^{(t,q)} &\simeq& 6.2 \times 10^{-3} \; \textrm{GeV}^{-1} \times y_t^2 \, v_u(z)^2\,,\\
\Gamma_M^{(b,q)} &\simeq&6.3 \times 10^{-3} \; \textrm{GeV}^{-1} \times y_t^2 \, v_u(z)^2\,,\\
\Gamma_M^{(\tau,\ell)} &\simeq&5.0 \times 10^{-5} \; \textrm{GeV}^{-1} \times y_t^2 \, v_u(z)^2\,.
\eea
\end{subequations}
As discussed earlier, we neglect $\Gamma_M^{(\widetilde t, \widetilde q)}$
and $\Gamma_H^{(H_1,H_2)}$, the relaxation rates for squarks and Higgs bosons,
respectively; we defer a calculation of these rates to future study.
Neglecting these relaxation rates will not have a large impact upon the final
BAU to the extent that (a) our analytical arguments from Sec.~\ref{sec:analytic}
hold true, and (b) these rates are non-resonant and therefore
much smaller than the relaxation rates for
Higgsinos and quarks that we have included.  The rates
$\Gamma_M^{(\widetilde{t},\widetilde{q})}, \Gamma_H^{(H_1,H_2)}$ affect
the BAU only through $\bar{\Gamma}$, which itself depends on the sum of all
relaxation rates (\ref{eq:gammabar}): 
\be
\bar{\Gamma} \propto \Gamma_M^{(\widetilde{t},\widetilde{q})} + \Gamma_M^{(t,q)} + \Gamma_H^{(\widetilde H, \widetilde V)} + \Gamma_H^{(H_1,H_2)}\; .
\ee
Therefore, a proper inclusion of squark and Higgs boson relaxation rates would
give only an
$\mathcal{O}\left[ \left(\Gamma_M^{(\widetilde{t},\widetilde{q})} + \Gamma_H^{(H_1,H_2)}\right) / \left(\Gamma_M^{(t,q)} + \Gamma_H^{(\widetilde H, \widetilde V)} \right) \right]$
correction to the analysis offered here.

\begin{table}[tbp] \centering
\begin{tabular}
%[c]{lclclllll}%
{|c|c|c|}
\hline
\quad Rate  \quad & \quad Benchmark Value (GeV) \quad &
\quad $\tau_{\textrm{diff}}/\tau_{\textrm{eq}}$ \quad\\
\hline
& &\\
\quad $\Gamma_{\widetilde V}^{(H_{1},\widetilde H)}$\quad & $5.3\times 10^{-1}$ & $3.4\times 10$\\
\quad $\Gamma_{\widetilde V}^{(H_{2},\widetilde H)}$\quad & $3.0\times 10^{-1}$ & $2.6\times 10$\\
\quad $\Gamma_{\widetilde V}^{(Q,\widetilde Q)}$\quad &  $1.8\times 10^{-5}$ & $6.4\times 10^{-4}$\\
\quad $\Gamma_{\widetilde V}^{(t ,\widetilde t)}$\quad &  $3.3\times 10^{-1}$ & $1.6\times 10$\\
\quad $\Gamma_{\widetilde V}^{(b ,\widetilde b)}$\quad &  $3.0\times 10^{-2}$ & $2.1$\\
\quad $\Gamma_{\widetilde V}^{(\ell ,\widetilde \ell)}$\quad &  $6.7\times 10^{-6}$ & $6.7\times 10^{-4}$\\
\quad $\Gamma_{\widetilde V}^{(\tau ,\widetilde \tau)}$\quad &  $1.6\times 10^{-1}$ & $2.8\times 10$\\
\quad $\Gamma_{Y}^{(\widetilde t, \widetilde Q ,  H_{1})}$\quad &  $7.6\times 10^{-7}$ & $3.6\times 10^{-5}$\\
\quad $\Gamma_{Y}^{(\widetilde b, \widetilde Q ,  H_{2})}$\quad &  $9.3\times 10^{-8}$ & $2.1\times 10^{-5}$\\
\quad $\Gamma_{Y}^{(\widetilde \tau, \widetilde \ell ,  H_{2})}$\quad &  $1.8\times 10^{-8}$ & $3.0\times 10^{-6}$\\
\quad $\Gamma_{Y}^{(\widetilde t, \widetilde Q ,  H_{2})}$\quad &  $3.3\times 10^{-7}$ & $1.6\times 10^{-5}$\\
\quad $\Gamma_{Y}^{(\widetilde b, \widetilde Q ,  H_{1})}$\quad &  $4.2\times 10^{-8}$ & $2.7\times 10^{-6}$\\
\quad $\Gamma_{Y}^{(\widetilde \tau, \widetilde \ell ,  H_{1})}$\quad &  $8.1\times 10^{-9}$ & $5.2\times 10^{-7}$\\
\quad $\Gamma_{Y}^{(\widetilde t,  Q ,\widetilde  H)}$\quad & $8.0$ & $2.8\times 10^2$\\
\quad $\Gamma_{Y}^{(\widetilde b,  Q ,\widetilde  H)}$\quad & $5.9\times 10^{-1}$ & $2.1\times 10$\\
\quad $\Gamma_{Y}^{(\widetilde \tau,  \ell,\widetilde  H)}$\quad & $8.9\times 10^{-2}$ & $7.6$\\
\quad $\Gamma_{Y}^{(t, Q , H_{1})}$\quad & $2.3$ & $8.2\times 10$\\
\quad $\Gamma_{Y}^{(b, Q , H_{2})}$\quad & $9.6\times 10^{-1}$ & $3.4\times 10$\\
\quad $\Gamma_{Y}^{(\tau, \ell , H_{2})}$\quad & $1.4\times 10^{-1}$ & $1.4\times 10$\\
\quad $\Gamma_{Y}^{(t, Q , H_{2})}$\quad & $1.8\times 10^{-2}$ & $6.5\times 10^{-1}$\\
\quad $\Gamma_{Y}^{(b, Q , H_{1})}$\quad & $0$ & $0$\\
\quad $\Gamma_{Y}^{(\tau, \ell , H_{1})}$\quad & $3.8\times 10^{-5}$ & $2.4\times 10^{-3}$\\
\quad $\Gamma_{Y}^{(t, \widetilde Q , \widetilde H)}$\quad &  $3.3\times 10^{-5}$ & $2.3\times 10^{-3}$\\
\quad $\Gamma_{Y}^{(b, \widetilde Q , \widetilde H)}$\quad &  $4.4\times 10^{-6}$ & $3.0\times 10^{-4}$\\
\quad $\Gamma_{Y}^{(\tau, \widetilde \ell , \widetilde H)}$\quad &  $7.9\times 10^{-7}$ & $6.8\times 10^{-5}$\\
\quad $\Gamma_{\rm ss}$\quad & $3.7 \times 10^{-1}$ & $1.3 \times 10^2$\\
& &\\
\hline
\end{tabular}
\caption{Yukawa, strong sphaleron, and supergauge interaction rates for the benchmark parameters and comparison to the diffusion time scale. \label{tab:benchmarkrates}}
\end{table}

The degree to which we should expect an agreement between analytical and numerical
solutions can be inferred from
Table~\ref{tab:benchmarkrates}, where we show each of the Yukawa and supergauge
rates for our benchmark scenario.  As we noted in Sec.~\ref{sec:analytic},
the appropriate sufficient condition that ensures that a particular rate leads to chemical
equilibrium is $\tau_{\textrm{diff}}/\tau_{\textrm{eq}} > 1$.  In the the
second column of Table \ref{tab:benchmarkrates}, we compute
$\tau_{\textrm{diff}}/\tau_{\textrm{eq}}$, where $\tau_{\textrm{eq}}$ is
determined through dividing the corresponding rate from the first column by
the appropriate $k$-factor, as per
Eq.~(\ref{eq:gammaeff}). The three
body rates in Table~\ref{tab:benchmarkrates} are all calculated
following the methods described in
Sections~\ref{section:Supergauge} and~\ref{section:YukTri}, with one
exception: $\Gamma_Y^{(t,q,H)}$.
After including thermal corrections, we have
$m_t\approx m_q\approx 65\,{\rm GeV}$ and $m_{H_1}\approx 50\,{\rm GeV}$,
such that on-shell scatterings of these three particles are
kinematically forbidden. Within the MSSM, we cannot alleviate this
kinematic suppression, since a negative mass square for the Higgs boson
needs to be present in the Lagrangian in order to ensure electroweak
symmetry breaking. Since the rate $\Gamma_Y^{(t,q,H)}$ plays a pivotal
role in the network Boltzmann equations~(\ref{eq:system}) that describe
diffusion, and it is in general non-zero due to the off-shell effects
and four body contributions, we treat it as an exception.
From Ref.~\cite{Joyce:1994zn}, we take
\begin{equation}
\Gamma_Y^{(t,q,H)}=0.129 \frac{g_3^2}{4\pi} T\,,
\end{equation}
which is an estimate of the four-body contributions. We re-emphasize however that a more detailed analysis of the four body and off-shell
contributions in the future would be desirable.

\input{fig-ipreducedfiducial}

In FIG.~\ref{fig:N:fiducial}, we display the numerical results for the
number densities of selected species and compare them to the analytical
predictions according to Section~\ref{sec:analytic}.
Numerically, the diffusion time scale is
\be
\tau_{\textrm{diff}} \equiv \frac{\bar{D}}{v_w^2} \simeq 2.0 \times 10^2 \; \textrm{GeV}^{-1} \sim 10^{-22} \; \textrm{s}.
\ee
Since $\tau_{\textrm{diff}}/\tau_{\textrm{eq}} > 1$ for those rates in
Table~\ref{tab:benchmarkrates} that do not involve the ultraheavy
sfermions $\widetilde q$ and $\widetilde \ell$ and that are not 
suppressed by $\sin\alpha$, our
numerical solutions match our analytical expectations to a good extent.
Regarding the
analytical solutions, the following additional comments are in order:
\begin{itemize}
\item
The overall normalization of the analytic estimate is slightly larger
than for the numerical result. This is because
the damping rates in the broken phase
are larger than the equilibration rates, and therefore the error of the
analytic approximation is not under control for $z\stackrel{>}{{}_\sim}0$.
\item
Sufficiently far ahead of the bubble wall, the prediction for the ratios
of the particular charge densities is accurate. In this
region, the analytical approximation is well under control. However, $n_{\mathrm{left}}$ close to the bubble
wall contributes significantly to $Y_B$, and in this region the analytic approximation is less reliable.
Close to the bubble wall and within the bubble, the relevant time-scale
is no longer $\tau _{\rm diff}$ in Eq.~(\ref{tau:diff}), but
$\bar \Gamma^{-1}$
[{\it cf.} Eqs~(\ref{eq:gammabar},~\ref{eq:amplitude},~\ref{eq:gammapm})].
Comparison of the numerical values for the relaxation
rates~(\ref{num:relax:rates}) to the equilibration rates in Table~\ref{tab:benchmarkrates}
shows that $\Gamma_{\widetilde V},\Gamma_{Y}\ll\Gamma_M,\Gamma_H$, such that
the analytical approximation is not justified in that regime.
In particular, since the densities $q$ and $\ell$ have opposite
sign [as  is generic for $m_{\widetilde b}>m_{\widetilde t}$, {\it cf.}
Eq.~(\ref{n:left})], the analytic
result for $n_{\rm left}$ is rather inaccurate. Here and
in the following, we therefore refrain from a direct comparison of the
analytical predictions for the baryon asymmetry to the numerical answer.
However, the analytical formulae are still very useful for understanding
the behavior of the numerical solutions qualitatively.
\item
The results in FIG.~\ref{fig:N:fiducial} do, indeed, reflect many of these qualitative features. In particular, in the right panel, we observe that the total first and second generation LH quark + squark densities are negligible compared to the third generation LH quark + squark and RH tau + stau densities shown in the left panel. This feature follows from the approximate vanishing of the third generation contribution to $N_5$, leading to the near absence of any induced first and second generation densities, as explained above. In addition, the relative signs of the $H$, $q_3$ and $\ell_3$ densities ahead of the wall follow closely the expectations based on Eqs.~(\ref{kappa:all}) (recall that $k_B < k_T$ for our fiducial parameter choice).
\end{itemize}

Besides, due to the choice of parameters
in  Table~\ref{table:input:fiducial},
the scenario presented in this Section shares the key features with those
that are presented in Refs.~\cite{Chung:2008aya,Chung:2009cb}. In order
to discuss these features, it is
instructive to consider certain combinations of chemical potentials,
that are displayed in
FIGs.~\ref{fig:supergauge},~\ref{fig:yukawa} under the label 
``$\tan\beta=15$''.
The following points are of importance:
\begin{itemize}
\item
Superequilibrium is maintained on diffusion time-scales $\Gamma_{\rm diff}^{-1}$,
as it is exhibited by the fact
that the chemical potentials for particles and their superpartners
are identical sufficiently far away from the bubble wall,
{\it cf.} FIG.~\ref{fig:supergauge} (``$\tan\beta=15$''). This feature
is crucial for the validity of the analytical approximation 
in Section~\ref{Sec:Analytical}.
The assumption of superequilibrium has
also been used in deriving the reduced set of Boltzmann equations,
which is presented in Ref.~\cite{Chung:2009cb} and follows from
Eqs.~(\ref{eq:system}).
\item
In deriving the analytical approximation, it is assumed that the relaxation
rates $\Gamma_Y$ are fast compared to the diffusion time-scale
$\Gamma_{\rm diff}$. This implies that the combinations of chemical potentials
that multiply these relaxation rates should vanish. From
FIG.~\ref{fig:yukawa} (``$\tan\beta=15$''), we see explicitly that Higgs bosons and
Standard Model fermions satisfy the corresponding equilibrium conditions. We have
checked that the same is also true for the additional supersymmetric
particles.
\item
The fact that Yukawa and triscalar interactions involving (s)bottoms
and (s)taus are in equilibrium leads to an important change in the flavor
dynamics ahead of the bubble wall, that has not been appreciated
in the literature~\cite{Huet:1995sh,Lee:2004we}
before our resent work~\cite{Chung:2008aya,Chung:2009cb}.
First, since we have a moderately large value of $\tan\beta$, also
lepton and slepton densities are induced and equilibrate
ahead of the bubble wall and
give an important contribution to electroweak baryogenesis. The same
is true for the (s)bottom-particles, as they have larger Yukawa and triscalar
interactions than the (s)tau. Due to the presence of the
strong sphaleron, the equilibrium of bottom Yukawa and triscalar interactions
has an additional consequence: The combination of chemical potentials
$N_5$ that
multiplies $\Gamma_{\rm ss}$ in the Boltzmann equations~(\ref{eq:system})
is vanishing for zero densities of first-generation quarks. Indeed, from
FIG.~\ref{fig:N:fiducial} we see that no asymmetry in first generation quarks
is diffusing ahead of
the bubble wall.
\end{itemize}

\input{fig-ipsupergauge}

\input{fig-ipyukawa}

\subsection{Dependence on $\tan\beta$}
\label{sec:NumTanBeta}

We now take the same parameters as given in
Table~\ref{table:input:fiducial} for our example point, but consider
values for $\tan\beta\in[1.5;20]$. The effect of this
on the resulting baryon asymmetry is displayed in FIG.~\ref{fig:BAU:TANBETA},
where we display the ratio of $Y_B$ obtained from the numerical simulations
and the observational value $Y_B^{\rm WMAP}$.
We see that $Y_B/Y_B^{\rm WMAP}$ increases for smaller values of $\tan\beta$.
This is because smaller values of $\tan\beta$ imply smaller
down-type Yukawa couplings. Therefore, a smaller lepton density is generated
ahead
of the wall, as can be seen when comparing FIG.~\ref{fig:N:tanbeta1pt5} to
FIG.~\ref{fig:N:fiducial}. Since quark and lepton asymmetries contribute with
opposite sign (provided $m_{\widetilde b}>m_{\widetilde t}$, as it is the case
here),
small values of $\tan\beta$ lead to a weaker cancellation in the left-handed
fermion density and therefore a larger baryon asymmetry.

\input{fig-tanbeta}
\input{fig-ipreducedtanbeta1pt5}

Besides, FIG.~\ref{fig:yukawa} (``$\tan\beta=1.5$'') exhibits that for $\tan\beta=1.5$,
neither the interactions mediated by
$y_b$  nor $y_\tau$ maintain equilibrium. However, even for values of
$\tan\beta$
close the lower bound that is theoretically allowed, a non-negligible density
in
$b$-quarks is produced. Apparently, the analytical approximation is only
reliable
for $\tan\beta\stackrel{>}{{}_\sim}15$, when all third generation Yukawa
interactions
are in equilibrium on diffusion time-scales. For
$\tan\beta\stackrel{<}{{}_\sim}15$, the interactions mediated by
$y_\tau$ are out of equilibrium on diffusion
time-scales, and for $\tan\beta\stackrel{<}{{}_\sim}5$
the same is true for the interactions mediated by $y_b$.
Yet, non-negligible densities of down-type fermions can occur in
general ahead of the bubble wall.
We also note that since for $\tan\beta=1.5$, bottom (s)quarks  do
not equilibrate ahead of the wall, a non-negligible
density of first-generation quarks is generated,
{\it cf.} FIG.~\ref{fig:N:tanbeta1pt5}. This density of left-handed
quarks of the first two generations is opposite to the third generation
density, and it has therefore the effect of suppressing the baryon
asymmetry ({\it cf.} the graph without taking account of leptons in 
FIG.~\ref{fig:BAU:TANBETA}). For large $y_b$, the equilibrium of axial charges that is
maintained by the strong sphaleron is satisfied by the relation
$\mu_b+\mu_t-\mu_q=0$ and $\mu_{q_{1,2}}=0$~\cite{Chung:2008aya,Chung:2009cb},
whereas for negligible $y_b$, one finds
$Q_{1,2}=2(Q+T)$ (where $Q$ and $T$ are of opposite sign and $|T|>|Q|$),
hence
$\mu_q$ and $\mu_{q_{1,2}}$ being opposite~\cite{Huet:1995sh,Lee:2004we}.
Note that the analytic formulae presented in Refs.~\cite{Huet:1995sh,Lee:2004we}
are not applicable for $\tan\beta=1.5$, since even though bottom quarks
do not completlely equilibrate, a sizeable density of them is yet present
ahead of the wall.

Comparison of the graphs with and without leptonic densities taken into account in FIG.~\ref{fig:BAU:TANBETA}, we also observe that for
values as small as $\tan\beta=1.5$, there is still a sizable leptonic
contribution to $n_{\rm left}$. We also see that fast interactions mediated by Binos and Winos still
maintain
superequilibrium ahead of the bubble wall, as exhibited in by the small $\tan\beta$ region shown in FIG.~\ref{fig:supergauge}~(``$\tan\beta=1.5$'').  

\input{fig-ipsupergauge-additional}

We now investigate the impact of the finite rate  of supergauge interactions.
For that purpose, in FIG.~\ref{fig:BAU:TANBETA}, also the results of a
simulation where superequilibrium is enforced (leading
to $\mu_x=\mu_{\widetilde x}$ everywhere) are displayed. The corresponding
plots of the chemical potentials can be seen in
FIG.~\ref{fig:supergauge-additional} ($\Gamma_{\widetilde V}\to\infty$).
At large $\tan\beta$, compared to the case with finite supergauge
interactions [FIG.~\ref{fig:supergauge} ($\tan\beta=15$)],
$\mu_\ell$ is enhanced, since it adapts
to some extent to $\mu_{\widetilde \ell}$
[{\it cf.} FIG.~\ref{fig:supergauge} ($\tan\beta=15$)]. On the other hand,
since $\widetilde q$ is superheavy ({\it i.e.} the density of 
$\widetilde q$ is very small), there is no corresponding enhancement of
$\mu_q$. In addition, imposing superequilibrium slightly favors the
production of leptons close to the bubble wall due to the summation
over the various supersymmetric production channels, whereas the quark production
rate is already comparably large. The combination of these effects leads
to an enhancement of the chiral lepton asymmetry which in turn suppresses
$Y_B$. For small $\tan\beta$ the situation is more complicated, since
now also first generation quarks play a role. Since for finite
supergauge interaction rates superequilibrium is still violated
to some extent close to the bubble wall, the apparent agreement
of the curves with finite and infinite supergauge interaction rates
in FIG.~\ref{fig:BAU:TANBETA} for small $\tan\beta$ is accidental.
Note also that for large $\tan\beta$, since the contributions of
$\ell$ and $q$  to $n_{\rm left}$ have opposite sign, the inaccuracy
incurred by assuming finite supergauge interactions is substantial,
up to the extent that the predicted value of $Y_B$ can flip sign.

\subsection{Absence of supergauge interactions}
\label{sec:NumNoSG}

We now investigate in more detail how superequilibrium can be maintained even in absence of
supergauge interactions, as discussed in Section~\ref{section:superconditions}.
We first note that since we have taken its mass to be $2\,{\rm TeV}$,
the interactions of the gluino are suppressed to an extent that they are negligible.
Now, we set in addition all interactions $\Gamma_{\widetilde V}$ appearing
in the Boltzmann equations~(\ref{eq:system}) to zero.
While this procedure could also be mimicked by taking the Bino and Wino masses
to be very heavy, we note that for electroweak baryogenesis within the MSSM,
it is at least required that either $M_1\simeq \mu$ or $M_2\simeq \mu$, in order
to have resonant $CP$-violation and to produce a large enough baryon asymmetry.
Therefore, the limit taken in this section may be considered as a theoretical
exercise. On the other hand, it may be conceivable that $M_2\gg 1\,{\rm TeV}$
and $M_1\simeq\mu$, but also
$\mu\simeq M_1\simeq m_{\widetilde t,\widetilde b,\widetilde \tau}$. In such a
case, three-body
interactions between the Higgsino and right handed fermions are kinematically
not allowed at zero temperature. Besides, if the $CP$-violating
source in supersymmetric scenarios different from the MSSM is not 
originating from Higgsino-gaugino mixing, there may be no obstacle
for successful EWB with heavy gauginos.

\input{fig-ipreducednosg}

For the purposes of this example, we again take the parameters
from Table~\ref{table:input:fiducial}, but we set
$M_{\widetilde \ell}=100\,{\rm GeV}$. For which species superequilibrium is
maintained and for which it is broken can now be inferred from
FIG.~\ref{fig:supergauge}~($\Gamma_{\widetilde V}=0$; $M_{\widetilde \ell}=100\,{\rm GeV}$): while $\{q,\,\widetilde q\}$ and
$\{t,\,\widetilde t\}$ do not satisfy superequilibrium,
$\{\ell,\,\widetilde \ell\}$ and $\{\tau,\,\widetilde \tau\}$ do.

To give an explanation of these observations, we first note
that since $\widetilde q$ is superheavy, the chain of equilibrium
conditions~(\ref{Yuk1}) is broken. To see this, we note that
$\mu_{\widetilde q}$ is sizeable, even far ahead of the bubble wall.
However, the number density $\widetilde q$ is small, since
we have taken the left-handed squark to be superheavy,
$m_{\widetilde q}=2\,{\rm TeV}$. This implies that for example
$\Gamma^{\widetilde q,\widetilde t,H_1}_Y/k_{\widetilde t}\ll\Gamma_{\rm diff}$ and
$\Gamma^{\widetilde q,\widetilde t,H_1}_Y/k_{H_1}\ll\Gamma_{\rm diff}$.
Therefore, a sizable value of $\mu_{\widetilde q}$ does not need to enforce a
large density of $H_1$, for the simple reason that the physical density
$\widetilde q$ is small. In FIG.~\ref{fig:yukawa}~($\Gamma_{\widetilde V}=0$; $M_{\widetilde \ell}=100\,{\rm GeV}$) it is exhibited, that
even though superequilibrium is violated, Yukawa equilibrium is still intact
for Standard Model fermions and Higgs bosons on diffusion time-scales.

In contrast, in the down-type sector, $\widetilde b$, $\widetilde \tau$ and
$\widetilde \ell$ are not heavy compared to $T$, such that these particles
can mediate the equilibration of $H_2$.
Consequently, superequilibrium is maintained here on diffusion time-scales
according to the argument given
in Section~\ref{section:superconditions}.

In FIG.~\ref{fig:BAU:TANBETA} (green dashed curve), we also show a simulation for
the parameter set as in  Table~\ref{table:input:fiducial} (but now
again with  $M_{\widetilde \ell}=2\,{\rm TeV}$), but with all supergauge
interaction rates set to zero.
The profiles of chemical potentials
are displayed in
FIG.~\ref{fig:supergauge-additional} ($\Gamma_{\widetilde V}\to 0$; $M_{\widetilde l}=2{\rm TeV}$).
We find that $Y_B$ is enhanced
when compared to the cases with finite or infinite $\Gamma_{\widetilde V}$.
Comparing to the plots in FIG.~\ref{fig:supergauge-additional} ($\Gamma_{\widetilde V}\to \infty$),
we see that in the absence of supergauge interactions due to the
$\widetilde q$ and $\widetilde \ell$ bottlenecks, superequilibrium
is broken in both the up and the down type sector. In addition, we observe:
\begin{itemize}
\item The signs of the $t$ and $\tau$ chemical potential are now reversed.
\item The magnitudes of the $t$, $b$, and $\ell$ densities are reduced.
\item The magnitudes of the $H_{1,2}$ are significantly suppressed.
\item The magntiude of the $q$ density increases. 
\end{itemize}

On general grounds, the absence of superequilibrium implies a degrading of the overall efficiency with which ${\widetilde{H}}$ density (induced by the CP-violating source) is transferred into the SM fermion densities. At the same time, the detailed balance between these densities and their net effect on $Y_B$ change substantially. Without a robust analytic framework for treating this case, we can only speculate on the reasons why these changes result in an increase in $Y_B$. Nonetheless, it is clear that the complete decoupling of supergauge interactions from the transport dynamics can have a substantial impact on the predicted baryon asymmetry. This siutation may be particularly relevant to extensions of the MSSM that can accommodate sizable CP-violating sources and heavy gauginos.

%The densities $H_{1,2}$ are suppressed, since they can only
%be produced along with $q$ and $\ell$. In order to
%satisfy the resulting non-local charge conservation
%$\int dz (q+\ell+H_1-H_2)=0$, the Higgs bosons diffuse away fast.
%The suppression of their local densities is therefore similar
%to the suppression of the density $R$ compared to $L$, described in
%Ref.~\cite{Chung:2009cb} and in Section~\ref{Sec:Analytical}.
%It turns out that for $\Gamma_{\widetilde V}\to 0$,
%$q$ is enhanced while $\ell$ is suppressed when compared to the case
%with $\Gamma_{\widetilde V}\to \infty$, such that a larger $Y_B$ results.

\section{Conclusions}
\label{sec:conclude}

In this work, we have generalized existing approaches
to EWB to account for finite supergauge interaction rates. We have developed
numerical solutions to the resulting Boltzmann equations that
describe the diffusion
processes. For particular illustrative points in parameter space,
we have presented numerical solutions.
When superequilibrium holds and all Yukawa interactions fully equilibrate,
these examples agree
with results published earlier~\cite{Chung:2008aya,Chung:2009cb}.
In turn, in the absence of superequilibrium
or when down-type Yukawa interactions only partially equilibrate,
our solutions show sizable deviations  from earlier results.

Regarding the consequences of finite supergauge interactions for
supersymmetric EWB, our conclusions are as follows:
\begin{itemize}
\item
In models for EWB with light gauginos ({\it e.g.} with masses not much heavier
than $T$), such as the MSSM with a Higgsino-gaugino
$CP$ violating source, superequilibrium is a robust assumption ahead of the
bubble wall.
\item
In models with heavy gauginos -- such as extensions of the MSSM that do not require light gauginos for the CP-violating sources --  superequilibrium may be restored
through the network of Yukawa and triscalar interactions. However, this chain of reactions may be broken when 
one of the superpartners becomes heavy compared to the temperature, thereby leading to a bottleneck.
We have presented an illustrative example of this situation, where
in particular the up-(s)quark sector violates superequilibrium, whereas
superequilibrium is maintained for down-type (s)quark and (s)leptons, {\it cf.} FIG.~\ref{fig:supergauge} ($\Gamma_{\widetilde V}=0$; $M_{\widetilde \ell}=100\,{\rm GeV}$). 
\item
More generally, the assumption of exact superequilibrium or the neglect of supergauge interactions may lead to significant errors in the prediction of $Y_B$, as illustrated, respectively, by the yellow and green curves in FIG.~\ref{fig:supergauge}. In the former case, which may apply in models with heavy gauginos, the assumption of superequilibrium may lead to larger left-handed particle densities than actually occur as one forces them by hand to match the corresponding sparticle densities. As indicated by  the yellow curve of FIG.~\ref{fig:supergauge}, the effect of this error is amplified because the LH squark densities are small, the LH slepton densities are relatively large, and the corresponding particle densities contribute to $n_\mathrm{left}$ with opposite sign. As a result, the enhanced negative lepton contribution suppresses the baryon asymmetry. The green curve illustrates the converse dynamics, wherein the neglect of supergauge interactions leads to a lepton contribution that is smaller in magnitude with a correspondingly larger $Y_B$. 
\item
When $m_{\widetilde t}<m_{\widetilde b}$ (as it is typically the
case in the MSSM with a strong first order phase transition), the
leptonic contribution to $n_{\rm left}$ is opposite in sign to the one
from quarks. The leptonic contribution is suppressed for small
values of $\tan\beta$, but yet needs to be taken into account for
values as small as $\tan\beta=1.5$ ({\it cf.} 
FIG.~\ref{fig:BAU:TANBETA}). In order to obtain a large asymmetry
in the MSSM, small
values of $\tan\beta$ are therefore needed from the solution of diffusion
equations. This points into the same direction (but it is a different effect)
as the suppression of the $CP$-violating source for large values of
$\tan\beta$~\cite{Moreno:1998bq,lightstopMSSM}. Note that we have
not taken the impact of $\tan\beta$ on the $CP$-violating source into
account in this work, in order to disentangle this effect from the
diffusion solutions.
However, large values of $\tan\beta$ are yet interesting
for EWB in the (M)SSM, since the Bino phase can be of order one from present EDM limits, which allows for a value of $Y_B$ in that region of paramater space~\cite{Li:2008kz,Li:2008ez}.
\end{itemize}

From our results, we can draw the following conclusions regarding the
relying on numerical solutions to the diffusion equations when one is interested in quantitatively reliable predictions for $Y_B$:
\begin{itemize}
\item
The numerical solutions are accurate close to the bubble wall and
within the bubble, where the analytic description is not under control.
This may have a sizable impact on the result for the BAU, which
is obtained from integrating over $n_{\rm left}$. Even in the presence of gauginos that are not heavy compared to
the temperature, it is advisable not to impose superequilibrium,
as it is usually broken close to the bubble wall. This is of particular
importance when particular densities that contribute to $n_{\rm left}$
cancel, as it is often the case.
\item
For $\tan\beta\stackrel{<}{{}_\sim}15$, $\tau$-Yukawa couplings or
both bottom- and $\tau$-couplings equilibrate incompletely, but there
is always a non-negligible density of $b$ and $\tau$. Since there
is no analytical method yet in order to describe this partial equilibration, the numerical solution is necessary for predictions
in this parametric region, which is of particular interest for EWB in the MSSM.
\end{itemize}

% In order to further improve the diffusion solutions for EWB, progress
% on the following points in the near future is desirable:
% \begin{itemize}
% \item
% {\bf ** We should leave this bullet out: it really has nothing to do with the present study and we already stated that this is an open problem **}
% The $CP$-violating source $S^{\cancel{CP}}$ and the relaxation
% rates $\Gamma_M$, $\Gamma_H$ need to be computed beyond the leading
% order vev-insertion approximation, possibly using a relaxation approach.
% The source $S^{\cancel{CP}}$ has been evaluated in a resummation
% approach in Refs.~\cite{Konstandin:2005cd}. However, in that formalism
% the detailed effects of the diffusion processes that are studied in this paper
% are not taken into account, and it has not been worked out yet how this
% can be achieved.
% \item
% In order to improve on the diffusion constants and chemical equilibration
% rates, off-shell effects and $2\to2$ rates need to be computed.
% \end{itemize}

In closing, we emphasize that our conclusions apply to any scenario of
supersymmetric EWB, beyond the MSSM.  Our discussion has been nearly
independent of the exact nature of the $CP$-violating source, which may arise
from Higgsinos, squarks, or an extended sector not present in the MSSM.
What is certain, however, is that in order for EWB to work, $CP$-violation
must be communicated to the left-handed matter fermion sector, which in general will give rise to
$CP$-violating asymmetries for squarks, quarks, sleptons and leptons
of various flavors  through
supergauge, Yukawa, triscalar and strong sphaleron processes.  The complete
spectrum of gauginos, squarks and sleptons as well as the value of $\tan\beta$
are relevant for the precise determination of the
baryon asymmetry in all supersymmetric EWB scenarios.

\vfill
\eject
\section*{Acknowledgements}
This work was supported in part by Department of Energy contracts
DE-FG02-08ER41531 and DE-FG02-95ER40896, and the
Wisconsin Alumni Research Foundation. MJRM thanks NORDITA, the Apsen Center for Physics, and TRIUMF, where part of this work was completed. 
\appendix 

\section{Numerical methods}
\label{appendix:numerical}

We have developed two independent numerical codes to solve the system of Boltzmann equations.  The first method, a modified version of the Euler method, is best illustrated for the differential equation
\be
y^{\prime\prime}(z) + a(z) \, y^\prime(z) + b(z) \, y(z) = s(z) \;, \label{ode}
\ee
where $a(z), \, b(z) \simeq$ constant and $s(z) \simeq 0$ far enough away from $z=0$, say, for $z \le z_1 < 0$ and $z \ge z_2 > 0$.  We exploit the fact that Eqn.~(\ref{ode}) is homogenous and exactly solvable for $z \le z_1$ and $z \ge z_2$.  Therefore, we can write
\be
y(z) = \left\{ \ba{c} A e^{\lambda_- z} \qquad z \le z_1 \\
     B e^{-\lambda_+ z} \qquad z \ge z_2 \ea \right.
\ee
where $\lambda_{\pm} > 0$.  Because $a,b$ are known functions, $\lambda_\pm$ are known as well; however, $A,B$ are unknown coefficients.  The characteristic equation resulting from Eqn.~(\ref{ode}) permits additional roots, which lead to solutions with diverge at $z \to \pm \infty$; we implement the boundary conditions that
\be
\lim_{z \to \pm \infty} y(z) \to 0
\ee
by discarding these solutions.  The next step is to use the Euler method to determine $y(z)$ in the interpolating region where $z_1 < z < z_2$.  We discretize this region into $N$ steps with length $\Delta$.  We begin with
\bea
y(z_1) &=& A e^{\lambda_- z_1} \\
y^\prime(z_1) &=& A \lambda_- e^{\lambda_- z_1} \\
y^{\prime\prime}(z_1) &=& A \lambda_-^2 e^{\lambda_- z_1} \; ;
\eea
then we iterate forward:
\bea
y(z_1+\Delta) &=& y(z_1) + y^\prime(z_1) \, \Delta + \; ... \\
y^\prime(z_1+\Delta) &=&  y^\prime(z_1) + y^{\prime\prime}(z_1) \, \Delta + \; ...\\
y^{\prime\prime}(z_1+\Delta) &=& - a(z_1 + \Delta) \, y^\prime(z_1+\Delta) - b(z_1 + \Delta) \, y(z_1 + \Delta) + s(z_1 + \Delta) \;  ,
\eea
where the ``...'' denotes the possibility of including higher order terms if needed.  Ultimately, after iterating from $z_1$ to $z_1 + N\Delta = z_2$, we obtain
\bea
y(z_2) &=& f_1(\lambda_-,z_1) \, A \\
y^\prime(z_2) &=& f_2(\lambda_-,z_1) \, A \;,
\eea
where $f_{1,2}$ are simply numbers (which depend on $\lambda_-$ and $z_1$).  Finally, we merely have to solve the equations
\bea
y(z_2) &=& f_1(\lambda_-,z_1) \, A = B e^{-\lambda_+ z_2} \\
y^\prime(z_2) &=& f_2(\lambda_-,z_1) \, A = B \lambda_+ e^{-\lambda_+ z_2}
\eea
to determine the unknown coefficients $A$ and $B$.  The application to the system of coupled densities (\ref{eq:tright}-\ref{eq:ditilderight}) follows by generalizing Eqn.~(\ref{ode}) to a matrix equation for the vector of densities $y = (t_R, \, \widetilde{t}_R, \, ... )$ and performing some diagonization gymnastics.

The second numerical way we employ to solve the diffusion equations
is a relaxation method, see {\it e.g.}~\cite{NRecipes}.
We decompose the $N$ diffusion equations into
$2N$ first order differential equations
\begin{equation}
\label{2NDiffeq}
y_i^\prime(z)+\gamma_{ij}(z)y_j(z)=s_i(z)\,,
\end{equation}
where $\gamma_{ij}(z)$ depends on the interaction rates, the wall velocity and the diffusion constants and $s_i(z)$ on the source. Among the $y_i$, $N$ components
represent the charge densities and $N$ the derivatives of these densities with
respect to $z$.

Then, we discretize~(\ref{2NDiffeq}) for $M$ interior points as
\begin{equation}
\label{Eik}
E^k_i=
y_i^k-y_i^{k-1}
+(z^k-z^{k-1})
\left[
\gamma_{ij}\left(\frac{z^k+z^{k-1}}2\right)
\frac{y^k_j+y_j^{k-1}}2
-s_i\left(\frac{z^k+z^{k-1}}2\right)
\right]
=0\,,
\end{equation}
where $z_k=z_{\rm min}+(z_{\rm max}-z_{\rm min})k/M$.
Two additional sets of equations at the exterior points $z_{\rm min}\ll 0$
and $z_{\rm max}\gg 0$ follow from the
boundary conditions. Here, we impose that the charge densities are vanishing far
away from the wall, that means for some pair of large negative and positive
values of $z_{\rm min}$ and $z_{\rm max}$.

We start with a initial guess $y_i^k=0$ for all $k$ and $i$. The $y_i^k=0$
are then updated by solving the linearized approximation
to~(\ref{Eik})
\begin{equation}
E^k_i(y^k+\Delta y^k,y^{k-1}+\Delta y^{k-1})\approx 
E^k_i(y^k,y^{k-1})
+\frac{\partial E^k_i(y^k,y^{k-1})}{\partial y_j^{k-1}}\Delta y_j^{k-1}
+\frac{\partial E^k_i(y^k,y^{k})}{\partial y_j^{k}}\Delta y_j^{k}=0
\end{equation}
subsequently for $\Delta y^{k-1}$ and $\Delta y^{k}$.
An improved approximation to the solution is then given by
$y^k\to y^k + a \Delta y^k$, where $a$ is a positive constant of order
{\it one}, to be chosen such that fast convergence is achieved.
For the present problem, a relative accuracy of one part in $10^{10}$
is typically attained after two or three iterations.

\end{document}

%% file: fig-ipreducedfiducial.tex
\begin{figure}[h]
\vskip.5cm
\begin{center}
\scalebox{1.}
{
\input{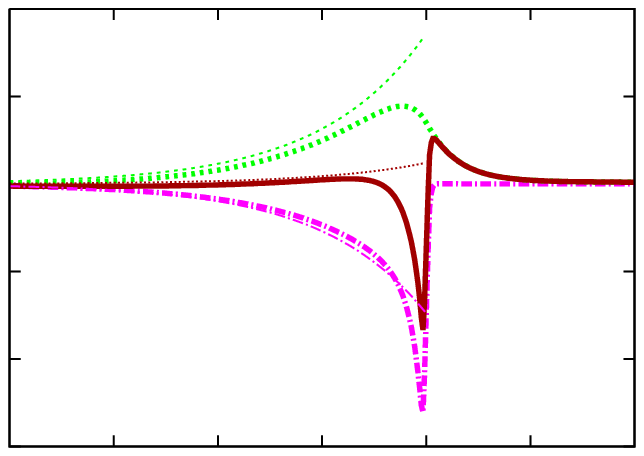}
}
\hskip1.5cm
\scalebox{1.}
{
\input{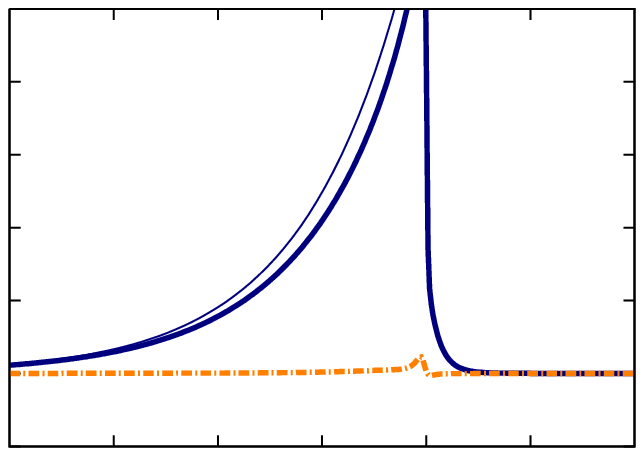}
}
\end{center}
\vskip.5cm
\caption{
\label{fig:N:fiducial}
Charge densities over $z$. Numerical results are represented by thick lines and analytical
results by thin lines. Left panel: $q_3$ (pink, dot-dashed), $\ell_3$ (green, dotted),
$n_{\rm left}$ (red, solid). Right panel: $H=H_1+H_2+\widetilde H$ (blue, solid),
$q_1+q_2$ (orange, dot-dashed).}
\end{figure}

%% file: fig-ipsupergauge.tex
\begin{figure}
\begin{center}
\vskip -.5cm
%\hskip-18cm
\scalebox{1.}
{
\input{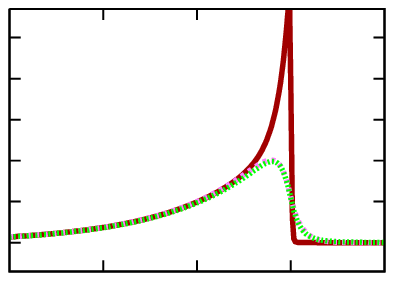}
}
\hskip1.25cm
\scalebox{1.}
{
\input{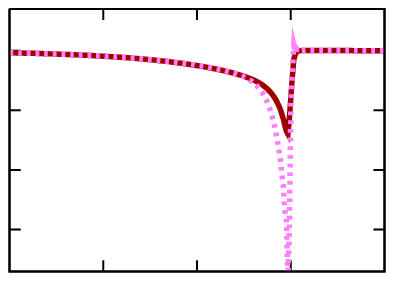}
}
\hskip1.25cm
\scalebox{1.}
{
\input{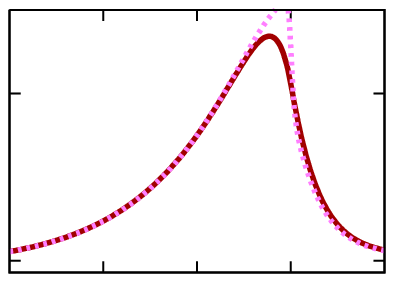}
}
\end{center}

\vskip0.6cm

\begin{center}
%\hskip-18cm
\scalebox{1.}
{
\input{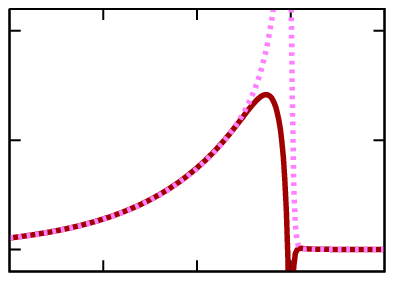}
}
\hskip1.25cm
\scalebox{1.}
{
\input{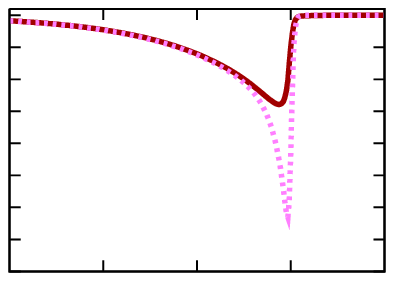}
}
\hskip1.25cm
\scalebox{1.}
{
\input{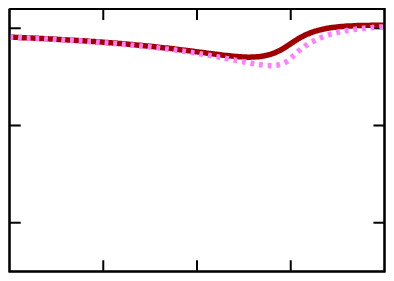}
}
\end{center}
\vskip.6cm
% \caption{
% \label{fig:supergauge:fiducial}
% Chemical potentials over $z$ at the fiducial point, illustrating
% supergauge-equilibrium. The key for the top right panel is
% $\mu_{H_1}$ (red, solid), $\mu_{H_2}$ (green, dashed),
% $\mu_{\widetilde H}$ (pink, dotted). For the other panels, the key is
% $\mu_{q,\ell,t,b,\tau}$ (red, solid),
% $\mu_{\widetilde q,\widetilde \ell,\widetilde t,\widetilde b,\widetilde \tau}$
% (pink, dotted).}
% \end{figure}
% 
% 
% \vskip.9cm
% \begin{figure}[h]
\begin{center}
%\hskip-18cm
\scalebox{1.}
{
\input{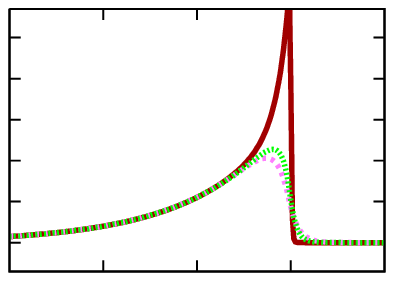}
}
\hskip1.25cm
\scalebox{1.}
{
\input{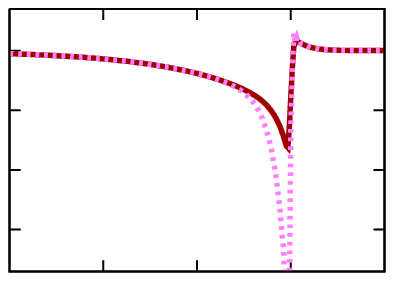}
}
\hskip1.25cm
\scalebox{1.}
{
\input{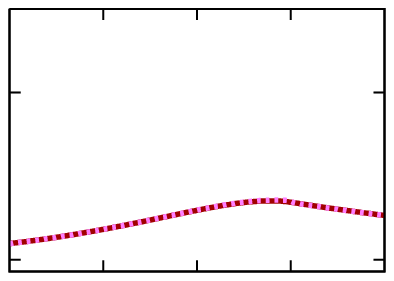}
}
\end{center}

\vskip0.6cm

\begin{center}
%\hskip-18cm
\scalebox{1.}
{
\input{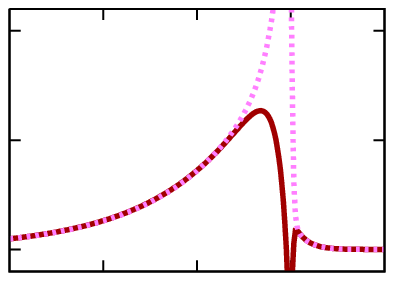}
}
\hskip1.25cm
\scalebox{1.}
{
\input{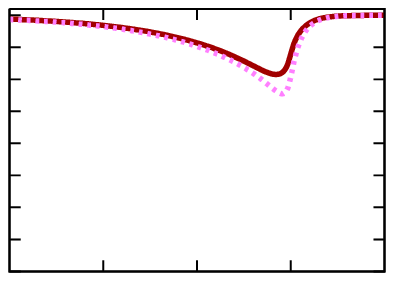}
}
\hskip1.25cm
\scalebox{1.}
{
\input{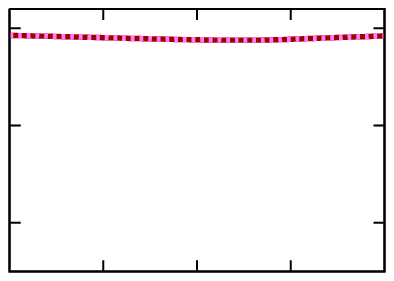}
}
\end{center}
\vskip.6cm
% \caption{
% \label{fig:supergauge:tanbeta1pt5}
% Chemical potentials over $z$ at the fiducial point but $\tan\beta=1.5$, illustrating
% supergauge-equilibrium. The key for the top right panel is
% $\mu_{H_1}$ (red, solid), $\mu_{H_2}$ (green, dashed),
% $\mu_{\widetilde H}$ (pink, dotted). For the other panels, the key is
% $\mu_{q,\ell,t,b,\tau}$ (red, solid),
% $\mu_{\widetilde q,\widetilde \ell,\widetilde t,\widetilde b,\widetilde \tau}$
% (pink, dotted).}
% \end{figure}
% 
% \vskip.9cm
% \begin{figure}[h]
\begin{center}
%\hskip-18cm
\scalebox{1.}
{
\input{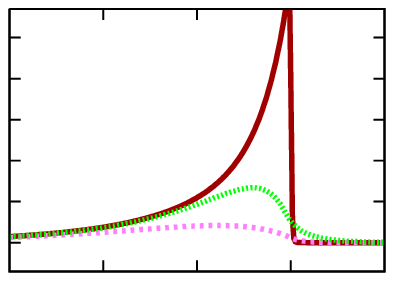}
}
\hskip1.25cm
\scalebox{1.}
{
\input{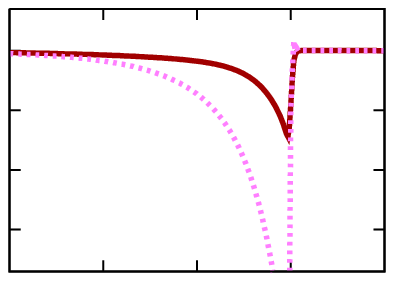}
}
\hskip1.25cm
\scalebox{1.}
{
\input{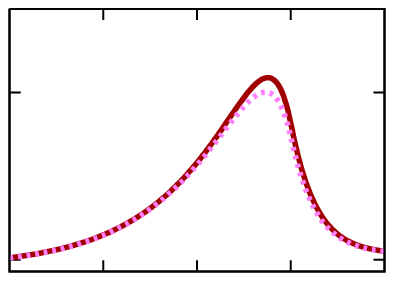}
}
\end{center}

\vskip0.6cm

\begin{center}
%\hskip-18cm
\scalebox{1.}
{
\input{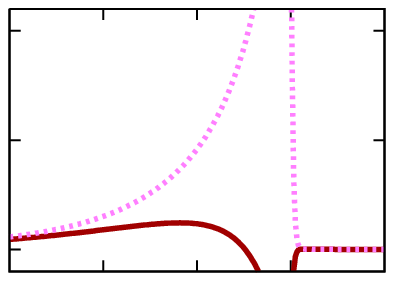}
}
\hskip1.25cm
\scalebox{1.}
{
\input{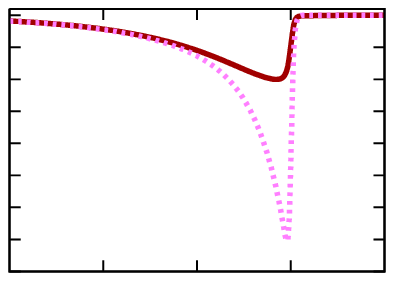}
}
\hskip1.25cm
\scalebox{1.}
{
\input{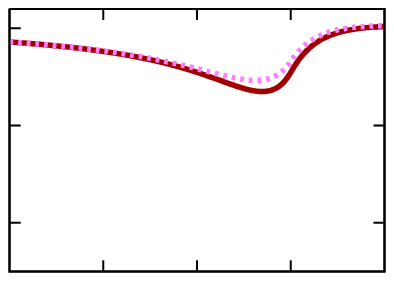}
}
\end{center}
\vskip.7cm
\caption{
\label{fig:supergauge}
Chemical potentials over $z$ 
illustrating supergauge-equilibrium. Key:
$\mu_{H_1}$ (pink, dotted), $\mu_{H_2}$ (green, dashed),
$\mu_{\widetilde H}$ (red, solid),
$\mu_{q,\ell,t,b,\tau}$ (red, solid),
$\mu_{\widetilde q,\widetilde \ell,\widetilde t,\widetilde b,\widetilde \tau}$
(pink, dotted).}
\end{figure}

%% file: fig-ipyukawa.tex
\begin{figure}
\vskip.9cm
\begin{center}
%\hskip-18cm
\scalebox{1.}
{
\input{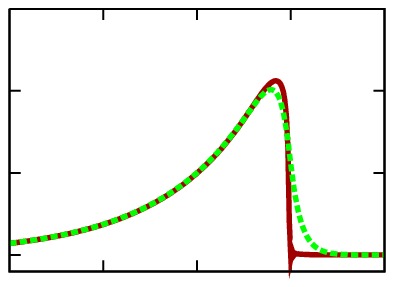}
}
\hskip1.25cm
\scalebox{1.}
{
\input{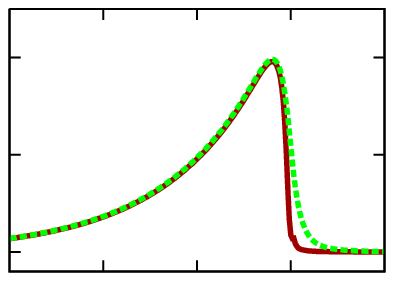}
}
\hskip1.25cm
\scalebox{1.}
{
\input{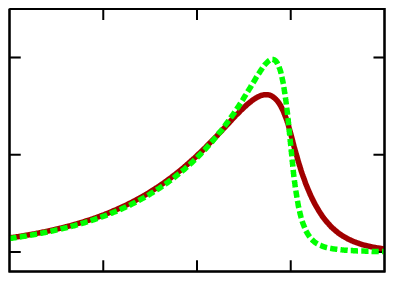}
}
\end{center}
\vskip.6cm
\begin{center}
%\hskip-18cm
\scalebox{1.}
{
\input{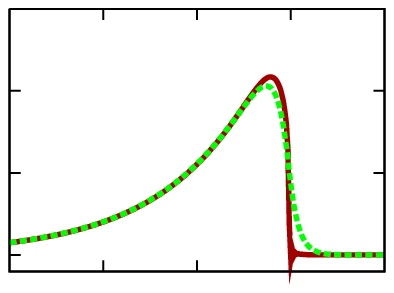}
}
\hskip1.25cm
\scalebox{1.}
{
\input{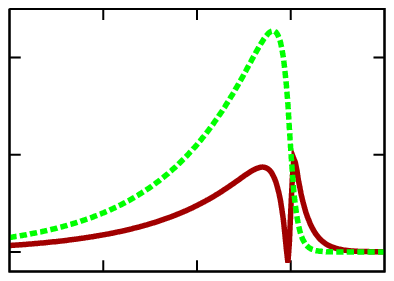}
}
\hskip1.25cm
\scalebox{1.}
{
\input{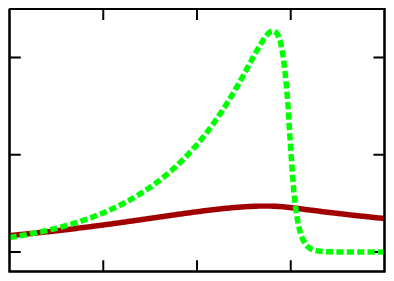}
}
\end{center}
\vskip.6cm
\begin{center}
%\hskip-18cm
\scalebox{1.}
{
\input{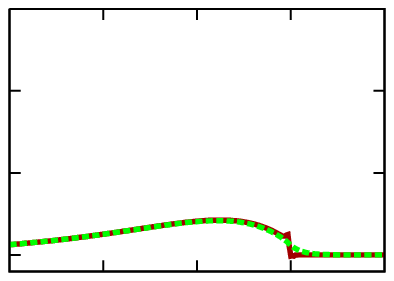}
}
\hskip1.25cm
\scalebox{1.}
{
\input{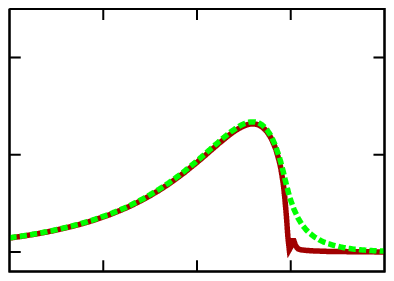}
}
\hskip1.25cm
\scalebox{1.}
{
\input{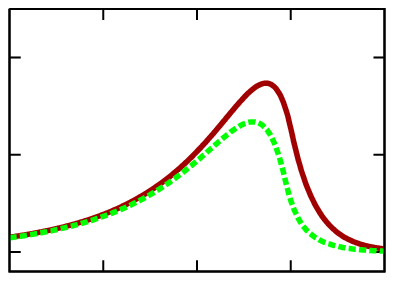}
}
\end{center}
\vskip.7cm
\caption{
\label{fig:yukawa}
Chemical potentials over $z$, illustrating
Yukawa-equilibrium. The key is $\mu_t-\mu_q$, $\mu_q-\mu_b$, $\mu_\ell-\mu_\tau$
(red, solid), $\mu_{H_1}$, $\mu_{H_2}$ (green, dashed).}
\end{figure}

%% file: fig-tanbeta.tex
\begin{figure}
\vskip.7cm
\begin{center}
%\hskip-18cm
\scalebox{1.}
{
\input{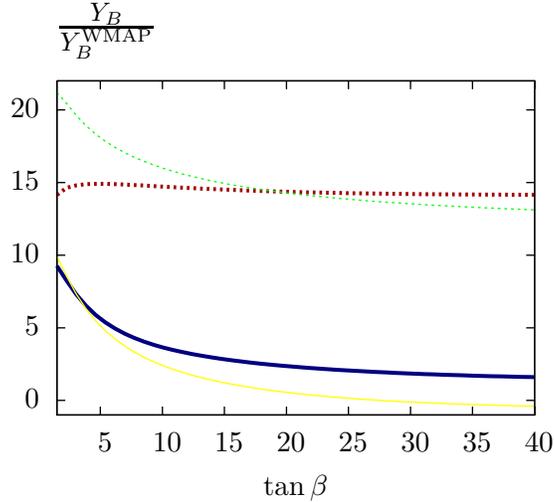}
}
\end{center}
\vskip.7cm
\caption{
\label{fig:BAU:TANBETA}
BAU over $\tan\beta$ taking account of all species (blue, thick, solid) and in
the hypothetical cases where lepton densities are not taken into account
(red, thick, dotted), where superequilibrium is enforced (yellow, thin, solid),
where supergauge interactions are not taken into account (green, thin, dotred).}
\end{figure}

%% file: fig-ipreducedtanbeta1pt5.tex
\begin{figure}[h]
\vskip.5cm
\begin{center}
\scalebox{1.}
{
\input{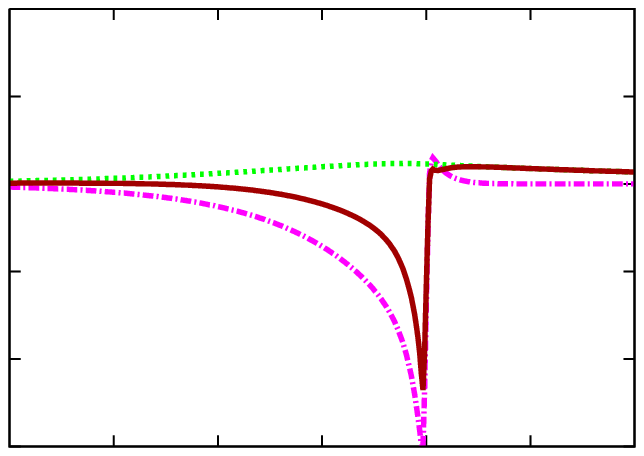}
}
\hskip1.5cm
\scalebox{1.}
{
\input{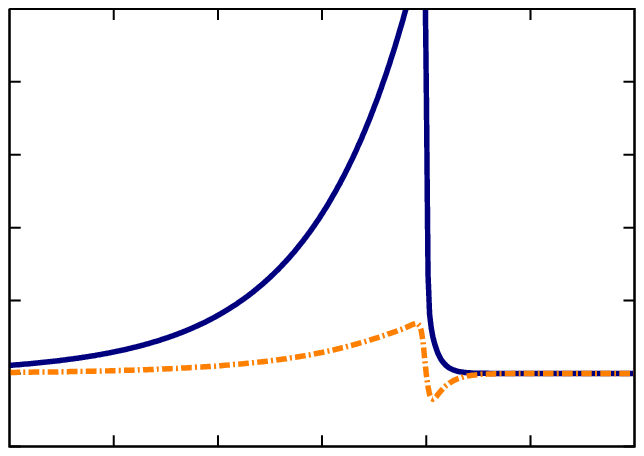}
}
\end{center}
\vskip.5cm
\caption{
\label{fig:N:tanbeta1pt5}
Charge densities over $z$, at the fiducial point but with $\tan\beta=1.5$. Left panel: $q_3$ (pink, dot-dashed), $\ell_3$ (green, dotted),
$n_{\rm left}$ (red, solid). Right panel: $H=H_1+H_2+\widetilde H$ (blue, solid),
$q_1+q_2$ (orange, dot-dashed).}
\end{figure}

%% file: fig-ipsupergauge-additional.tex
\begin{figure}
\vskip -.5cm
\begin{center}
%\hskip-18cm
\scalebox{1.}
{
\input{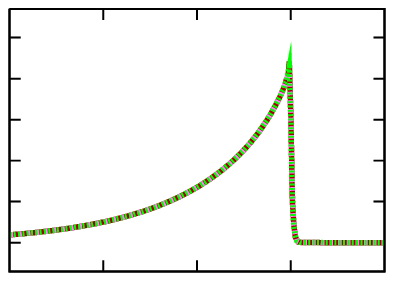}
}
\hskip1.25cm
\scalebox{1.}
{
\input{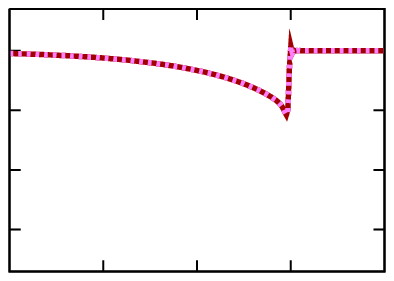}
}
\hskip1.25cm
\scalebox{1.}
{
\input{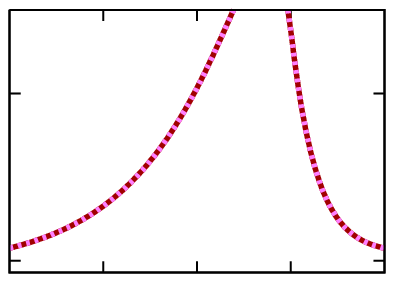}
}
\end{center}

\vskip0.6cm

\begin{center}
%\hskip-18cm
\scalebox{1.}
{
\input{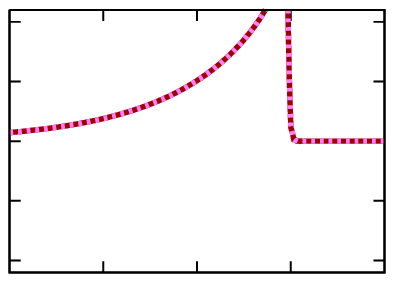}
}
\hskip1.25cm
\scalebox{1.}
{
\input{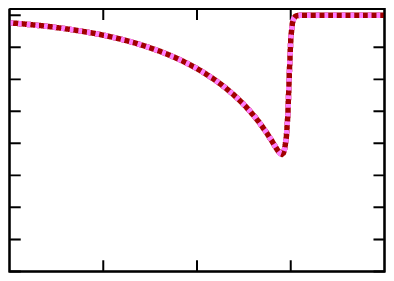}
}
\hskip1.25cm
\scalebox{1.}
{
\input{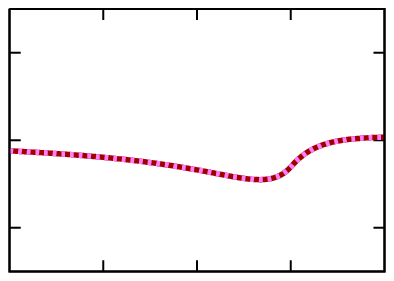}
}
\end{center}
\vskip.6cm
\begin{center}
%\hskip-18cm
\scalebox{1.}
{
\input{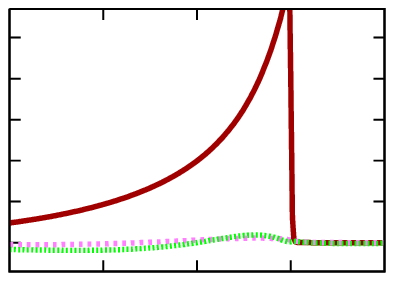}
}
\hskip1.25cm
\scalebox{1.}
{
\input{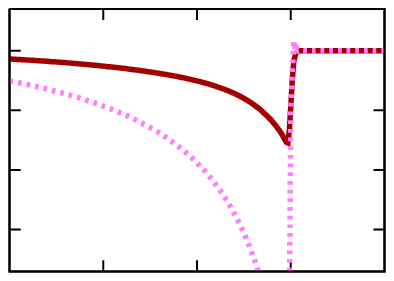}
}
\hskip1.25cm
\scalebox{1.}
{
\input{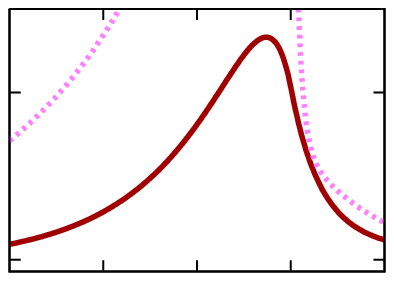}
}
\end{center}

\vskip0.6cm

\begin{center}
%\hskip-18cm
\scalebox{1.}
{
\input{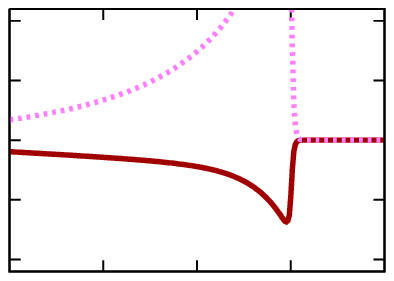}
}
\hskip1.25cm
\scalebox{1.}
{
\input{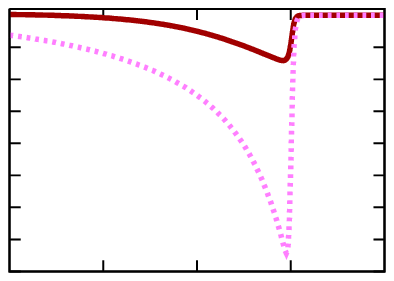}
}
\hskip1.25cm
\scalebox{1.}
{
\input{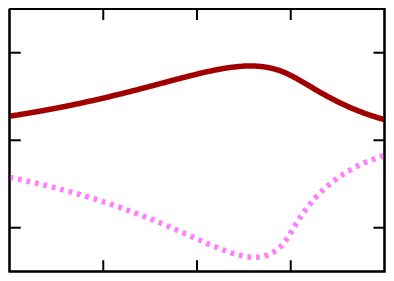}
}
\end{center}
\vskip.7cm
\caption{
\label{fig:supergauge-additional}
Chemical potentials over $z$ 
illustrating supergauge-(non-)equilibrium for the case when supergauge
interactions are infinitely slow and infinitely fast, respectively. Key:
$\mu_{H_1}$ (pink, dotted), $\mu_{H_2}$ (green, dashed),
$\mu_{\widetilde H}$ (red, solid),
$\mu_{q,\ell,t,b,\tau}$ (red, solid),
$\mu_{\widetilde q,\widetilde \ell,\widetilde t,\widetilde b,\widetilde \tau}$
(pink, dotted).}
\end{figure}

%% file: fig-ipreducednosg.tex
\begin{figure}[h]
\vskip.5cm
\begin{center}
\scalebox{1.}
{
\input{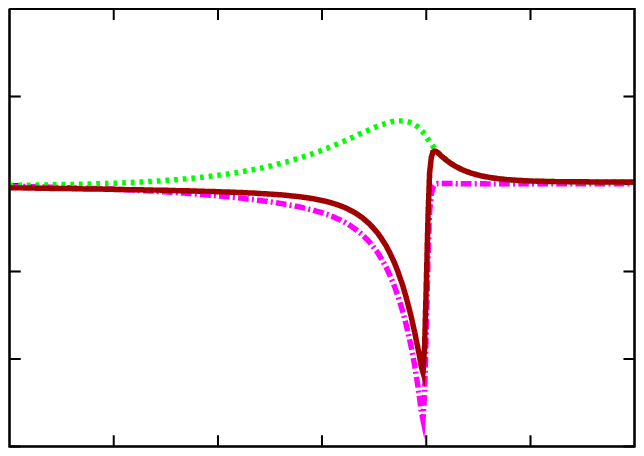}
}
\hskip1.5cm
\scalebox{1.}
{
\input{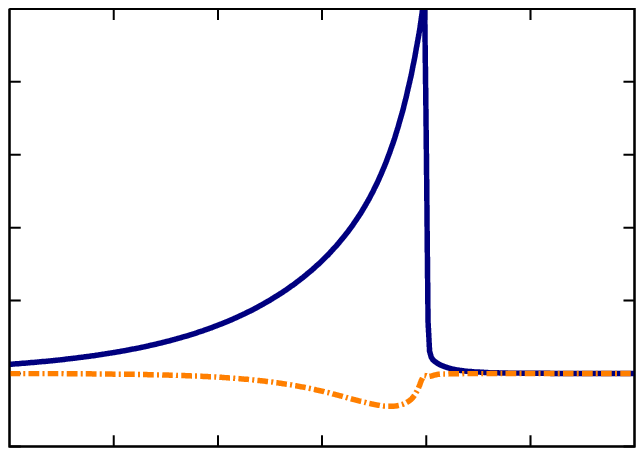}
}
\end{center}
\vskip.5cm
\caption{
\label{fig:N:nosg}
Charge densities over $z$, at the fiducial point but without supergauge interactions. Left panel: $q_3$ (pink, dot-dashed), $\ell_3$ (green, dotted),
$n_{\rm left}$ (red, solid). Right panel: $H=H_1+H_2+\widetilde H$ (blue, solid),
$q_1+q_2$ (orange, dot-dashed).}
\end{figure}